\documentclass[prb,reprint,twocolumn,amsmath,amssymb,superscriptaddress]{revtex4-1}
\usepackage{graphicx}
\usepackage{graphics}
\usepackage{multirow}
\usepackage{fancyhdr, ifpdf}
\usepackage{amsxtra}
\usepackage{stmaryrd}
\usepackage{mathrsfs}
\usepackage{psfrag}
\usepackage{epsfig}
\usepackage{rotating}
\usepackage{setspace}
\usepackage{float}
\usepackage{amsfonts}
\usepackage{amsbsy}
\usepackage{amscd}
\usepackage{amsthm}
\usepackage{color}
\usepackage{tabularx}
\usepackage{caption}
\usepackage{subcaption}
\usepackage{sidecap}
\usepackage{dcolumn}
\usepackage{esint}
\usepackage{mathalfa}
\usepackage[ruled]{algorithm2e}
\usepackage{algorithmic}
\usepackage{ulem}
\usepackage[explicit]{titlesec}
\newcommand{\parder}[2]{\frac{\partial #1}{\partial #2}}

\newcommand{\bzero}{\mathbf{0}}
\newcommand{\del}{\nabla}
\newcommand{\bPsi}{\boldsymbol{\Psi}}

\newcommand{\btau}{\boldsymbol{\tau}}
\newcommand{\bb}{\boldsymbol{b}}

\newcommand{\bff}{\boldsymbol{f}}

\newcommand{\bk}{\boldsymbol{k}}

\newcommand{\bn}{\boldsymbol{n}}

\newcommand{\br}{\boldsymbol{\textbf{r}}}
\newcommand{\bs}{\boldsymbol{s}}

\newcommand{\bx}{\boldsymbol{\textbf{x}}}
\newcommand{\by}{\boldsymbol{\textbf{y}}}

\newcommand{\bC}{\boldsymbol{C}}
\newcommand{\bE}{\boldsymbol{\textbf{E}}}
\newcommand{\bF}{\boldsymbol{F}}
\newcommand{\bH}{\boldsymbol{\textbf{H}}}
\newcommand{\bI}{\boldsymbol{I}}
\newcommand{\bK}{\boldsymbol{\textbf{K}}}

\newcommand{\bffbar}{\bar{\bff}}
\newcommand{\fbar}{\bar{f}}
\newcommand{\bM}{\boldsymbol{\textbf{M}}}
\newcommand{\bL}{\boldsymbol{\textbf{L}}}

\newcommand{\bQ}{\boldsymbol{Q}}
\newcommand{\bR}{\boldsymbol{\textbf{R}}}
\newcommand{\bS}{\boldsymbol{\textbf{S}}}
\newcommand{\bT}{\boldsymbol{T}}
\newcommand{\bU}{\boldsymbol{U}}
\newcommand{\bUbar}{\boldsymbol{\bar{U}}}
\newcommand{\bV}{\boldsymbol{V}}

\newcommand{\bPhi}{\boldsymbol{\Phi}}
\newcommand{\bPhibar}{\bar{\boldsymbol{\Phi}}}
\newcommand{\bPsibar}{\bar{\boldsymbol{\Psi}}}

\newcommand{\bGam}{\boldsymbol{\Gamma}}
\newcommand{\bGambar}{\boldsymbol{\bar{\Gamma}}}
\newcommand{\bStrain}{\boldsymbol{\mathcal{E}}}
\newcommand{\Gammabar}{\bar{\Gamma}}
\newcommand{\varphibar}{\bar{\varphi}}

\newcommand{\ubarn}{\bar{u}}

\newcommand{\bkappa}{\boldsymbol{\kappa}}
\newcommand{\phibar}{\bar{\phi}}
\newcommand{\psibar}{\bar{\psi}}

\newcommand{\rhobar}{\bar{\rho}}

\newcommand{\Vbar}{\bar{V}}

\newcommand{\intd}{\int \displaylimits}
\newcommand{\fintd}{\fint \displaylimits}

\newcommand{\Omegaper}{\Omega_\text{p}}
\newcommand{\dx}{\,d\bx}
\newcommand{\dy}{\,d\by}

\newcommand{\dk}{\,d\bk}

\newcommand{\Rthree}{ {\mathbb{R}^{3}} }

\newcommand{\order}{\mathcal{O}}

\newcommand{\ket}[1]{\left| #1 \right>} 
\newcommand{\bra}[1]{\left< #1 \right|} 
\newcommand{\chiepsx}{\,\btau^{\varepsilon}(\bx)\,}
\newcommand{\chiepsy}{\,\btau^{\varepsilon}(\by)\,}
\newcommand{\chieps}{\,\btau^{\varepsilon}}
\newcommand{\bdir}{\boldsymbol{\Upsilon}}
\newcommand{\dir}{\Upsilon}
\newcommand{\derveps}{\frac{d}{d\varepsilon}}
\newcommand{\atzero}{|_{\varepsilon = 0}}
\newcommand{\atzerob}{\bigg|_{\varepsilon = 0}}
\newcommand{\atzerobb}{\Bigg|_{\varepsilon = 0}}

\newcommand{\detb}[1]{\,\det\left(#1\right)}
\newcommand{\vself}[1]{\bar{V}^{#1}_{\tilde{\delta}}}
\newcommand{\invSzero}[1]{S^{0^{{\scriptstyle -1}}}_{#1}}
\newcommand{\invSeps}[1]{S^{\eps^{{\scriptstyle -1}}}_{#1}}
\newcommand{\invbSeps}{\bS^{\eps^{{\scriptstyle -1}}}}
\newcommand{\phiepsconj}[1]{\phibar^{\eps^{\scriptstyle *}}_{#1}}
\newcommand{\uepsconj}[1]{\ubarn^{\eps^{\scriptstyle *}}_{#1}}
\newcommand{\uzeroconj}[1]{\ubarn^{0^{\scriptstyle *}}_{#1}}
\newcommand{\phizeroconj}[1]{\phibar^{0^{\scriptstyle *}}_{#1}}
\newcommand{\gammaphieps}[1]{\Gammabar^{\phi^{\scriptstyle \eps}}_{#1}}
\newcommand{\gammaphizero}[1]{\Gammabar^{\phi^{\scriptstyle 0}}_{#1}}
\newcommand{\eps}{\varepsilon}
\newcommand{\dirac}{\tilde{\delta}}
\setcitestyle{citesep={,}}

\definecolor{hellgruen}{rgb}{0.2,0.7,0.2}

\titleformat{\paragraph}[runin]
  {\normalfont\normalsize\bfseries}{}{11pt}{\theparagraph\hspace*{1em}#1:}
\titleformat{name=\paragraph,numberless}[runin]
  {\normalfont\normalsize\bfseries}{}{11pt}{#1:}
\bibpunct{[}{]}{;}{n}{}{}
\begin{document}
\preprint{}
 \title{Configurational forces in electronic structure calculations using Kohn-Sham density functional theory}
\author{Phani Motamarri}
\affiliation{Department of Mechanical Engineering, University of Michigan, Ann Arbor, MI 48109, USA}
\author{Vikram Gavini}
\affiliation{Department of Mechanical Engineering, University of Michigan, Ann Arbor, MI 48109, USA}
\affiliation{Department of Materials Science and Engineering, University of Michigan, Ann Arbor, MI 48109, USA}
\begin{abstract}
We derive the expressions for configurational forces in Kohn-Sham density functional theory, which correspond to the generalized variational force computed as the derivative of the Kohn-Sham energy functional with respect to the position of a material point $\bx$. These configurational forces that result from the inner variations of the Kohn-Sham energy functional provide a unified framework to compute atomic forces as well as stress tensor for geometry optimization. Importantly, owing to the variational nature of the formulation, these configurational forces inherently account for the Pulay corrections. The formulation presented in this work treats both pseudopotential and all-electron calculations in single framework, and employs a local variational real-space formulation of Kohn-Sham DFT expressed in terms of the non-orthogonal wavefunctions that is amenable to reduced-order scaling techniques. We demonstrate the accuracy and performance of the proposed configurational force approach on benchmark  all-electron and pseudopotential calculations conducted using higher-order finite-element discretization. To this end, we examine the rates of convergence of the finite-element discretization in the computed forces and stresses for various materials systems, and, further, verify the accuracy from finite-differencing the energy. Wherever applicable, we also compare the forces and stresses with those obtained from Kohn-Sham DFT calculations employing plane-wave basis (pseudopotential calculations) and Gaussian basis (all-electron calculations). Finally, we verify the accuracy of the forces on large materials systems involving a metallic aluminum nanocluster containing 666 atoms and an alkane chain containing 902 atoms, where the Kohn-Sham electronic ground state is computed using a reduced-order scaling subspace projection technique~\cite{motam2014}.

\end{abstract}
\maketitle
\section{Introduction}\label{sec:intro}
Electronic structure calculations based on density functional theory (DFT) have played an important role in providing significant insights into a wide variety of materials properties. The Kohn-Sham approach~\cite{kohn64, kohn65} to DFT provides an efficient formulation to compute the ground-state properties of materials systems by reducing the many-body Schr\"odinger problem of interacting electrons into an equivalent problem of noninteracting electrons in an effective mean field that is governed by the electron-density. While this effective single-electron formulation has no approximations, and is exact, in principle, the quantum mechanical interactions between electrons manifest in the form of an unknown exchange-correlation functional, which is modeled in practice. While improving the accuracy of the exchange-correlation functional is still an active area of research, the widely used models for the exchange-correlation functional~\cite{XCReview2005} have shown to predict a range of materials properties across various materials systems with good accuracy.

An important aspect of electronic structure calculations using DFT is the determination of accurate quantum mechanical forces on the atomic nuclei and stress tensor in periodic systems. The forces and stresses are used in many aspects of electronic structure calculations, including geometry optimization, calculation of the dynamical matrix and phonon spectra, determination of elastic properties and ab-initio molecular dynamics calculations. In DFT, the atomic force on a nuclei can be computed by taking recourse to the Hellmann-Feynman theorem~\cite{hell,feyn}, which relates the force acting on a nucleus to the electrostatic field at the nucleus due to the electron-density and other nuclei~\cite{rmartin}. However, in practical DFT calculations, the variational force can differ from the Hellmann-Feynman force if the numerical basis sets employed in the DFT calculations explicitly depend on atomic positions ~\cite{pulay1987,fahnle1995,bendt1983} (Pulay force correction due to incomplete basis-set), and, moreover, the Hellmann-Feynman theorem does not provide elastic stresses on periodic systems. Stress is computed as the derivative of the energy with respect to strain, and the practical numerical implementation of these derivatives relies very much on the particular basis set type employed in the DFT calculation and has to account for possible strain dependence on the basis sets employed, which manifest as Pulay stresses. To this end, there have been significant efforts in the past few decades to implement variationally consistent forces~\cite{izc79,VASP96,svb85,fas89,Qchem,SIESTA,FHI-aims,ss92,mv93,sw89,sw90,sw93,ysk91,gm92,PARSEC1,linlin2017} and  stresses~\cite{no2011,nm83,nm85,tas2002,kj99,tfbzg2008,ks2000,doll2004,SIESTA} in ab-initio codes employing various basis sets. 

We note that the general approach to computing atomic forces or stresses in the existing numerical implementations rely on the outer variations of the Kohn-Sham energy functional with respect to the change in the position of atoms (for computing forces) or the lattice vectors (for computing stresses). Thus, most of the approaches in the literature rely on separate formulations for computing forces and stresses. In this work, we propose a configurational force approach to compute both atomic forces and stresses in periodic systems under a single framework using Kohn-Sham density functional theory. The idea of configurational force dates back to Eshelby's formulation~\cite{eshelby1951}, and is widely used in continuum physics~\cite{gurtin2000} to study problems involving material inhomogenities like coherent phase interfaces, evolving surfaces, cracks etc., in materials. In contrast to the existing methods employed in electronic structure calculations, the proposed approach is based on the inner variations of the Kohn-Sham energy functional resulting in configurational forces. These forces correspond to the generalized variational derivative of the Kohn-Sham energy functional with respect to the position of a material point $\bx$. Hence, the proposed approach provides a generalized force with respect to internal positions of atoms as well as the periodic cell, in the case of periodic calculations, thus providing a unified framework to compute atomic forces as well as stresses for geometry optimization. Moreover, the proposed formulation allows us to treat both pseudopotential and all-electron calculations in a single framework. Furthermore, we note that the configurational forces inherently account for Pulay corrections owing to the variational nature of the formulation, and no separate treatment is required to account for these corrections.

The proposed configurational force approach provides a natural way to evaluate atomic forces and stresses in the context of finite-element discretization. The finite-element basis ~\cite{Brenner-Scott}, a piecewise polynomial basis, offers some unique advantages for Kohn-Sham DFT calculations. Finite-element basis can handle arbitrary boundary conditions, which enables the consideration of isolated, semi-periodic, as well as, periodic systems under a single framework. Further, the finite-element basis is amenable to adaptive spatial resolution, which can be exploited for handling all-electron DFT calculations~\cite{bylaska,lehtovaara,motam2013,motam2014,motam2017} as well as the development of coarse-graining techniques~\cite{QCOFDFT,Bala2010,Ponga2016}. Moreover, the locality of these basis functions provides good scalability on parallel computing platforms. Though significant efforts have been focused in the recent past to develop finite element based methodologies for conducting Kohn-Sham DFT calculations~\cite{Hermansson,white,tsuchida1995,tsuchida1996,ACRES,tsuchida1998,pask1999,Batcho2000,pask2001,pask2005,Zhou2008,PUFE,bylaska,lehtovaara,surya2010,Lin2012,Zhou2012,Bao2012,Masud2012,motam2012,motam2013,motam2014,motam2017,bikash2017}, most of these efforts have been confined to computing the electronic ground-state. To the best of our knowledge, expressions for computing both stresses and forces that are variationally consistent have not been developed for the finite-element discretization. The proposed work fills this important gap to enable large-scale electronic structure calculations using finite-element basis.  Further, the developed configurational forces also provide the generalized forces corresponding to the nodal positions of the underlying finite-element triangulation, which can in turn be effectively used to determine the optimal positions of the finite element nodes--- $r$-adaptivity, an {\it a posteriori} mesh adaption technique ~\cite{thoutireddy}. 

In this work, the proposed configurational force approach is presented in the context of spectral finite-element discretization. However, the method presented is general, and applicable to any real-space basis employed in the solution of the Kohn-Sham DFT problem. Keeping in mind the reduced-order scaling methods~\cite{carlos,motam2014}, the formulation has been developed using non-orthogonal electronic wavefunctions. We begin by considering the formulation of the Kohn-Sham density functional theory at finite temperature in terms of non-orthogonal wavefunctions, where the electronic ground state, for a given position of atoms, is governed by a variational problem involving the minimization of the Kohn-Sham free energy functional with respect to the wavefunctions and the density matrix subject to the constraint on the total number of electrons in the system. Subsequently, we discuss the local variational real-space formulation of Kohn-Sham DFT, where the computation of the Kohn-Sham electronic ground-state for a given position of atoms can be formulated as a local variational problem involving wavefunctions, density matrix and electrostatic potentials. We then derive the expressions for the configurational forces corresponding to geometry optimization. Using the approach of inner variations, we evaluate the generalized forces from the locally reformulated energy functional corresponding to perturbations of the underlying space. We note that the final form of the configurational force expression involves integrals of Eshelby tensors contracted with the gradient of the generator ($\bdir$) associated with the perturbation of the underlying space. The derived configurational forces provide a unified framework to compute both forces on atoms and stresses in periodic systems. To elaborate, the force on any given atom is given by the configurational force computed using a generator whose compact support includes only the atom of interest, while the stress tensor is computed by using a generator corresponding to affine deformations of the underlying space. Subsequently, we present the details involved in the numerical implementation of the configurational force expressions within the framework of finite-element discretization, and thereby discuss a computationally efficient strategy to conduct geometry optimization using finite-element basis. 

Finally, we present the accuracy of the proposed configurational force approach on sample benchmark problems. To this end, we consider both non-periodic and periodic material systems, and conduct both pseudopotential and all-electron calculations. We begin by examining the rates of convergence of the finite-element approximation in the computed forces and stresses. We observe a close to $\order(h^{2k-1})$ convergence in all benchmark problems. Further, in order to asses the variational nature of the computed forces and stresses, we compare these with those obtained from finite differencing energies, which are in excellent agreement. As a further test, we compute the forces and stresses as a function of bond length or lattice parameter and compare these with the derivative of polynomial fits to the energy dependence on these parameters. These extensive tests have ascertained the variational nature of the configurational forces in our numerical implementation. We also compare the accuracies of forces and stresses obtained from the proposed configurational force approach with those obtained using external DFT packages, and find very good agreement. Finally, we consider large materials systems comprising of an aluminum nanocluster containing $5 \times 5 \times 5$ unit-cells ($666$ atoms) and an alkane chain containing $902$ atoms, where the finite-element discretized Kohn-Sham problem is solved using the reduced order scaling subspace projection method~\cite{motam2014} involving non-orthogonal localized wavefunctions. We find that the configurational forces computed using the reduced-order scaling method are in very good agreement with those computed without the additional approximations used in the reduced-order scaling method, with differences being well within the desired chemical accuracy. 

The remainder of the paper is organized as follows: Section II describes the  formulation of the Kohn-Sham DFT problem in the context of non-orthogonal wavefunctions, followed by the local real-space reformulation of electrostatic interactions in Section III. The configurational force expressions for both non-periodic and periodic cases are presented in Section IV, with the details of the derivation presented in the Appendix. Section V describes the numerical implementation of forces and stresses in the context of finite-element discretization. Section VI presents the results on various systems demonstrating the accuracy and variational nature of the computed forces and stresses. Finally, we conclude with a summary and outlook in Section VII.

\section{Kohn-Sham density functional theory}\label{sec:formulation}
Consider a materials system consisting of $N_e$ electrons with $N_a$ nuclei whose position vectors are denoted by $\bR=\{\bR_1, \bR_2, \bR_3, \cdots \bR_{N_a}\}$. Neglecting spin, the free energy of the system in Kohn-Sham density functional theory(DFT) ~\cite{kohn64,kohn65} at finite temperature ~\cite{mermin65} can be written as 
\begin{equation}\label{free_energy}
\begin{split}
&\mathcal{F}(\bGam^{\phi},\bPhi,\bR) \\
&= T_{\text{s}}(\bGam^{\phi},\bPhi) + E_{\text{xc}}(\rho) + E_{\text{el}}(\bGam^{\phi},\bPhi,\bR) - E_{\text{ent}}(\bGam^{\phi}) \,,
\end{split}
\end{equation}
where $\bPhi = \{\phi_1(\bx),\phi_2(\bx),\cdots,\phi_{N}(\bx)\}$ ($N>N_e/2$) denotes the set of electronic wavefunctions. In all generality, we assume that these wavefunctions are non-orthogonal in order to realize a reduced-order scaling numerical implementation of DFT~\cite{carlos,motam2014}. Here $\bGam^{\phi}$ denotes the matrix corresponding to the single particle density operator ($\hat{\Gamma}$) expressed in the non-orthogonal basis $\bPhi$, i.e. $\Gamma^{\phi}_{ij}=\sum_{k=1}^{N}S^{-1}_{ik}\bra{\phi_{k}}\hat{\Gamma}\ket{\phi_{j}}$, where $S^{-1}_{ik}$ are the matrix elements of the inverse of the overlap matrix given by
\begin{equation}
S_{ij} = \int \phi^{*}_i(\bx)\phi_j(\bx) \dx \,.
\end{equation}
We note that the superscript * in the above, and all the equations subsequently, denotes the complex conjugate. The electron-density $\rho$ in equation~\eqref{free_energy} can be expressed in terms of the density matrix and the non-orthogonal wavefunctions as
\begin{equation}\label{rhoexp}
\rho(\bx) = 2 \sum_{i,j,k=1}^{N}\Gamma^{\phi}_{ij} \,S^{-1}_{jk} \,\phi_{k}^{*}(\bx)\, \phi_i(\bx) \,.
\end{equation}
Further, $T_{\text{s}}$, which denotes the kinetic energy of non-interacting electrons, is given by
\begin{equation}\label{kin}
T_{\text{s}}(\bGam^{\phi},\bPhi) = 2\,\sum_{i,j,k=1}^{N}\int \Gamma^{\phi}_{ij}\, S^{-1}_{jk}\, \phi_{k}^{*}(\bx) \left(-\frac{1}{2} \del^{2} \right)\,\phi_i(\bx) \dx \,.
\end{equation}
The exchange-correlation energy, which incorporates all the quantum-mechanical interactions, is denoted by $E_{\text{xc}}(\rho)$. While the explicit form of $E_{\text{xc}}(\rho)$ remains elusive, various approximations have been developed over the past decades, with the local density approximation (LDA)~\cite{rmartin} and generalized gradient approximation (GGA)~\cite{gga1,gga2} being adopted across a range of materials systems. In the present work, LDA exchange-correlation energy is adopted which has the following functional form:
\begin{equation}\label{exc}
E_{\text{xc}}(\rho) = \int F(\rho) \dx = \int \varepsilon_{\text{xc}}(\rho) \rho(\bx) \dx \,,
\end{equation}
where $\varepsilon_{\text{xc}}(\rho) = \varepsilon_x(\rho) + \varepsilon_c(\rho)$. In particular, we employ the Slater exchange and Perdew-Zunger~\cite{alder,perdew} form of the correlation functional. 

The electrostatic interaction energy, $E_{\text{el}}$, represents the interactions between electrons and nuclei, which can be further decomposed as
\begin{align}\label{electro}
E_{\text{el}}(\bGam^{\phi},\bPhi,\bR) = E_{\text{H}}(\rho) + E_{\text{ext}}(\bGam^{\phi},\bPhi,\bR)  + E_{\text{zz}}(\bR) \,,
\end{align}
where the first term $E_{\text{H}}$ is the Hartree energy representing the electrostatic interaction energy between electrons, and $E_{\text{zz}}$ denotes the repulsive energy between nuclei.  These are given by
\begin{equation}\label{EHartRep}
E_{\text{H}} = \frac{1}{2}\int\int\frac{\rho(\bx)\rho(\by)}{|\bx - \by|} \,\dx\,\dy ,\;\; E_{\text{zz}}= \frac{1}{2}\sum_{\substack{I,J \neq I}} \frac{Z_I Z_J}{|\bR_I-\bR_J|}\,,
\end{equation}
with $Z_I$ denoting the charge on the $I^{th}$ nucleus. $E_{\text{ext}}$ in equation~\eqref{electro} denotes the classical interaction energy between electrons and nuclei, and is given by
\begin{equation}\label{allElecExt}
E_{\text{ext}} = - \sum_{J}\int \rho(\bx)  \frac{Z_J}{|\bx-\bR_J|}\dx\,.
\end{equation}
As chemical bonding in many material systems is not influenced by the tightly bound core electrons close to the nucleus of an atom, these core electrons may not play a significant role in governing many materials properties. Hence, the pseudopotential approach is commonly adopted, where only wavefunctions for the valence electrons are computed in an effective potential of the nucleus and core electrons given by a pseudopotential. The pseudopotential is often defined by the operator $\mathcal{V}_{\text{PS}} = \mathcal{V}_{\text{loc}} + \mathcal{V}_{\text{nl}}$, where $\mathcal{V}_{\text{loc}}$ is the local part of the pseudopotential operator and $\mathcal{V}_{\text{nl}}$ is the nonlocal part of the operator. In pseudopotential Kohn-Sham DFT, $E_{\text{ext}}$ is given by
\begin{equation}\label{pseudoEext}
E_{\text{ext}} = 2 \sum_{i,j,k=1}^{N} \int \int \Gamma^{\phi}_{ij} \,S^{-1}_{jk} \,\phi_{k}^{*}(\bx)V_{\text{PS}}(\bx,\by,\bR)\, \phi_i(\by) \dy\dx\,.
\end{equation}
The norm-conserving Troullier-Martins pseudopotential~\cite{tm91} in the Kleinman-Bylander form~\cite{bylander82} is employed in this work, where the action of the operator $\mathcal{V}_{\text{PS}}$ on the wavefunction $\phi_k(\bx)$ is given by
\begin{align}
&\int V_{\text{loc}}(\bx,\by,\bR) \phi_i(\by)\dy = \sum_{J} V^{J}_{\text{loc}}(|\bx-\bR_J|)\phi_i(\bx)\,,\label{locPS}\\
&\int V_{\text{nl}}(\bx,\by,\bR)\phi_i(\by)\dy \notag\\
&= \sum_{J} \sum_{lm} C^{J,i}_{lm}\,V^{J}_{lm}\zeta^{J}_{lm}(\bx,\bR_J)\Delta V^{J}_{l}(|\bx-\bR_J|)\,,\label{nlps}
\end{align}
with
\begin{gather*}
\Delta V^{J}_{l}(|\bx - \bR_J|) = V^{J}_{l}(|\bx - \bR_J|) - V^{J}_{\text{loc}}(|\bx - \bR_J|)\,,\\
C^{J,i}_{lm} = {\int \zeta^{J}_{lm}(\by,\bR_J) \Delta V^{J}_{l}(|\by-\bR_J|) \phi_i(\by) \dy}\,, \label{Clm} \\
\frac{1}{V^{J}_{lm}} = {\int \zeta^{J}_{lm}(\by,\bR_J) \Delta V^{J}_{l}(|\by-\bR_J|) \zeta^{J}_{lm}(\by,\bR_J)\dy}\,\, \label{Vlm}.
\end{gather*}
In the above, $V^{J}_{\text{loc}}(|\bx-\bR_J|)$ is the local potential of atom $J$, $V^{J}_{l}(|\bx - \bR_J|)$ denotes the pseudopotential component of atom $J$ corresponding to the azimuthal quantum number $l$, and $\zeta^{J}_{lm}(\bx,\bR_J)$ is the corresponding single-atom pseudo-wavefunction with azimuthal quantum number $l$ and magnetic quantum number $m$.

In a non-periodic setting, representing an isolated atomic system, all integrals in equations ~\eqref{kin}-\eqref{nlps} are over $\Rthree$ and the summations include all atoms in the system. In the case of an infinite periodic crystal, all integrals involving $\bx$ in equations ~\eqref{kin}-\eqref{nlps} are over the unit-cell, whereas the integrals involving $\by$ are over $\Rthree$. Further, the summation over $I$ is on atoms in the unit-cell, and the summation over $J$ extends over all lattice sites.

The electronic entropy contribution, $E_{\text{ent}}$ in equation~\eqref{free_energy}, is given by
\begin{equation}\label{Eent}
E_{\text{ent}} = -2\,\sigma\; \text{tr}\left[\bGam^{\phi} \ln \bGam^{\phi} + (\bI - \bGam^{\phi})\ln(\bI - \bGam^{\phi})\right]\,,
\end{equation}
where $\sigma=k_B\,T$ with $k_B$ denoting the Boltzmann constant and $T$ denoting the electronic temperature.

Finally, the ground state in Kohn-Sham DFT is governed by the variational problem
\begin{equation}\label{groundstate}
 \min_{\bR} \min_{\bGam^{\phi},\bPhi} \mathcal{F}_c(\bGam^{\phi},\bPhi,\bR)\,,
\end{equation}
where $\mathcal{F}_c = \mathcal{F} - \mu \left[2\,\text{tr}(\bGam^{\phi}) - N_e\right]$
with $\mu$ denoting the Lagrange multiplier (Fermi-energy) enforcing the constraint on number of electrons, i.e., $2\,\text{tr}(\bGam^{\phi}) = N_e$. Equation~\eqref{groundstate} indicates that the electronic ground-state needs to be computed for every configuration of the nuclei encountered during the minimization procedure over the nuclear positions.

\section{Real-space formulation}\label{sec:RS}
In this section, we present the local variational real-space formulation of Kohn-Sham DFT, which is subsequently used to derive the expressions for configurational forces associated with internal atomic relaxations and cell relaxation. We note that the various components of the electrostatic interaction energy (cf. equation~\eqref{electro}) are non-local in real-space. These extended interactions are typically computed in Fourier space, using Fourier transforms.  However, Fourier space formulations employing plane-waves provide only uniform spatial resolution and restrict simulation domains to periodic geometries and boundary conditions. Thus, they are not well suited for material systems involving molecules, nano-clusters, or, systems containing defects. Further, the non-locality of the plane-wave basis in real-space limits the scalability of computations on parallel computing platforms.

The real-space formulation discussed here is devoid of the aforementioned limitations of a Fourier space formulation. The proposed formulation extends the recent efforts in the local reformulation of Kohn-Sham DFT~\cite{pask1999,surya2010,motam2013}, and differs from prior efforts in the way the extended electrostatic interactions have been treated. In particular, the proposed formulation is crucial to developing a unified framework to compute the configurational forces associated with both atomic relaxations and cell relaxations, discussed subsequently. We note that the treatment of extended electrostatic interactions is similar to Das et al.~\cite{sambit2015}, which was proposed in the context of orbital-free DFT formulation. Here, we provide the relevant details in the context of Kohn-Sham DFT.

Let $\tilde{\delta}(\bx - \bR_I)$ denote a regularized Dirac distribution located at $\bR_I$. Thus, the charge distribution of the $I^{\text{th}}$ nuclear charge is given by $-Z_I\tilde{\delta}(\bx - \bR_I)$. We define the nuclear charge distribution $b(\bx,\bR) = - \sum_{I}Z_I\tilde{\delta}(|\bx - \bR_I|)$ and $b(\by,\bR) = - \sum_{J} Z_J\tilde{\delta}(|\by - \bR_J|)$ to reformulate the repulsive energy $E_{zz}$ as
\begin{equation}\label{repulsive}
E_{zz} = \frac{1}{2}\int \int \frac{b(\bx,\bR)\,b(\by,\bR)}{|\bx - \by|}\dx\dy\, - E_{\text{self}} \,,
\end{equation}
where $E_{\text{self}}$ denotes the self energy of the nuclear charges which depends only on the nuclear charge distribution and can be expressed as
\begin{equation}\label{eself}
E_{\text{self}} =  -\frac{1}{2}\sum_{I}\int Z_I\tilde{\delta}(|\bx-\bR_I|)\vself I(\bx)\dx\,,
\end{equation}
with $\vself I(\bx)$ denoting the electrostatic potential corresponding to the $I^{\text{th}}$ nuclear charge $-Z_I\tilde{\delta}(\bx - 
\bR_I)$ and is given by
\begin{equation}
\vself I(\bx) =\int \frac{-Z_I\tilde{\delta}(|\by-\bR_I|)}{|\bx-\by|} \dy\,.
\end{equation}
Since the kernel corresponding to extended electrostatic interactions is the Green's function of the Laplace operator, the evaluation of self energy as well as the potential $\vself I(\bx)$ can be reformulated as the following local variational problem:
\begin{equation}\label{vselfReformulation}
\begin{split}
E_{\text{self}}&= - \sum_{I} \min_{V^{I} \in H^{1}(\Rthree)} \Bigl\{\frac{1}{8\pi} \int |\del V^{I}(\bx)|^2 \dx \\&+ \int Z_I \tilde{\delta}(|\bx - \bR_I|) V^{I}(\bx) \dx \,\Bigr\}\,,
\end{split}
\end{equation}
with $\vself I(\bx) $ being the minimizer of the variational problem in equation~\eqref{vselfReformulation} associated with the $I^{\text{th}}$ nucleus. In the preceding equation, $H^{1}(\Rthree)$ denotes the Hilbert space of functions such that the functions and their first-order derivatives are square integrable on $\Rthree$. Further, the local part of $E_{\text{ext}}$ in equation~\eqref{pseudoEext} can be rewritten as 
\begin{align}
&E_{\text{ext}}^{\text{loc}}(\rho,\bR) = \int \int \frac{\rho(\bx)b(\by,\bR)}{|\bx - \by|} \dx \dy \notag\\&+\sum_{J} \int \left(V^{J}_{\text{loc}}(|\bx-\bR_J|) - \vself J(|\bx-\bR_J|\right)\rho(\bx)\dx \,.
\end{align}
We note that $Z_{J}$ in $b(\by,\bR)$ (and $Z_{I}$ in $b(\bx,\bR)$) denote the valence charges of the $J^{\text{th}}$ (and $I^{\text{th}}$) nucleus in the case of pseudopotential calculations. While, for all-electron calculations, they denote the atomic number with $V^{J}_{\text{loc}} =   \vself J$.  The kernel corresponding to the extended electrostatic interactions in the expressions for $E_{\text{H}}$, $E^{\text{loc}}_{\text{ext}}$ and $E_{\text{zz}}$  is the Green's function of the Laplace operator. Thus, $E_{\text{H}} + E^{\text{loc}}_{\text{ext}} + E_{\text{zz}}$ can be rewritten as the following local variational problem:
\begin{widetext}
\begin{align}\label{elReformulation}
&\int\int\Biggl[\frac{1}{2}\frac{\rho(\bx)\rho(\by)}{|\bx - \by|}  
+\frac{\rho(\bx)b(\by,\bR)}{|\bx - \by|} + \frac{1}{2}\frac{b(\bx,\bR)b(\by,\bR)}{|\bx - \by|}\Biggr] \dx\dy
+\sum_{J} \int \left(V^{J}_{\text{loc}}(|\bx-\bR_J|) - \vself J(|\bx-\bR_J|)\right)\rho(\bx)\dx - E_{\text{self}} \notag \\
&= -\min_{\varphi \in \mathcal{Y}} \left\{\frac{1}{8\pi}\int |\del \varphi(\bx)|^2 \dx - \int (\rho(\bx) + b(\bx,\bR))\varphi(\bx)\dx\right\} 
+\sum_{J} \int \left(V^{J}_{\text{loc}}(|\bx-\bR_J|) - \vself J(|\bx-\bR_J|)\right)\rho(\bx)\dx - E_{\text{self}} \,,
\end{align}
\end{widetext}
where $\varphi(\bx)$ denotes the trial function for the total electrostatic potential due to the electron-density and the nuclear charge distribution, and $\mathcal{Y}$ is a suitable function space corresponding to the boundary conditions of the problem, discussed subsequently.   Using the local reformulation in equations ~\eqref{vselfReformulation} and ~\eqref{elReformulation}, the electrostatic interaction energy $E_{\text{el}}$ in DFT can now be expressed as the following variational problem:
\begin{equation}\label{electroLocal}
E_{\text{el}} = E_{\text{H}} + E_{\text{ext}} + E_{\text{zz}} = \max_{\varphi \in \mathcal{Y}}\min_{V^{I} \in H^{1}(\Rthree)} \mathcal{L}_{\text{el}}(\bGam^{\phi},\bPhi,\varphi,\mathcal{V},\bR)\,,
\end{equation}
where
\begin{equation*}
\mathcal{L}_{\text{el}}(\bGam^{\phi},\bPhi,\varphi,\mathcal{V},\bR) = \mathcal{L}^{\text{ALL}}_{\text{el}}(\rho,\varphi,\mathcal{V},\bR) + \mathcal{L}^{\text{PSP}}_{\text{el}}(\bGam^{\phi},\bPhi,\bR)
\end{equation*}
with
\begin{align}
&\mathcal{L}^{\text{ALL}}_{\text{el}}(\rho,\varphi,\mathcal{V},\bR) = \notag\\
&\int\left[-\frac{1}{8\pi} |\del \varphi(\bx)|^2 + (\rho(\bx) + b(\bx,\bR))\varphi(\bx)\right]\dx\notag\\
&+ \sum_{I} \int \left[\frac{1}{8\pi} |\del V^{I}(\bx)|^2  +  Z_I \tilde{\delta}(|\bx - \bR_I|) V^{I}(\bx) \right]\dx\,,
\end{align}
\begin{equation}
\mathcal{L}^{\text{PSP}}_{\text{el}}(\bGam^{\phi},\bPhi,\bR) = \mathcal{L}^{\text{loc}}(\rho,\bR) + \mathcal{L}^{\text{nl}}(\bGam^{\phi},\bPhi,\bR)\,, \label{lagPSP}
\end{equation}
\begin{align}
&\mathcal{L}^{\text{loc}}= \sum_{J} \int \left(V^{J}_{\text{loc}}(|\bx-\bR_J|) - \vself J(|\bx-\bR_J|)\right)\rho(\bx)\dx \notag\\ 
&\mathcal{L}^{\text{nl}} =  2 \sum_{i,j,k=1}^{N} \int \int \Gamma^{\phi}_{ij} \,S^{-1}_{jk} \,\phi_{k}^{*}(\bx)V_{\text{nl}}(\bx,\by,\bR)\, \phi_i(\by)\dx \dy \,,\notag
\end{align}
with $\mathcal{V} = \{V^{1},V^{2},\cdots,V^{N_a}\}$ denoting the vector containing the trial electrostatic potentials corresponding to all nuclear charges in the simulation domain. Further, the minimization over $V^{I}$ in equation~\eqref{electroLocal} refers to a simultaneous minimization over all these electrostatic potentials in $\mathcal{V}$. We further note that $\mathcal{L}^{\text{PSP}}_{\text{el}}(\bGam^{\phi},\bPhi,\bR) = 0$ in the case of all-electron Kohn-Sham DFT calculations. Thus, this local reformulation also provides a unified framework for both pseudopotential as well as all-electron DFT calculations.

Finally, using the local reformulation of the extended electrostatic interaction energy, the computation of  the electronic ground-state  for a given position of atoms in equation ~\eqref{groundstate} can be formulated as the following local variational problem in wavefunctions, density matrix and electrostatic potentials:
\begin{equation}\label{saddlepoint}
\mathcal{F}_0(\bR) = \min_{\bGam^{\phi}}\min_{\bPhi \in (\mathcal{Y})^{N}} \;\max_{\varphi \in \mathcal{Y}}\; \mathcal{L}(\bGam^{\phi},\bPhi,\varphi;\bR)\,,
\end{equation}
\vspace{-0.25in}
\begin{align*}
\text{where}\;\;\;\mathcal{L}(\bGam^{\phi},\bPhi,\varphi;\bR) &= \widetilde{\mathcal{L}}(\bGam^{\phi},\bPhi) 
+ \mathcal{L}_c(\bGam^{\phi})\\
&+ \min_{V^{I} \in H^{1}(\Rthree)}\mathcal{L}_{\text{el}}(\bGam^{\phi},\bPhi,\varphi,\mathcal{V};\bR)
\end{align*}
\vspace{-0.25in}
\begin{align}
\text{with}\;\;\widetilde{\mathcal{L}}(\bGam^{\phi},\bPhi) &= T_{\text{s}}(\bGam^{\phi},\bPhi)+ E_{\text{xc}}(\rho) + E_{\text{ent}}(\bGam^{\phi})\,,\label{lagrangian}\\  
\mathcal{L}_c(\bGam^{\phi}) &= -\mu \left[2\,\text{tr}(\bGam^{\phi}) - N_e\right]\,.
\end{align}
In the above, $\mathcal{Y}$ denotes a suitable function space that guarantees the existence of minimizers. Further, we remark that numerical computations involve the use of bounded domains, which in non-periodic calculations correspond to a large enough domain containing the compact support of the wavefunctions, and, in periodic calculations, correspond to the supercell. Denoting such an appropriate bounded domain by $\Omega$ subsequently, $\mathcal{Y}=H^{1}_{0}(\Omega)$ in the case of non-periodic problems, and $\mathcal{Y}=H^{1}_{per}(\Omega)$ in the case of periodic problems. We note that, in practice, the solution to the variational problem in equation~\eqref{saddlepoint} is computed by taking recourse to the Kohn-Sham equations, which constitutes a non-linear eigenvalue problem solved using self-consistent field iteration on the electron density (cf.~\cite{surya2010,motam2013}).

\section{Configurational forces}\label{sec:ConfigForces}
We now derive the expressions for the configurational forces corresponding to geometry optimization. To this end, we employ the approach of inner variations, where we evaluate the generalized forces corresponding to perturbations of underlying space, which provides a unified expression for the generalized force corresponding to the geometry of the simulation cell---internal atomic positions, as well as, the simulation domain boundary. Further, owing to the real-space formulation in Section~\ref{sec:RS}, the derived expressions are applicable to both pseudopotential and all-electron calculations. 

In order to derive the expressions for the configurational forces, we first define a bijective mapping $\chieps : \Rthree \rightarrow \Rthree$  which represents the infinitesimal perturbation of the underlying space, mapping a material point $\bx$ to a new point $\bx'$ such that ${\btau}^0 = \textbf{I}$. Further, we define the generator of this mapping as $\bdir = \derveps\chiepsx\atzero$. We note that the mapping $\chieps$ should be constrained to rigid body deformations in the compact support of the regularized nuclear charge distribution $\bb(\bx)$ in order to preserve the integral constraint $\int \tilde{\delta}(\bx - \bR_I) = 1$. To this end, $\chiepsx = \bQ_{I}^{\varepsilon}\bx + \bT_{I}^{\varepsilon}$ in the compact support of $\tilde{\delta}(\bx - \bR_I)$ for $I = 1 \cdots N_a$, where $\bQ_{I}^{\varepsilon}$ is unitary, and $\bQ_{I}^{\varepsilon}$, $\bT_{I}^{\varepsilon}$ are independent of $\bx$ denoting rotation and translation operations, respectively.
\vspace{0.1in}
\subsection{Non-periodic DFT calculations}\hspace{0.01in}
 We first discuss the case of non-periodic problems with bounded domains $\Omega$, where $\Omega$ is large enough that the values of wavefunctions are negligible outside of $\Omega$, i.e., we assume the wavefunctions have a compact support on $\Omega$. We will subsequently discuss the case of periodic calculations with periodic boundary conditions. In view of the $C^{0}$ finite-element basis employed to discretize the DFT functional subsequently, we recast $T_s(\bGam^{\phi},\bPhi)$ in equation~\eqref{lagrangian} as the following:
\begin{equation*}
\begin{split}
T_{\text{s}}(\bGam^{\phi},\bPhi) &= 2\,\sum_{i,j,k=1}^{N}\int_{\Omega} \Gamma^{\phi}_{ij}\, S^{-1}_{jk}\, \phi_{k}^{*}(\bx) \left(-\frac{1}{2} \del^{2} \right)\,\phi_i(\bx) \dx \\
&= 2\sum_{i,j,k=1}^{N}\frac{1}{2}\int_{\Omega} \Gamma^{\phi}_{ij}\, S^{-1}_{jk}\, \del \phi_{k}^{*}(\bx).\del \phi_i(\bx) \dx\,.
\end{split}
\end{equation*}
In the above, we employ the divergence theorem and make use of the fact that $\bPhi \in (H^{1}_{0}(\Omega))^N$ for non-periodic problems.

Let us consider the perturbation of the underlying space given by $\chieps$, which maps a material point $\bx$ in $\Omega$ to $\bx' = \chieps(\bx)$ in $\Omega'$ (which denotes the image of $\Omega$ under the perturbation). The ground-state energy in the perturbed space is given by
\begin{align}\label{freenergynonper}
\mathcal{F}_0(\chieps) &= \mathcal{L}^{\eps}({\bGambar^{\phi}}^{\eps},\bPhibar^{\eps}, \varphibar^{\eps};\bR^{\eps})\notag\\
&= \widetilde{\mathcal{L}}^{\eps}({\bGambar^{\phi}}^{\eps},\bPhibar^{\eps}) + \mathcal{L}_c^{\eps}({\bGambar^{\phi}}^{\eps}) + \mathcal{L}^{\eps}_{\text{el}}({\bGambar^{\phi}}^{\eps},\bPhibar^{\eps},\varphibar^{\eps},\bar{\mathcal{V}}_{\dirac}^{\eps};\bR^{\eps})\,,
\end{align}
where $\bPhibar^{\eps}\in (H^1_0(\Omega'))^{N}$, $\varphibar^{\eps}\in H^1_0(\Omega')$, $\bar{\mathcal{V}}_{\dirac}^{\eps}$ and $\bGambar^{\phi^{\scriptstyle \eps}}$ are the solutions of the saddle point problem ~\eqref{saddlepoint} on the perturbed space. We now evaluate the configurational force by computing the G\^ateaux derivative of $\mathcal{F}_0(\chieps)$, i.e., $\derveps \mathcal{F}_0(\chieps) \atzero = \derveps (\widetilde{\mathcal{L}}^{\eps} + \mathcal{L}_c^{\eps} + \mathcal{L}^{\eps}_{\text{el}}) \atzero$, to arrive at (cf. Appendix for details of the derivation):
\begin{widetext}
\begin{equation}\label{Eshelbyforce}
\frac{d\mathcal{F}_0(\chieps)}{d\varepsilon}\atzerob = \int\displaylimits_{\Omega} \bE:\del \bdir (\bx) \dx + \sum_{I} \int\displaylimits_{\Rthree} \bE'^I:\del \bdir (\bx) \dx \,+\, \text{F}^{\text{PSP}}\,,
\end{equation}
where $\bE = \bE^{\text{loc}} + \bE^{\text{nl}}$ and $\bE'$ denote Eshelby tensors whose expressions in terms of the solutions of the saddle point problem ~\eqref{saddlepoint} on the original space ($\bGambar^{\phi^0}$, $\bPhibar^{0}$, $\varphibar^0$, $\bar{\mathcal{V}}_{\dirac}^{0}$) are provided below. We note that in the above expressions, and subsequently, the outer product between two vectors is denoted by ``$\otimes$'', the dot product between two vectors by ``.'' and dot product between two tensors by ``:''. Dropping the superscript $0$ on the electronic fields for notational convenience, the Eshelby tensors are given by:
\begin{gather}
\begin{split}
\bE^{\text{loc}} &= \Biggl(\sum_{i,j,k=1}^{N}{\Gammabar^{\phi}_{ij}} S^{-1}_{jk} \del \phibar_{k}^{*}(\bx)\cdot\del \phibar_i(\bx) - 2 \sum_{i,j,p,q=1}^{N} {\Gammabar^{\phi}_{ij}} S^{-1}_{jp} \phibar^{*}_p(\bx)\; \phibar_q(\bx) H^{\text{loc}}_{qi} + \varepsilon_{\text{xc}}(\rhobar)\rhobar(\bx)  -\frac{1}{8\pi} |\del \varphibar(\bx)|^2 + \rhobar(\bx) \varphibar(\bx) \notag \\
&+ \sum_J (V_{\text{loc}}^{J} - \vself J) \rhobar(\bx) \Biggr)\bI - \sum_{i,j,k=1}^{N}{\Gammabar^{\phi}_{ij}} S^{-1}_{jk}\left[\del \phibar_{k}^{*}(\bx) \otimes \del \phibar_i(\bx) + \del {\phibar_i}(\bx) \otimes \del {\phibar_{k}}^{*}(\bx)\right] + \frac{1}{4\pi}\del \varphibar(\bx) \otimes \del \varphibar(\bx)\,,
\end{split} \\
\bE^{\text{nl}} = \Biggl(- 2 \sum_{i,j,p,q=1}^{N} {\Gammabar^{\phi}_{ij}} S^{-1}_{jp} \phibar^{*}_p(\bx)\; \phibar_q(\bx) H^{\text{nl}}_{qi} + 2 \sum_{i,j,k=1}^{N}  {\Gammabar^{\phi}_{ij}} S^{-1}_{jk}\left(\text{P}^{\text{nl}}_{ki} + {\text{P}^{\text{nl}}_{ik}}^{*}\right)\Biggr)\bI\notag
\end{gather}
where
\begin{gather*}
\rhobar(\bx) = 2 \sum_{i,j,k=1}^{N}\Gammabar^{\phi}_{ij} \,S^{-1}_{jk} \,\phibar_{k}^{*}(\bx)\, \phibar_i(\bx) \,, \qquad S_{jk} = \int\displaylimits_{\Omega} \phibar_{j}^{*}(\bx)\phibar_k(\bx) \dx \,,\\
H^{\text{loc}}_{qi} = \int\displaylimits_{\Omega} \sum_{k=1}^{N} S_{qk}^{-1}\left(\frac{1}{2} \del \phibar_{k}^{*}(\bx)\cdot\del \phibar_i(\bx) + \phibar_{k}^{*}(\bx)\; \phibar_i(\bx) V^{\text{loc}}_{\text{eff}}(\rhobar)\right)\dx\,,\;\;\;\;
H^{\text{nl}}_{qi} = \int\displaylimits_{\Omega}\int\displaylimits_{\Omega} \sum_{k=1}^{N} S_{qk}^{-1}{\phibar_{k}}^{*}(\bx) V_{\text{nl}}(\bx,\by,\bR) \phibar_i(\by)\dy\dx,\\
\begin{split}
\text{P}^{\text{nl}}_{ki} =  \sum_{J} \sum_{lm} V_{lm}^{J}\,\phibar_{k}^{*}(\bx)\zeta^{J}_{lm}(\bx,\bR_J)\Delta V_{l}^{J}(|\bx-\bR_J|)\int\displaylimits_{\Omega}\zeta^{J}_{lm}(\by,\bR_J)\Delta V_{l}^{J}(|\by-\bR_J|)\phibar_{i}(\by)\dy \,. \notag
\end{split}
\end{gather*}
Further, the other terms in equation ~\eqref{Eshelbyforce} are given by
\begin{gather*}
\bE'^{I} = \frac{1}{8\pi} |\del \Vbar_{\dirac}^{I}(\bx)|^2 \bI - \frac{1}{4\pi}\del \Vbar_{\dirac}^{I}(\bx) \otimes \del \Vbar_{\dirac}^{I} (\bx)\,,\\[0.05in]
\text{F}^{\text{PSP}} = \sum_{J} \int\displaylimits_{\Omega} \rhobar(\bx)\left(\del \left(V^{J}_{\text{loc}}(|\bx-\bR_J|) - \vself J(|\bx-\bR_J|)\right)\right)\cdot \left(\bdir(\bx) - \bdir(\bR_J) \right)\dx  \,+ 2 \sum_{i,j,k=1}^{N}  {\Gammabar^{\phi}_{ij}} S^{-1}_{jk}\left(\text{F}^{\text{nl}}_{ki} + {\text{F}^{\text{nl}}_{ik}}^{*}\right)
\end{gather*}
where
\begin{align*}
 \text{F}^{\text{nl}}_{ki} =  \sum_{J} \sum_{lm} V_{lm}^{J}\Bigg[\int\displaylimits_{\Omega} \phibar_{k}^{*}(\bx) \del \left(\zeta^{J}_{lm}(\bx,\bR_J)\Delta V_{l}^{J}(|\bx-\bR_J|)\right)\cdot\left(\bdir(\bx) - \bdir(\bR_J)\right)\dx\Biggr]\Biggl[\int\displaylimits_{\Omega}\zeta^{J}_{lm}(\by,\bR_J)\Delta V_{l}^{J}(|\by-\bR_J|)\phibar_{i}(\by) \dy\Biggr].
\end{align*}
\end{widetext}
We note that the terms $\varphi\,b$ and $\tilde{\delta}(|\bx - \bR_I|)\,\Vbar_{\dirac}^{I}(\bx)$ do not appear in the expressions for $\bE^{\text{loc}}$ and $\bE'_I$, respectively, owing to the restriction that $\tau^{\eps}(\bx)$ corresponds to rigid body deformations in the compact support of the nuclear charge distribution $b(\bx)$. Hence, in the compact support of $b$, we have $\nabla.\bdir = 0$. We note that the second term in equation~\eqref{Eshelbyforce} involves an integral over $\Rthree$ which appears intractable. However, we can split this integral on a bounded domain $Q$, containing the compact support of $\tilde{\delta}(|\bx - \bR_I|)$, and its complement $\Rthree/Q$ which can in turn be computed as a surface integral. Thus,
\begin{equation*}
\int\displaylimits_{\Rthree} \bE'^I : \del \bdir \dx = \int\displaylimits_{Q} \bE'^I:\del \bdir \dx + \int\displaylimits_{\Rthree/Q} \bE'^I:\del \bdir \dx 
\end{equation*}
which can be written as the following equation:
\begin{equation}
\int\displaylimits_{\Rthree} \bE'^I : \del \bdir \dx = \int\displaylimits_{Q} \bE'^I:\del \bdir \dx - \int\displaylimits_{\partial Q} \bE'^I:\hat{\bn} \otimes \bdir \;d\bs \,, \label{surfinteg}
\end{equation}
where $\hat{\bn}$ denotes the outward normal to the surface $\partial Q$. The last equality follows from the fact that $\del^{2}\Vbar_{\dirac}^{I} = 0$ on $\Rthree/Q$. 
Though the expression for configurational force in equation~\eqref{Eshelbyforce} is derived in the case of pseudopotential calculations, the expression for all-electron calculations is obtained by using $V^{J}_{\text{loc}}(|\bx-\bR_J|) = \vself J(|\bx-\bR_J|)$,  $\bE^{\text{nl}} = \bzero$ and $\text{F}^\text{PSP} = 0$. Finally, we note that the force on any given atom is computed by choosing $\bdir$ such that its compact support only contains the atom of interest. We refer to the Appendix for the detailed derivation of equation~\eqref{Eshelbyforce}.

The computation of electronic ground-state in equation~\eqref{saddlepoint} is equivalent to finding the occupied eigen subspace $\mathbb{V}^N$ spanned by the eigenfunctions (canonical wavefunctions) corresponding to the smallest $N$ eigenvalues of the Kohn-Sham self-consistent Hamiltonian. Consequently, if $\bPsibar_{\perp} = \{\psibar_1,\psibar_2,\psibar_3 \cdots \psibar_N \}$ represent the orthonormal canonical eigenfunctions spanning the occupied subspace $\mathbb{V}^{N}$ of the Kohn-Sham Hamiltonian, the expressions for the Eshelby tensor $\bE$ and $\text{F}^\text{PSP}$ in equation ~\eqref{Eshelbyforce} reduce to the following form in terms of $\bPsibar_{\perp}$, $\fbar_i$ (orbital occupancy functions), $\bar{\epsilon}_i$ (Kohn-Sham eigenvalues) and $\varphibar$ (electrostatic potential):
\begin{widetext}
\begin{align}\label{EshelbyforceOrth}
&\bE = \Biggl(\sum_{i=1}^{N}\Bigl(\fbar_i\del \psibar_{i}^{*}(\bx)\cdot\del \psibar_i(\bx) - 2 \, \fbar_i\, \bar{\epsilon}_i\,\psibar^{*}_i(\bx)\; \psibar_i(\bx)\Bigr)  + \varepsilon_{\text{xc}}(\rhobar)\rhobar(\bx)  -\frac{1}{8\pi} |\del \varphibar(\bx)|^2 + \rhobar(\bx) \varphibar(\bx) \,\notag\\
&+ \sum_J (V_{\text{loc}}^{J} - \vself J)\rhobar(\bx) \,+  \text{E}^{\text{nl}} \,+ {\text{E}^{\text{nl}}}^{*} \Biggr)\bI - \sum_{i=1}^{N}\fbar_i\Bigl[\del \psibar_{i}^{*}(\bx) \otimes \del \psibar_i(\bx)
 + \del {\psibar_i}(\bx) \otimes \del \psibar_{i}^{*}(\bx)\Bigr] + \frac{1}{4\pi}\del \varphibar(\bx) \otimes \del \varphibar(\bx)
\end{align}
where
\begin{gather}
\rhobar(\bx) = 2\sum_{i} \fbar_{i}\,\psibar^{*}_i(\bx)\,\psibar_i(\bx)\,,\;\;\;\;\fbar_i = \frac{1}{1 + \exp(\frac{\bar{\epsilon}_i -\mu}{k_B\,T})} \,,\notag \\
\text{E}^{\text{nl}} = 2\;\sum_{i=1}^{N} \sum_{J} \sum_{lm} \fbar_i\; V_{lm}^{J} \,\psibar_{i}^{*}(\bx)\;\zeta^{J}_{lm}(\bx,\bR_J)\Delta V_{l}^{J}(|\bx-\bR_J|) {\int\displaylimits_{\Omega} \zeta^{J}_{lm}(\by,\bR_J)\, \Delta V^{J}_{l}(|\by-\bR_J|) \psibar_i(\by) \dy} \,.\notag 
\end{gather}
Further,
\begin{equation}
\text{F}^{\text{PSP}} = \sum_{J} \int\displaylimits_{\Omega} \rhobar(\bx)\left(\del \left(V^{J}_{\text{loc}}(|\bx-\bR_J|) - \vself J(|\bx-\bR_J|)\right)\right)\cdot \left(\bdir(\bx) - \bdir(\bR_J) \right)\dx  \,+\, \text{F}^\text{nl} \,+\,  {\text{F}^\text{nl}}^{*}\,,\notag
\end{equation}
where
\begin{align*}
\text{F}_\text{nl} &= 2\;\sum_{i=1}^{N} \sum_{J} \sum_{lm} \fbar_i V_{lm}^{J}\Bigg[\int\displaylimits_{\Omega} \psibar_{i}^{*}(\bx) \del \left(\zeta^{J}_{lm}(\bx,\bR_J)\Delta V_{l}^{J}(|\bx-\bR_J|)\right)\cdotp\left(\bdir(\bx) - \bdir(\bR_J)\right)\dx\Biggr]\\
&\times\Biggl[\int\displaylimits_{\Omega}\zeta^{J}_{lm}(\by,\bR_J)\Delta V_{l}^{J}(|\by-\bR_J|)\psibar_{i}(\by) \dy\Biggr] \,.
\end{align*}
\end{widetext}

In the next section, we present the configurational forces for the case of DFT calculations involving periodic geometries and boundary conditions. 

\subsection{Periodic DFT calculations}\hspace{0.01in}
In the case of infinite periodic crystals, the Kohn-Sham eigenfunctions are given by the Bloch theorem~\cite{ashcroft}, and the Bloch-periodic Kohn-Sham problem on an infinite crystal can be reduced to a periodic problem on a unit-cell.  In numerical simulations involving periodic calculations, it is computationally efficient to deal with unit-cells, which are much smaller than the supercells, and the computation of electron-density, kinetic energy and the electrostatic interaction energy in the case of pseudopotentials involves an additional integration over the Brillouin zone that is computed using numerical quadratures~\cite{rmartin}. 

Consider a unit cell $\Omega_\text{p}$ (periodic domain) containing $N_a$ atoms and $N_e$ electrons with nuclei positioned at $\bR=\{\bR_1, \bR_2, \bR_3, \cdots \bR_{N_a}\}$. To this end, the kinetic energy of non-interacting electrons per unit cell, in terms of the orthonormal wavefunctions and the orbital occupancy functions can be written as
\begin{equation}\label{kinbloch}
\begin{split}
&T_s(\bff,\bPsi_{\perp}) \\
&= 2\sum_{n=1}^{N}\fintd_{BZ} \int\displaylimits_{\Omegaper} f_{n}(\bk)\,\psi_{n}^{*}(\bx,\bk) \left(-\frac{1}{2} \del^{2} \right)\,\psi_{n}(\bx,\bk) \dx \dk\,,
\end{split}
\end{equation}
where $\fintd_{BZ}$ denotes the volume average of the integral over the Brillouin zone corresponding to $\Omegaper$. In the above,  $\bPsi_{\perp} = \{\psi_1(\bx,\bk), \psi_2(\bx,\bk), \cdots \psi_N(\bx,\bk);\, \forall \,\bk\in BZ\}$ and $\bff$ denotes the vector of orbital occupancy functions, i.e., $\bff = \{f_{1}(\bk),f_{2}(\bk),\cdots f_{N}(\bk);\, \forall \,\bk\in BZ\}$, while $N>N_e/2$ denotes the number of states of interest for any given point in the Brillouin zone. Using Bloch theorem, $\psi_{n}(\bx,\bk)$ can be expressed as
\begin{equation}\label{bloch}
\psi_{n}(\bx,\bk) = e^{i\bk\cdotp\bx} u_{n}(\bx,\bk) \,,
\end{equation}
where $i = \sqrt{-1}$ and $u_n(\bx,\bk)$ is a function that is periodic on the unit-cell. Denoting $u_{n}(\bx,\bk)$ and $f_n(\bk)$ as $u_{n\bk}(\bx)$ and $f_{n\bk}$ subsequently, $T_s$ in equation ~\eqref{kinbloch} can be rewritten using equation~\eqref{bloch} as
\begin{equation}
\begin{split}
&T_s(\bff,\bU) \\
& = 2 \sum_{n=1}^{N} \fintd_{BZ} \int\displaylimits_{\Omegaper}f_{n\bk} u^{*}_{n\bk}\left(-\frac{1}{2}\del^2 - i \bk\cdotp\del + \frac{1}{2}|\bk|^2 \right) u_{n\bk}\dx\dk \,, \label{kinEngyBloch1}
\end{split}
\end{equation} 
where $\bU = \{U_{\bk};\, \forall \,\bk\in BZ\}$ and $U_{\bk}=\{u_{1\bk}(\bx), u_{2\bk}(\bx) \cdots u_{N\bk}(\bx)\}$. In view of the $C^{0}$ finite-element basis employed subsequently to discretize the Kohn-Sham functional, it is convenient to recast the above equation to the following form involving partial derivatives of only first order:
\begin{align}
T_s(\bff,\bU) &= 2 \sum_{n=1}^{N} \fintd_{BZ} \int\displaylimits_{\Omegaper}f_{n\bk}\Bigl( \frac{1}{2}|\del u_{n\bk}|^2 - i u^{*}_{n\bk} \bk\cdotp \del u_{n\bk} \notag\\
&+ \frac{1}{2} |\bk|^2 |u_{n\bk}|^2\Bigr) \dx\dk \,. \label{kinEngyBloch2}
\end{align}
To arrive at the above equation, we employ the divergence theorem and use the fact $u_{n\bk}(\bx) \in H_{per}^{1}(\Omegaper)$. 

The contribution to the electrostatic interaction energy from the non-local pseudopotential is given by
\begin{align}\label{nonlocpspEngy}
& \mathcal{L}^{\text{nl}}(\bff,\bPsi_{\perp}) = \notag\\
& 2\sum_{n=1}^N  \fintd_{BZ} f_{n\bk}\Biggl(\;\int\displaylimits_{\Omegaper}\psi_{n}^{*}(\bx,\bk)\Biggl(\;\int\displaylimits_{\Rthree}\; V_{\text{nl}}(\bx,\by,\bR)\psi_{n}(\by,\bk)\dy\Biggr)\dx\Biggr)\dk\,.
\end{align}
\begin{widetext}
\noindent Using Bloch theorem, equation~\eqref{nonlocpspEngy} is rewritten as 
\begin{equation}\label{Vnl}
\mathcal{L}^{\text{nl}}(\bff,\bU) = 2\sum_{n=1}^N \fintd_{BZ}f_{n\bk}\Biggl(\;\int\displaylimits_{\Omegaper} u^{*}_{n\bk}(\bx) e^{-i\bk\cdotp\bx}\Biggl(\;\int\displaylimits_{\Rthree}V_{\text{nl}}(\bx,\by,\bR)e^{i\bk\cdotp\by}\,u_{n\bk}(\by) \dy\Biggr) \dx\Biggr)\dk \,.
\end{equation}
In the above equation, within the Kleinman-Bylander setting we have
\begin{equation}
e^{-i\bk\cdotp\bx}\int\displaylimits_{\Rthree} V_{\text{nl}}(\bx,\by,\bR) e^{i\bk\cdotp\by}\,u_{n\bk}(\by) \dy = \sum_{a,l,m} \sum_{r}\,e^{-i\bk.(\bx - \bL_r)}C_{lm}^{a,n\bk}\,V_{lm}^{a}\,\zeta^{a}_{lm}(\bx,\bR_a+\bL_r) \Delta V_l^{a}(|\bx - (\bR_a + \bL_r)|) \,,
\end{equation}
where the summation over $r$ runs on all lattice points in the periodic crystal and  $a$ runs on all the $N_a$ atoms in the unit-cell. We further note that $C_{lm}^{a}$ and $V_{lm}^{a}$ in the above equation have the following form:
\begin{align*}
&C_{lm}^{a,n\bk} =  \int \displaylimits_{\Omegaper}\sum_{r}e^{i\bk.(\by - \bL_{r})} \zeta^{a}_{lm}(\by,\bR_a+\bL_{r}) \Delta V_l^{a}(|\by - (\bR_a + \bL_{r})|)u_{n\bk}(\by) \dy \,,\\
&\frac{1}{V_{lm}^{a}} = \int\displaylimits_{\Omegaper}\sum_{r} \zeta^{a}_{lm}(\by,\bR_a+\bL_{r})\,\Delta V_l^{a}(|\by - (\bR_a + \bL_{r})|) \zeta^{a}_{lm}(\by,\bR_a+\bL_{r})\dy \,.
\end{align*}
Finally, using the local formulation of the extended electrostatic interaction energy presented in Section III, the computation of the electronic ground-state, for a given position of atoms, in the context of the periodic DFT calculations is governed by the following variational problem:
\begin{gather}
\mathbb{E}_0(\bR) =  \min_{f_{\bk} \in \mathbb{R}/{\mathbb{R}^{-}}}\min_{U_{\bk}\in (\mathcal{Y})^N} \max_{\varphi \in \mathcal{Y}}\;\; \mathcal{L}(\bff,\bU,\varphi;\bR)\;\;\;
\text{such that}\;\;\;\int\displaylimits_{\Omegaper}u_{i\bk}(\bx) u_{j\bk}(\bx) \dx = \delta_{ij},\,\;\;\;\;2\sum_{n=1}^{N}\fintd_{BZ}f_{n\bk}\dk = N_e\,, \label{minmaxper}\\
\text{where}\;\;\mathcal{L}(\bff,\bU,\varphi;\bR) = T_s(\bff,\bU) + E_{\text{xc}}(\rho) - E_{\text{ent}}(\bff) + \min_{V^{I} \in H^{1}(\Rthree)}\mathcal{L}_{\text{el}}(\bff,\bU,\varphi,\mathcal{V};\bR) \,,\notag
\end{gather}
\begin{gather}
\mathcal{L}_{\text{el}}(\bff,\bU,\varphi,\mathcal{V};\bR) = \intd_{\Omegaper}\left[-\frac{1}{8\pi} |\del \varphi(\bx)|^2 + (\rho(\bx) + b(\bx,\bR))\varphi(\bx)\right]\dx
+ \sum_{I} \intd_{\Rthree} \left[\frac{1}{8\pi} |\del V^{I}(\bx)|^2  +  Z_I \tilde{\delta}(|\bx - \bR_I|) V^{I}(\bx) \right]\dx\notag\\
+ \sum_{a} \sum_{r}\intd_{\Omegaper} \left(V^{a}_{\text{loc}}(|\bx-(\bR_a+\bL_r)|) - \vself a(|\bx-(\bR_a+\bL_r)|)\right)\rho(\bx)\dx + \mathcal{L}^{\text{nl}}(\bff,\bU)\,, \notag\\
\rho(\bx) = 2\sum_{n=1}^{N}\fintd_{BZ}f_{n\bk} |u_{n\bk}(\bx)|^2 \dk,\;\;\;\;E_{\text{ent}}(\bff) = -2k_BT\sum_{n=1}^{N}\fintd_{BZ}\left(f_{n\bk} \log f_{n\bk} + (1 - f_{n\bk})\log(1 - f_{n\bk})\right) \dk \,.\notag
\end{gather}
\end{widetext}
In the above we note that $\mathcal{Y} = H^{1}_{per}(\Omegaper)$. In order to compute the configurational force, we follow a similar procedure as in the case of non-periodic calculations. Let $\bx$ denote a point in $\Omegaper$, whose image in $\Omegaper' = \chieps(\Omegaper)$  is $\bx' = \chieps(\bx)$ with $\bdir = \derveps \chieps\atzero$ being the generator of the underlying deformation. We restrict  $\chieps$ to deformations that preserve the periodic geometry, i.e., $\Omegaper'$ represents a periodic domain, given that $\Omegaper$ is a periodic domain. Further, let  $\bk' = \bkappa^{\eps}(\bk)$ correspond to the bijective mapping representing the infinitesimal perturbation of the  reciprocal space due to the underlying deformation of the real-space. We now evaluate the configurational force in the periodic setting by computing the G\^{a}teaux derivative of $\mathbb{E}_0(\chieps) = \mathcal{L}^{\eps}(\bffbar^{\eps},\bUbar^{\eps},\varphibar^{\eps};\bR^{\eps })$, where $\bar{U}_{\bk'}^{\eps}\in (H_{per}^{1}(\Omegaper'))^{N}$, $\varphibar^{\eps}\in H_{per}^{1}(\Omegaper')$, $\bffbar^{\eps}$ are the solutions of the saddle point variational problem in equation~\eqref{minmaxper}. The configurational force is given by (cf. Appendix for detailed derivation)
\begin{widetext}
\begin{equation}\label{EshelbyforceOrthPer}
\begin{split}
 \frac{d\,\mathbb{E}_0(\chieps)}{d\varepsilon}\atzerob = \int\displaylimits_{\Omega_p} \bE:\del \bdir (\bx) \dx + \sum_{I} \int\displaylimits_{\Rthree} \bE'^I:\del \bdir (\bx) \dx \,
+\, \text{F}^{\text{PSP}}  + \text{F}^{K}
\end{split}
\end{equation}
where $\bE$ and $\bE'$ denote Eshelby tensors whose expressions in terms of the solutions of the saddle point problem ~\eqref{minmaxper} ($\bffbar^0$, $\bUbar^0$, $\varphibar^0$, $\mathcal{V}_{\dirac}^{0}$) solved over $\Omega_p$ are given below. Dropping the superscript $0$ on the electronic fields for notational convenience, we have:
\begin{align*}
&\bE = \Biggl(\sum_{n=1}^{N}\fintd_{BZ} \fbar_{n\bk}\Bigl(|\del \ubarn_{nk}(\bx)|^2 -2\,i\,  \ubarn^{*}_{n\bk}(\bx) \,\bk\cdotp \del \ubarn_{n\bk}(\bx) + (|\bk|^2 - 2\,\bar{\epsilon}_{n\bk}) \,|\ubarn_{n\bk}(\bx)|^2 \Bigr)\dk  + \varepsilon_{\text{xc}}(\rhobar)\rhobar(\bx)  -\frac{1}{8\pi} |\del \varphibar(\bx)|^2 + \rhobar(\bx) \varphibar(\bx) \,\notag\\
& +\sum_a \sum_r \left(V^{a}_{\text{loc}}(|\bx-(\bR_a+\bL_r)|) - \vself a(|\bx-(\bR_a+\bL_r)|)\right)\rhobar(\bx) +\, \text{E}^{\text{nl}} \,+\, {\text{E}^{\text{nl}}}^{*} \Biggr)\bI  \\
&- \sum_{n=1}^{N}\fintd_{BZ}\fbar_{n\bk}\Bigl[\del  \ubarn^{*}_{n\bk}(\bx) \otimes \del \ubarn_{n\bk}(\bx) + \del \ubarn_{n\bk}(\bx) \otimes \del \ubarn^{*}_{n\bk}(\bx)- 2\,i \,\ubarn^{*}_{n\bk}(\del \ubarn_{n\bk} \otimes \bk)\Bigr]\dk + \frac{1}{4\pi}\del \varphibar(\bx) \otimes \del \varphibar(\bx)
\end{align*}
where
\begin{align*}
E^{\text{nl}} = 2\sum_{n=1}^N \,\sum_{a,l,m} \fintd_{BZ} \fbar_{n\bk}\,\sum_{r}\;\ubarn_{n\bk}^{*}(\bx)\,e^{-i\bk\cdotp(\bx - \bL_r)}\,C_{lm}^{a,n\bk}\,V_{lm}^{a}\, \zeta_{lm}^{a}(\bx,\bR_a + \bL_{r}) \Delta V_{l}^{a}(|\bx - (\bR_a + \bL_{r})|)\dk\,.
\end{align*}
Further, the other terms in equation~\eqref{EshelbyforceOrthPer} are given by
\begin{equation*}
\bE'^{I} = \frac{1}{8\pi} |\del \Vbar_{\dirac}^{I}(\bx)|^2 \bI - \frac{1}{4\pi}\del \Vbar_{\dirac}^{I}(\bx) \otimes \del \Vbar_{\dirac}^{I} (\bx)\,,
\end{equation*}
\begin{equation*}
\text{F}^{\text{PSP}} = \sum_{a} \sum_{r}\intd_{\Omegaper}\rhobar(\bx) \left[\del \left(V^{a}_{\text{loc}}(|\bx-(\bR_a+\bL_r)|) - \vself a(|\bx-(\bR_a+\bL_r)|)\right)\right]\cdotp \left(\bdir(\bx) - \bdir(\bR_a + \bL_r) \right)\dx  \,+\, \text{F}^\text{nl} \,+\,  {\text{F}^\text{nl}}^{*}\,,
\end{equation*}
where
\begin{equation*}
\begin{split}
&\text{F}^\text{nl} = 2\sum_{n=1}^N \;\sum_{a,l,m}\fintd_{BZ}\;\intd_{\Omegaper}\fbar_{n\bk}\sum_{r} \ubarn^{*}_{n\bk}(\bx)e^{-i\bk\cdotp(\bx - \bL_r)}C_{lm}^{a,n\bk}\,V_{lm}^{a}\,\Bigl[\del (\zeta_{lm}^{a}(\bx,\bR_a + \bL_{r}) \Delta V_{l}^{a}(|\bx - (\bR_a + \bL_{r})|))\cdotp\left(\bdir(\bx) - \bdir(\bR_a + \bL_r)\right) \\
&- \zeta_{lm}^{a}(\bx,\bR_a + \bL_{r}) \Delta V_{l}^{a}(|\bx - (\bR_a + \bL_{r})|)\Bigl(i\bk\cdotp(\bdir(\bx) - (\bdir(\bR_a + \bL_r) - \bdir(\bR_a)))\Bigr)\Bigr]\dx\, \dk\,,
\end{split}
\end{equation*}
and finally we have
\begin{equation*}
\begin{split}
\text{F}^{K} &= \sum_{n=1}^{N}\fintd_{BZ} \int\displaylimits_{\Omegaper}  \fbar_{n\bk}\Biggl(-2\,i\,\ubarn^{*}_{n\bk}(\bx)\derveps\bkappa^{\eps}(\bk)\atzerob\cdotp \del \ubarn_{n\bk}(\bx) + \derveps|\bkappa^{\eps}(\bk)|^2\atzerob\,|\ubarn_{n\bk}(\bx)|^2\Biggr)\dx \dk \\
&+ 2\,\sum_{n=1}^N \;\sum_{a,l,m}\fintd_{BZ} \fbar_{n\bk}V_{lm}^{a} \derveps \Biggl\{\Biggl[\int\displaylimits_{\Omegaper} \sum_{r} \ubarn^{*}_{n\bk}(\bx)\,e^{-i\bkappa^{\eps}(\bk).(\bx - \bL_{r})}\, \zeta^{a}_{lm}(\bx,\bR_a+\bL_r) \,\Delta V_l^{a}(|\bx - (\bR_a + \bL_r)|) \dx\Biggr]\\
&\times \Biggl[\int \displaylimits_{\Omegaper}\sum_{s}e^{i\bkappa^{\eps}(\bk).(\by - \bL_{s})} \,\zeta^{a}_{lm}(\by,\bR_a+\bL_{s}) \,\Delta V_l^{a}(|\by - (\bR_a + \bL_{s})|)\,u_{nk}(\by) \dy\Biggr]\Biggr\}\atzerobb \dk\,.
\end{split}
\end{equation*}
\end{widetext}
In the above, $\bar{\epsilon}_{n\bk}$ $(1 \leq n \leq N)$ denote the smallest $N$ eigenvalues in the Kohn-Sham eigenvalue problem, which arises from the Euler-Lagrange equations of the variational problem in equation~\eqref{minmaxper}. The second term in equation ~\eqref{EshelbyforceOrthPer}, involving an integral over $\Rthree$, is evaluated as explained in equation ~\eqref{surfinteg} of the previous subsection. We note that the expression for configurational force in equation~\eqref{EshelbyforceOrthPer} is equally applicable for all-electron calculations by using $V^{a}_{\text{loc}} = \vself a$,  $E^{\text{nl}} = 0$, $\text{F}^\text{PSP} = 0$. The force on any given atom in $\Omegaper$ is computed via equation~\eqref{EshelbyforceOrthPer} by choosing $\bdir$ such that its compact support lies within $\Omegaper$ and contains only the atom of interest. In this case, we note that $\bk' = \bk$, and, thus, $\text{F}^{K} = 0$.

\paragraph*{Computation of stress tensor}\hspace{0.01in} We now discuss the evaluation of the stress tensor, associated with cell relaxations, using the derived configurational force expression. Geometry optimization of the unit cell is realized by application of affine deformations, which preserves the periodicity of the cell. Thus, in order to compute the stress tensor, we chose an affine perturbation of the underlying space:
\begin{align}\label{affine}
(\chieps(\bx))_{i} = {x'}_{i}= x_i \,+\, \varepsilon C_{ij}x_j  \;\;\implies \dir_i = C_{ij} x_j \notag \\
\;\;\implies \;\; \bdir = \bC\bx \,,
\end{align}
where $\bC$ is independent of $\bx$. The stress tensor is the derivative of the energy density with respect to the infinitesimal strain tensor $\bStrain$, and is given by
\begin{equation}\label{sigma}
\sigma_{\text{ij}} = \frac{1}{\Omegaper}\frac{\partial \mathbb{E}_0}{\partial \mathcal{E}_{\text{ij}}}\,.
\end{equation}
Using equation~\eqref{affine}, the deformation gradient $\bF = \frac{\partial \bx'}{\partial \bx}$ is given by $\bF = \bI + \eps \bC$, where $\eps$ is a small, denoting an infinitesimal perturbation. Thus, the strain tensor $\boldsymbol{\mathcal{E}}$ is given by
\begin{equation}\label{strain}
\boldsymbol{\mathcal{E}} = \frac{1}{2} (\bF^T\bF - \bI) = \frac{1}{2}\eps\,(\bC^T + \bC) +\order(\eps^2)\,.
\end{equation}
We note that the energy $\mathbb{E}_0(\bStrain(\eps))$ can be expanded about $\eps=0$ as:
\begin{equation*}
\mathbb{E}_0(\bStrain(\eps)) = \mathbb{E}_0(\bStrain(\eps=0)) \,+\, \left(\frac{\partial \mathbb{E}_0}{\partial \bStrain}\bigg|_{\bStrain(\eps=0)}\boldsymbol{:}\;\bStrain\right) \,+\, \order(\eps^2) \,,
\end{equation*}
which, in turn, can be written using equation ~\eqref{sigma} and ~\eqref{strain} as
\begin{equation*}
\mathbb{E}_0(\bStrain(\eps)) = \mathbb{E}_0(\bStrain(\eps=0)) \,+ \Omegaper \left(\boldsymbol{\sigma}\boldsymbol{:}\,\frac{1}{2}\eps\,(\bC^T + \bC)\right) + \order(\eps^2)\,.
\end{equation*}
Thus, the G\^ateaux derivative $\frac{d\,\mathbb{E}_0(\chieps)}{d\varepsilon}\atzerob$, which is the configurational force, can be written in terms of stress tensor $\boldsymbol{\sigma}$ as \begin{equation}\label{configForceForStress1}
\frac{d\,\mathbb{E}_0(\chieps)}{d\varepsilon}\atzerob = \Omegaper \,\frac{1}{2} \left((\bC + \bC^T): \boldsymbol{\sigma} \right) =  \Omegaper\,(\bC : \boldsymbol{\sigma}  ) \,.
\end{equation}
The second equality in the above equation results from the symmetry of the stress tensor $\boldsymbol{\sigma}$ . 

Next, we consider the configurational force associated with affine perturbation, given by equation~\eqref{affine}, evaluated using equation~\eqref{EshelbyforceOrthPer}. To this end, we note that in the reciprocal space $\bk' = \boldsymbol{\kappa}^{\eps}(\bk) = (\bI - \varepsilon \bC^{T})\bk$, when $\bdir(\bx)$ is given by equation~\eqref{affine}. Substituting $\bdir = \bC\bx$ and $\boldsymbol{\kappa}^{\eps}(\bk) = (\bI - \varepsilon \bC^{T})\bk$ in ~\eqref{EshelbyforceOrthPer} we have
\begin{equation}\label{configForceForStress2}
\frac{d\,\mathbb{E}_0(\chieps)}{d\varepsilon}\atzerob = \bC : \left(\intd_{\Omegaper} (\bE + \widetilde{\bE}) \dx + \intd_{\Rthree} \sum_{I}\bE'^I \dx\right)
\end{equation}
where $\bE$ and $\bE'^I$ are defined as in  ~\eqref{EshelbyforceOrthPer}, and $\widetilde{\bE}$  is the tensor  arising out of the term $F^{\text{PSP}} + F^{\bk}$ in equation ~\eqref{EshelbyforceOrthPer}. Finally, comparing equation ~\eqref{configForceForStress1} and ~\eqref{configForceForStress2}, we arrive at the following expression for the stress tensor $\boldsymbol{\sigma}$:
\begin{equation}\label{eshelbyStress}
\boldsymbol{\sigma} = \frac{1}{\Omegaper}\left(\intd_{\Omegaper} (\bE + \widetilde{\bE}) \dx + \intd_{\Rthree} \sum_{I}\bE'^I \dx\right)\,.
\end{equation}
$\widetilde{\bE}$ in the above equation can be expressed as:
\begin{equation}\label{eshelbyStressPart}
\widetilde{\bE} = \bE^{\text{PSP}} + \bE^{K}
\end{equation}
and the expressions for $\bE^{\text{PSP}}$  and $\bE^{K}$ in the above are given below:
\begin{widetext}
\begin{equation}
\bE^{\text{PSP}} = \sum_{a} \sum_{r}\intd_{\Omegaper}\rhobar(\bx) \left[\del \left(V^{a}_{\text{loc}}(|\bx-(\bR_a+\bL_r|)) - \vself a(|\bx-(\bR_a+\bL_r|))\right)\right]\otimes \left(\bx - (\bR_a + \bL_r) \right)\dx  \,+\, \bE^\text{nl} \,+\,  {\bE^\text{nl}}^{*}\,,
\end{equation}
where
\begin{equation}
\begin{split}
&\bE^\text{nl} = 2\sum_{n=1}^N \sum_{a,l,m}\fintd_{BZ}\intd_{\Omegaper}\fbar_{n\bk} \sum_{r} \ubarn^{*}_{n\bk}(\bx)e^{-i\bk\cdotp(\bx - \bL_r)}C_{lm}^{a,n\bk}\,V_{lm}^{a}\del\Bigl(\zeta_{lm}^{a}(\bx,\bR_a + \bL_{r}) \Delta V_{l}^{a}(|\bx - (\bR_a + \bL_{r})|)\Bigr)\otimes\left(\bx - (\bR_a + \bL_r)\right) \dx \dk.
\end{split}
\end{equation}
and finally we have
\begin{equation}
\bE^{K} = \sum_{n=1}^{N}\fintd_{BZ} \int\displaylimits_{\Omegaper} \fbar_{n\bk}\Bigl(2\,i\,\ubarn^{*}_{n\bk}(\bx)\left(\bk \otimes \del \ubarn_{n\bk}(\bx)\right) - 2 (\bk \otimes \bk)|\ubarn_{n\bk}(\bx)|^2 \Bigr)\dx \dk
\end{equation}
\end{widetext}

\section{Numerical Implementation}
In this section, we present the details of the numerical implementation of configurational forces within the framework of finite-element discretization. Subsequently, we discuss a computationally efficient and robust strategy to conduct atomic relaxations and cell-relaxations using the finite-element basis.
\subsection{Finite-element discretization}
Though the finite-element (FE) basis offers some unique advantages for electronic structure calculations, initial studies~\cite{bylaska,Hermansson,surya2010} which employed linear finite-element basis functions suggested that they require a large number of basis functions---of the order of 100,000 basis functions per atom, or more, were required to achieve chemical accuracy in electronic structure calculations. This compared poorly with plane-wave basis and atomic-orbital type basis. A recent investigation ~\cite{motam2013} has indicated that the use of adaptive higher-order spectral finite-elements can significantly improve the computational efficiency of real-space electronic structure calculations. In particular, staggering computational savings of the order of 1000-fold relative to linear finite-elements for both all-electron and pseudopotential calculations have been realized. Further, for accuracies commensurate with chemical accuracy, it was demonstrated that the computational efficiency afforded by higher-order finite-element discretizations is competing with plane-wave discretizations for pseudopotential calculations~\cite{motam2013} and with Gaussian basis for all-electron calculations using enrichment finite-element basis~\cite{bikash2017}. Efficient computational strategies~\cite{motam2013,motam2014,motam2017,bikash2017} in conjunction with finite-element discretization have enabled large-scale real-space Kohn-Sham DFT calculations on material systems containing up to ten of thousand atoms in the case of pseudopotential calculations and ten thousand electrons in the case of all-electron calculations.

We note that the expressions derived for evaluating the configurational forces require that electronic fields satisfy the Euler-Lagrange equations corresponding to the variational problem ~\eqref{saddlepoint} in the case of non-periodic problems, and~\eqref{minmaxper} in the case of periodic calculations, i.e., the electronic fields are at their ground-state corresponding to the Kohn-Sham energy functional.  To this end, the Kohn-Sham ground state for a given position of atoms is obtained by computing self-consistently the occupied eigen-subspace of the Kohn-Sham eigenvalue problem while evaluating the electrostatic potentials $\varphi$ and $\vself I$ -via- the Poisson problems in equations~\eqref{electroLocal}.  In the present work, we employ higher-order finite-element basis for all the electronic fields involved in the Kohn-Sham problem. Let $\mathbb{V}^{M}_h$ represents the $M$-dimensional subspace spanned by the finite-element basis $N_j(\bx): 1 \leq j \leq M$, a piecewise polynomial basis generated from a finite-element discretization ~\cite{Brenner-Scott} with characteristic mesh-size $h$. The representation of the various electronic fields in the Kohn-Sham problem---the wavefunctions and the electrostatic potential---in the finite-element basis is given by 
\begin{equation}
g^{h}(\bx) = \sum_{i=1}^{M} N_i(\bx)g_i \,,
\end{equation}
where $g^{h}(\bx)$ denotes the finite-element discretized electronic field with $g_i$ denoting the coefficients in the expansion. We note that $g_i$ also corresponds to the nodal values at the $i^{th}$ node on the finite-element mesh.  The finite-element discretization of the Poisson problems corresponding to equation~\eqref{electroLocal} 
results in the following linear system of equations:
\begin{equation}
\bK \boldsymbol{\varphi} = \br^{\text{tot}}\,;\;\;\;\;\bK \bV^{I} = \br^{I}\,,
\end{equation}
where $K_{jk} = (1/4\pi)\int \del N_j(\bx) \cdotp \del N_k(\bx) \dx$, $r^{\text{tot}}_j = \int (\rho^{h}(\bx) + b(\bx,\bR)) N_j(\bx) \dx$ and further we have $r^{I}_j = \int Z_I \delta(\bx - \bR_I) N_j(\bx) \dx$. $\boldsymbol{\varphi}$ denotes the discrete electrostatic potential corresponding to the sum of electron-density $\rho^h(\bx)$ and the nuclear charge distribution $b(\bx,\bR) = \sum_I Z_I \delta(\bx-\bR_I)$. The discrete nuclear potential corresponding to the $I^{th}$ nuclear charge $Z_I \delta(\bx - \bR_I)$, computed in the finite-element basis, is denoted as $\bV^{I}$. We note that the nuclear charges are located on the nodes of the finite-element triangulation, and are treated as point charges ($Z_I \delta(\bx - \bR_I)$), and the discreteness of the finite-element triangulation provides a regularization of the potential fields. 

We next discuss the discretization of the Kohn-Sham eigenvalue problem. Using the L\"{o}wdin orthonormalized finite-element basis ($q_j(\bx) = \sum_{k=1}^M M_{jk}^{-1/2} N_k(\bx)$), the Kohn-Sham eigenvalue problem reduces to the following standard eigenvalue problem~\cite{motam2013}:
\begin{equation}\label{femEigen}
\bar{\bH}\bar{\bPsi}_i = \epsilon_i^h \bar{\bPsi}_i
\end{equation}
where $\bar{\bH} = \bM^{-1/2}\bH\bM^{-1/2}$, with 
\begin{align} 
H_{jk} = &\frac{1}{2} \int \del N_j(\bx) \cdotp \del N_k(\bx) \dx \notag\\
&+ \int N_j(\bx)(V^{\text{loc},h}_{\text{eff}} + V^{h}_{\text{nl}})\,N_k(\bx) \dx\,.\notag
\end{align} 
The discrete Kohn-Sham eigenvalue problem ~\eqref{femEigen} is solved using a self-consistent field iteration (SCF) with Anderson mixing scheme by employing either  ``ChFSI-FE'' technique~\cite{motam2013} or ``SubPJ-FE''\cite{motam2014} method. In each SCF iteration, the ``ChFSI-FE'' technique~\cite{motam2013} comprises of employing Chebyshev filtering~\cite{saad2006} to efficiently compute the occupied eigenspace using spectral finite-elements in conjunction with Gauss-Lobatto quadrature for numerical integration. Upon computing the approximate eigenspace, the Chebyshev filtered vectors spanning the eigenspace are orthonormalized, and, subsequently, the projection of the Hamiltonian into this orthonormal basis is computed. Finally, the projected Hamiltonian is diagonalized to compute the eigenvalues and eigenvectors, which, in turn, are used in the computation of the electron-density. On the other hand, in the ``SubPJ-FE'' method~\cite{motam2014}, in each SCF iteration, the Chebyshev filter is applied to a subspace (spanned by localized functions) to compute an approximation to the occupied eigenspace. Then, a localization procedure is employed to construct nonorthogonal localized wavefunctions spanning the Chebyshev filtered space. The localized functions are truncated using a truncation tolerance, below which they are set to zero, to provide a compact support for these functions. The Kohn-Sham Hamiltonian is then projected into this localized basis, and a Fermi-operator expansion in terms of the projected Hamiltonian is employed to compute the electron-density. We note that ``SubPJ-FE'' has enabled large-scale electronic structure calculations using spectral finite-element discretizations with reduced order scaling, and, importantly, the method treats both metallic and insulating systems on a similar footing.

Upon computing the Kohn-Sham electronic ground-state for a given position of atoms in the discrete setting, in order to compute the forces on atoms we make use of the expressions for configurational forces in Section~\ref{sec:ConfigForces}. Note that the configurational force expressions have been derived in the continuum setting, and the force can be evaluated at every material point $\bx$ in the domain with an appropriate choice of $\bdir(\bx)$. In the discrete finite-element setting, we note that the ground-state energy is not only a function of position of atoms, but also a function of position of the nodes contained in the underlying finite-element triangulation. Hence, in addition to the physical force associated with the atoms, there is an additional mesh force (Pulay force) on all the nodes of the FE triangulation. The configurational force expressions presented in Section~\ref{sec:ConfigForces}, which is a variational force, inherently accounts for these contributions.

In order to evaluate the configurational force in the discrete setting, we discretize the generator $\bdir(\bx)$, representing the perturbation of the underlying space to lie in the finite-dimensional subspace $\mathbb{V}^{\widetilde{M}}_h \subset \mathbb{V}^{M}_h$  spanned by the $\widetilde{M}$ linear finite-element basis functions $\{\widetilde{N}_k(\bx)\}$. We note that the sub-parametric linear interpolation of the geometry in our FE implementation restricts the perturbation of the underlying geometry $\bdir^h(\bx)$ to lie in the space $\mathbb{V}^{\widetilde{M}}_h$ spanned by the linear finite-element basis functions as:
\begin{equation}\label{discreteGen}
\dir^h_j(\bx) = \sum_{i} \dir_{i,j} \widetilde{N}_{i}(\bx) \quad\,\,i = 1 \cdots \widetilde{M}\,\,\text{and}\,\,j = 1,2,3 \,\,,
\end{equation}
where $\widetilde{N}_{i}(\bx)$ is the linear finite-element basis function associated with node $i$, and $\dir_{i,j}$ is the nodal value of the generator associated with the node $i$ in the $j^{th}$ direction. To evaluate the force $\mathfrak{f}^h_{i,j}$ acting on the $i^{th}$ node in the $j^{th}$ direction, we use ~\eqref{discreteGen} with $\dir_{i,j} = 1$ for the $i^{th}$ node and $j^{th}$ direction, and $\dir_{i,j} = 0$ otherwise. Finally, the expression for the configurational force in ~\eqref{Eshelbyforce} reduces to
\begin{equation}\label{discreteEshelbyForce}
\begin{split}
\mathfrak{f}^h_{i,j} &=  \sum_{p=1}^3 \left(\int\displaylimits_{\Omega} E^h_{jp}\,\,\frac{\partial \widetilde{N}_i(\bx)}{\partial x_p} \dx + \sum_{I} \int\displaylimits_{\Rthree} {E'^{h}_{jp}}^{I}\,\, \frac{\partial \widetilde{N}_i(\bx)}{\partial x_p}  \dx \right)\\
&+\, {\text{F}^{h^\text{\scriptsize{PSP}}}_{i,j}}\,\,,\;\;\;\text{where}\;\;\;j = 1,2,3 \;\;\text{and}\;\; i = 1\cdots\widetilde{M}\,.
\end{split}
\end{equation}
In the above equation~\eqref{discreteEshelbyForce}, $E^h_{jp}$ and ${E'^{h}_{jp}}^{I}$ represent the $(j,p)$ component of the Eshelby tensors $\bE$ and $\bE'^{I}$ in their discrete form involving the discrete wavefunctions---$\bar{\bPhi}^{h}$ for the non-periodic case and $\bar{\bU}^h$ for the periodic case---and the discrete electrostatic potentials $\bar{\varphi}^h$, $\vself {I^h}$. We note that $\text{F}^{h^\text{\scriptsize{PSP}}}_{i,j}=0$ in the case of all-electron calculations. In the case of pseudopotential calculations, the term $\text{F}^{\text{PSP}}$ in equation ~\eqref{Eshelbyforce} involves expressions of the form $\del p(\bx)\cdotp(\bdir(\bx) - \bdir(\bR_J))$ which reduce to the form $\frac{\partial p}{\partial x_j} (\widetilde{N}_i(\bx) - 1)$ or  $\frac{\partial p}{\partial x_j} (\widetilde{N}_i(\bx))$ in $\text{F}^{h^\text{\scriptsize{PSP}}}_{i,j}$, depending on whether a nucleus lies on the finite-element node $i$ or not. 
\vspace{0.1in}
\paragraph{Atomic relaxations}\hspace{0.01in} We recall from equation ~\eqref{groundstate} that the overall ground-state in DFT is obtained by minimizing the Kohn-Sham energy functional over the nuclear positions, while obtaining the electronic ground-state for every configuration of nuclei encountered during the minimization over nuclear positions. This involves driving the forces associated with the atoms to zero, in order to obtain the equilibrium configuration of atoms in a given material system. We now discuss a computationally efficient strategy to evaluate the forces on atoms within the finite-element framework. 

It is evident from equation ~\eqref{discreteEshelbyForce} that the configurational force can be computed on every node of the underlying FE triangulation. An ideal approach for atomic relaxations is to drive the configurational forces on all $\widetilde{M}$ nodes below a prescribed tolerance. However, this approach is not computationally efficient as the number of degrees of freedom involved in the atomic relaxation will scale with the number of nodes in the underlying FE mesh. Furthermore, those nodes with atomic nuclei can encounter large displacements during the iterative process of atomic relaxations, which can result in a deterioration of the quality of the FE mesh. To  circumvent the above issues, we move a ball of nodes around every nucleus to design a robust atomic relaxation procedure. To this end, we compute the configurational force associated with an atom $I$ using the generator $\dir_j(\bx)$ to be of the form $\Theta(\bx - \bR_{I})$, which is a localized function around the atom of interest. In the current work, we choose $\Theta(\bx - \bR_{I})$ to be a Gaussian function which has the following form:
\begin{equation}
\Theta(\bx - \bR_{I}) = e^{-\alpha |\bx -\bR_I|^2}\,,
\end{equation}
where the parameter $\alpha>0$ determines the radius of influence of the Gaussian function. Using the finite-element discretization of $\bdir(\bx)$, the function $\Theta(\bx - \bR_{I})$ is interpolated to obtain
\begin{equation}\label{discreteGaussian}
\dir^h_j(\bx) = \Theta^h(\bx - \bR_{I}) = \sum_{i=1}^{\widetilde{M}} \dir_{i,j} \widetilde{N}_{i}(\bx)\;\;\; j = 1,2,3\,,
\end{equation}
where $\dir_{i,j} = \exp(-\alpha|\bx_i - \bR_I|^2)$ with $\bx_i$ denoting the position coordinate of $i^{th}$ linear finite-element node. Finally, using equation~\eqref{discreteEshelbyForce}, the configurational force associated with the generator $\Theta^h(\bx - \bR_{I})$ is given by
\begin{equation}\label{forceAtoms}
\mathfrak{f}^h_{I,j} = \sum_{i=1}^{\widetilde{M}} \mathfrak{f}^h_{i,j} \, e^{-\alpha|\bx_i - \bR_I|^2} \,.
\end{equation}
It is evident from equation~\eqref{forceAtoms} that the total degrees of freedom in the atomic relaxation procedure is now commensurate with the number of atoms. We determine the equilibrium configuration of atoms by driving $\mathfrak{f}^{h}_{I,j}$ below a prescribed tolerance.
\vspace{0.1in}
\paragraph{Periodic cell relaxations}\hspace{0.01in}
The configurational force derived in section~\ref{sec:ConfigForces} provides the generalized variational force with respect to both the internal positions of atoms as well as the external cell domain in the case of periodic calculations. Geometry optimization involving external cell relaxation is performed by application of affine deformations that change the shape of the cell, while preserving the periodicity of its faces. Hence, we restrict $\bdir$ to affine deformations to compute the stresses, as discussed in Section IV. We note that affine functions are exactly represented by linear FE basis functions, and, thus, the stress on a unit-cell, in the discrete setting, is given by
\begin{equation}\label{discretestresses}
\boldsymbol{\sigma} = \frac{1}{\Omegaper}\left(\intd_{\Omegaper} (\bE^h + \widetilde{\bE}^h) \dx + \intd_{\Rthree} \sum_{I}{\bE'^{h}}^{I} \dx\right)\,,
\end{equation}
where $\bE^h$ and $\widetilde{\bE}^h$ are the Eshelby tensors in equation~\eqref{eshelbyStress} expressed in their discrete form, involving the discrete wavefunctions $\bar{\bU}^h$ and the discrete electrostatic potentials $\bar{\varphi}^h$, $\vself {I^h}$.

\section{Results and discussion}
In this section, we discuss the accuracy and performance of the proposed configurational force approach on benchmark problems within the framework of spectral finite-element discretization. 

We first consider non-periodic systems involving CO, CH$_4$, N$_2$ and SiF$_4$ molecules. All-electron calculations have been performed on CO and CH$_4$ molecules, while norm-conserving Troullier-Martins pseudopotential in the Kleinman-Bylander form~\cite{tm91,bylander82} has been employed in the studies on N$_2$ and SiF$_4$ molecules. In these non-periodic benchmark examples, we examine the numerical rates of convergence of the finite-element approximation in the atomic forces for a given position of atoms, using a sequence of meshes with decreasing mesh sizes employing higher order finite-element interpolating polynomials. Further, the computed Kohn-Sham DFT ground-state energies as a function of bond lengths are fit using polynomial functions, and the derivative of these curves are compared with the computed atomic forces using the configurational force approach. 

We then consider all-electron and norm-conserving Troullier-Martins pseudopotential periodic calculations on systems comprising of Al face-centered cubic unit cell and Li body-centered cubic unit cell. In these periodic systems, we examine the numerical rates of convergence of the finite-element approximation in the atomic forces on perturbed atoms and the unit cell stresses, using a sequence of meshes with decreasing mesh sizes. Further, similar to the non-periodic studies, the Kohn-Sham ground-state energies are plotted as a function of the atomic displacement in the unit-cells and the derivatives of these curves are compared with the computed forces. The Kohn-Sham ground-state energies are also plotted as a function of the lattice parameter, and the derivative of these curves are compared with the computed stresses using the proposed configurational force approach. Wherever applicable, we benchmark the accuracies of forces and stresses obtained with calculations conducted using external DFT packages. We note that in all the above benchmark studies, the finite-element discretized Kohn-Sham problem is solved using the Chebyshev filtered subspace iteration method~\cite{motam2013} (ChFSI-FE), and the forces and stresses are evaluated in terms of discretized electronic fields involving orthogonal canonical eigenfunctions and electrostatic potentials obtained using ChFSI-FE procedure. 

Finally, we consider benchmark studies involving larger materials systems---aluminum nanocluster containing $5 \times 5 \times 5$ unit-cells ($666$ atoms) and an alkane chain containing $902$ atoms. Here, the finite-element discretized Kohn-Sham problem is solved using the subspace projection method (SubPJ-FE)~\cite{motam2014}, and the forces on all the atoms are computed in terms of discretized electronic fields involving non-orthogonal localized wavefunctions. These forces are then compared with those obtained using ChFSI-FE procedure involving orthogonal wavefunctions, thus quantifying the accuracy afforded by SubPJ-FE in atomic forces.

In all our simulations, we use the $n$-stage Anderson mixing scheme~\cite{andmix65} on the electron-density in self-consistent field iteration of the Kohn-Sham problem with a stopping criterion of $10^{-8}$ in the $L^{2}$ norm of the change in electron-density in two successive iterations. Further, all calculations involving atomic relaxations are conducted until atomic forces are below threshold values of $5 \times 10^{-6}$~Ha/Bohr.

\subsection{Non-periodic systems}
In this subsection, we present all-electron Kohn-Sham DFT calculations on isolated material systems involving carbon monoxide (CO) and methane molecule (CH$_4$), and pseudopotential calculations on nitrogen molecule (N$_2$) and silicon tetrafluoride molecule (SiF$_4$). We being by examining the convergence of the configurational force on atoms with respect to the finite element discretization, and assess the accuracy of the computed forces. Here and subsequently, we use the \textit{a priori} mesh adaption techniques developed by Motamarri et al.~\cite{motam2013} to construct the finite-element meshes, and refer to this prior work for the details.

We use quadratic (HEX27) and quartic (HEX125SPECT) spectral finite elements to study the numerical rates of convergence of the forces. To this end, a sequence of finite-element meshes is generated with increasingly smaller element sizes by uniformly subdividing the initial coarse mesh. Denoting by $h$, the measure of the size of finite-element, the magnitude of the discrete configurational force associated with one of the atoms, $\mathfrak{f}^h = |\boldsymbol{\mathfrak{f}}^{I^h}|$ (see equation~\eqref{forceAtoms}), is computed using the above sequence of meshes. The extrapolation procedure adopted in ~\cite{motam2013} is used to estimate the force magnitude in the limit as $h \rightarrow 0$, and is denoted by $\mathfrak{f}^0$. To this end, $\mathfrak{f}^h$ computed from the sequence of meshes using HEX125SPECT finite-elements is used to fit
\begin{equation}\label{fit}
|\mathfrak{f}^h - \mathfrak{f}^0| = C_f\left(\frac{1}{N_{el}}\right)^{p/3} \,,
\end{equation}
to determine $C_f$, $p$ and $\mathfrak{f}^0$. Here $N_{el}$ denotes the number of elements in the finite-element mesh. We use  $\mathfrak{f}^0$ as the reference value of the configurational force, and the relative error $|\mathfrak{f}^h - \mathfrak{f}^0|/|\mathfrak{f}^0|$ is plotted against $(1/N_{el})^{1/3}$. The obtained reference value, $\mathfrak{f}^0$, is also compared with that obtained from NWCHEM~\cite{nwchem} package for all-electron calculations and ABINIT~\cite{ABINIT} for pseudopotential calculations. We further verify the accuracy and variational nature of the computed forces using the finite difference test. To this end, we displace one of the atoms $I$ by small perturbations $+d$ and $-d$ in each of the spatial directions ($x$, $y$ and $z$) with $d$ chosen to be $1 \times 10^{-4}$ a.u.. The finite-difference force in each of the directions is computed to be $(E^{+d} - E^{-d})/(2*d)$, where $E^{+d}$ and $E^{-d}$ denote the discrete ground-state energies with the atom $I$ displaced by $+d$ and $-d$, respectively. Further, we fit the computed ground-state energies as a function of bond length with polynomials, and compare the derivative of these energy curves with the computed forces. In all the studies below, we choose a HEX125SPECT finite-element mesh with relative discretization error of around $10^{-5}$ in forces to verify the variational nature of the computed force. We now discuss the specific details of these calculations in each of the benchmark examples considered.
\subsubsection*{\normalsize{\textbf{All-electron calculations}}}
\begin{figure}[htbp]
\centering
\includegraphics[width=0.45\textwidth]{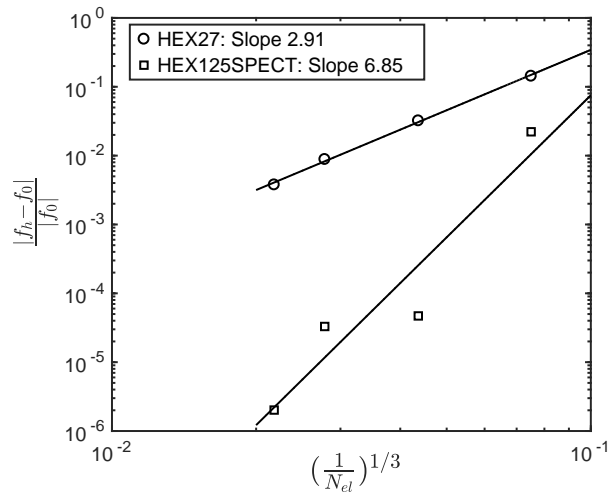}
\vspace{-0.8in}
\caption{\small{Convergence of the finite-element approximation in the force computed on the oxygen atom of CO molecule with a bond length of 2.4 Bohr.}}
\label{fig:convStudyCO}
\end{figure}

\paragraph{Carbon-monoxide}\hspace{0.01in}
We consider carbon-monoxide molecule with a C-O bond length of 2.4 a.u. enclosed within a simulation domain size of 80 a.u.. We first conduct the numerical convergence study on the force acting on the oxygen atom with quadratic (HEX27) and quartic (HEX125SPECT) spectral finite-elements. Figure~\ref{fig:convStudyCO} shows the relative errors in the force plotted against $(\frac{1}{N_{el}})^{1/3}$, which represents a measure of the mesh size $h$.  The value of $\mathfrak{f}^0$ computed from equation~\eqref{fit} is 0.202802087~Ha/Bohr, and is used to compute the relative errors in the force plot. Further, we note that the magnitude of atomic force computed by performing an all-electron calculation using pc-4 Gaussian basis~\cite{Jensen} with the NWCHEM~\cite{nwchem} package is 0.202830~Ha/Bohr. The slopes of the linear fit to the computed force in Fig.~\ref{fig:convStudyCO} provide the rates of convergence of the finite-element approximation for the force, and these results show close to $\order(h^{2k-1})$ convergence in the forces. 

Further, the force on oxygen atom is also computed using a finite-difference test, and the $l^2$ norm of the difference between the finite difference force and the computed force is $7.8 \times 10^{-6}$ Ha/Bohr. Moreover, the computed atomic force (negative of the configurational force) on oxygen atom is plotted as a function of interatomic distance in Fig. ~\ref{fig:forceBondlengthAllElec}, and compared with the negative derivative of the quartic polynomial fit to the ground-state energy plot (cf. Fig. 1 in Supplemental Material for the energy plot). These results confirm that the computed forces are variational.

Finally, we perturb the atomic positions such that interatomic distance differs by up to 20 percent from the equilibrium bond length and conduct geometry optimization to find the equilibrium position of atoms. The equilibrium bond length of carbon-monoxide obtained from atomic relaxation procedure is 2.1296 a.u., which is in very good agreement with the equilibrium bond length of 2.1294 a.u. obtained using NWCHEM.
\vspace{-0.6in}

\begin{figure}[h]
\centering
\includegraphics[width=0.45\textwidth]{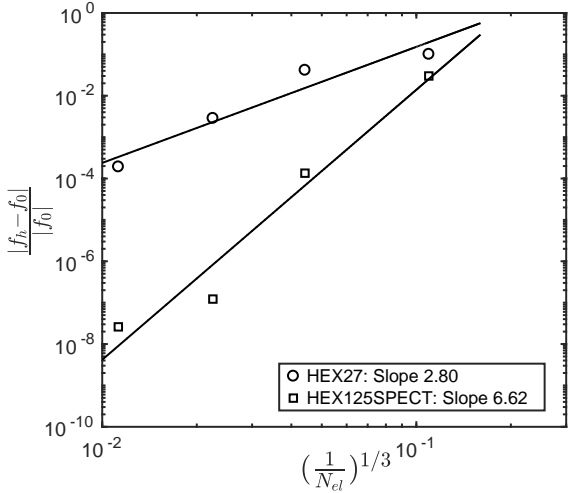}
\vspace{-0.8in}
\caption{\small{Convergence of the finite-element approximation in the magnitude of force computed on the hydrogen atom of CH$_4$ molecule with a C-H bond length of 2.424 Bohr.}}
\label{fig:convStudyCH4}
\end{figure}

\begin{figure}[h]
\centering
\includegraphics[width=0.43\textwidth]{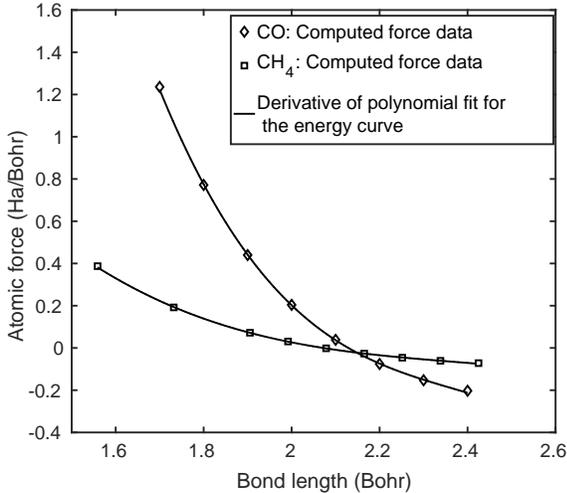}
\vspace{-0.8in}
\caption{\small{Comparison of the computed atomic force and the derivative of ground-state energy plot as a function of C-O and C-H bond lengths - Case study: CO and CH$_4$ molecules}}\label{fig:forceBondlengthAllElec}
\end{figure}

\vspace{0.1in}
\paragraph{Methane}\hspace{0.01in}
We consider methane molecule with C-H bond length of 2.424 a.u. and C-H-C tetrahedral of 109.47$^{0}$ enclosed within a simulation domain size of 80 a.u.. We first study the convergence of the magnitude of force acting on the hydrogen atom  with quadratic (HEX27) and quartic (HEX125SPECT) spectral finite-elements. Figure~\ref{fig:convStudyCH4}  shows the relative errors in the force magnitudes plotted against $(\frac{1}{N_{el}})^{1/3}$. The value of $\mathfrak{f}^0$ computed from equation~\eqref{fit} is 0.0718449~Ha/Bohr, which is used to compute the relative errors in the force. The magnitude of atomic force computed by performing an all-electron calculation using pc-4 Gaussian basis~\cite{Jensen} with the NWCHEM~\cite{nwchem} package is 0.0718628~Ha/Bohr. The rates of convergence of the finite-element approximation in forces for this system is also close to $\order(h^{2k-1})$.

In order to further test the accuracy and variational nature of the computed forces, we compute the finite-difference force obtained from finite differencing energies. The $l^2$ norm of the difference between the finite-difference force and the computed force is $8.3 \times 10^{-6}$ Ha/Bohr. Further, the Kohn-Sham DFT ground-state energies computed for various C-H bond lengths are fit to a fourth order polynomial (cf. Fig. 2 in the Supplemental Material), and the negative derivative of this energy curve is compared with the atomic force acting on the hydrogen atom along the C-H bond direction, which is shown in Fig. \ref{fig:forceBondlengthAllElec}.

Finally, we perturb the atomic positions such that the four C-H bond lengths differ by up to 20 percent from the equilibrium bond length and conduct geometry optimization to find the equilibrium position of atoms. The computed equilibrium bond length of C-H from atomic relaxation is 2.0735 a.u., which is in very good agreement with the equilibrium bond length of 2.0733 a.u. obtained using NWCHEM.
\vspace{-0.7in}
\vspace{-0.7in}

\subsubsection*{\normalsize{\textbf{Pseudopotential calculations}}}
\begin{figure}[h]
\centering
\includegraphics[width=0.45\textwidth]{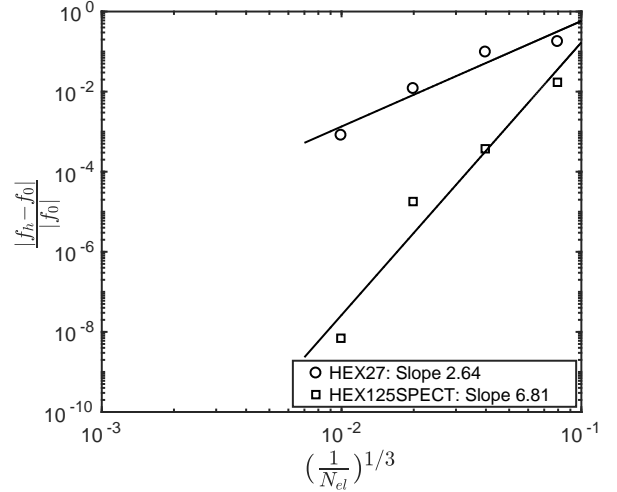}
\vspace{-0.8in}
\caption{\small{Convergence of the finite-element approximation in the force computed on the nitrogen atom of N$_2$ molecule with a bond length of 2.4 Bohr.}}
\label{fig:convStudyN2}
\end{figure}

\paragraph{Nitrogen molecule}\hspace{0.01in}
We consider nitrogen molecule with a N-N bond length of 2.4 a.u. enclosed within a simulation domain size of 80 a.u..  We consider the $2s$, $2p$, $3d$ angular momentum components to compute the projector, while the $4f$ component is chosen to be the local part of the non-local pseudopotential expressed in the Kleinman-Bylander form. The pseudopotentials for this system, as well as all other subsequent systems with pseudopotential calculations, are generated using the fhi98pp~\cite{fhi98pp} software. The default cut-off radius is used for $2s$, $2p$ and $3d$ components, which is 1.45 a.u.. The convergence of the force acting on the nitrogen atoms is studied using quadratic (HEX27) and quartic (HEX125SPECT) spectral finite-elements. Figure~\ref{fig:convStudyN2} shows the relative errors in the force plotted against $(\frac{1}{N_{el}})^{1/3}$, and the value of $\mathfrak{f}^0$ computed from equation~\eqref{fit} is 0.2779956~Ha/Bohr. We note that the force computed by performing a pseudopotential DFT calculation using plane-wave basis with the ABINIT~\cite{ABINIT} package is 0.2779960~Ha/Bohr. The slopes of the linear fits in Fig.~\ref{fig:convStudyN2}, which provide the rates of convergence of the finite-element approximation for the force, are close to $\order(h^{2k-1})$ as in the case of all-electron calculations.
\vspace{-0.7in}

The force on nitrogen atom is also computed from finite differencing the energies, and the $l^2$ norm of the difference between the finite-difference and the computed force is $6.7 \times 10^{-6}$ Ha/Bohr. Further, the computed atomic force on the nitrogen atom for different interatomic distances (N-N bond length) is provided in Fig. ~\ref{fig:forceBondlengthPSP}. Comparing these with the negative derivative of the quintic polynomial fit to the ground-state energies (cf. Fig. 3 in the Supplemental Material for the energy curve), confirms the variational nature of the computed forces.

\begin{figure}[h]
\centering
\includegraphics[width=0.45\textwidth]{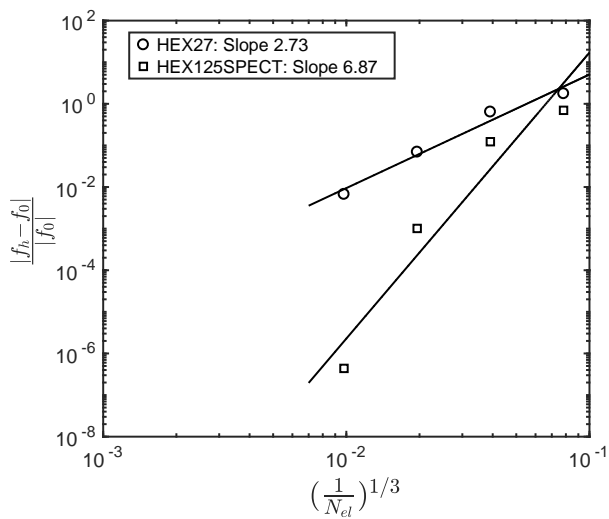}
\vspace{-0.7in}
\caption{\small{Convergence of the finite-element approximation in the magnitude of force computed on the fluorine atom of SiF$_4$ molecule with a  Si-F bond length of 3.464 Bohr.}}
\label{fig:convStudySiF4}
\end{figure}

\begin{figure}[h]
\centering
\includegraphics[width=0.45\textwidth]{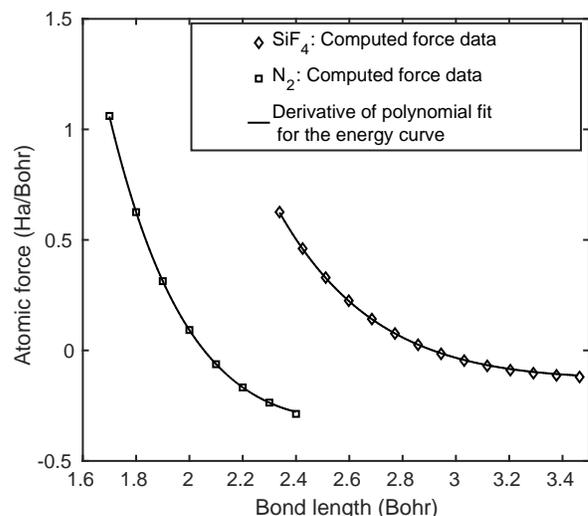}
\vspace{-0.8in}
\caption{\small{Comparison of the computed atomic force and the derivative of ground-state energy plot as a function of N-N and Si-F bond lengths - Case study: N$_2$ and SiF$_4$ molecules}}\label{fig:forceBondlengthPSP}
\end{figure}

Finally, we perturb the atomic positions such that interatomic distance differs by up to 20 percent from the equilibrium bond length and conduct geometry optimization to find the equilibrium position of atoms. The equilibrium bond length of nitrogen molecule obtained from atomic relaxation is 2.05509  a.u., which is in very good agreement with the equilibrium bond length of 2.05513 a.u. obtained using ABINIT.
\vspace{0.1in}
\paragraph{Silicon tetrafluoride}\hspace{0.01in}
We now consider silicon tetrafluoride molecule with Si-F bond length of 3.464 a.u. and F-Si-F tetrahedral angle of 109.47$^{0}$ enclosed within a simulation domain size of 80 a.u.. In the case of silicon, we consider the $3s$, $3p$, $3d$ angular momentum components to compute the projector, while the $4f$ component is chosen to be the local part of the nonlocal pseudopotential expressed in the Kleinman-Bylander form. The default cut-off radii used for $3s$, $3p$ and $3d$ components are 1.73 a.u., 1.9 a.u. and 2.03 a.u., respectively. In the case of fluorine, we consider the $2s$, $2p$, $3d$ components to compute the projector, while the $4f$ component is chosen to be the local part of the pseudopotential. The default cut-off radius is used for $2s$, $2p$ and $3d$ components, which is equal to 1.35 a.u. for all the components. The numerical convergence of the magnitude of force acting on one of the fluorine atoms with quadratic (HEX27) and quartic (HEX125SPECT) spectral finite-elements is shown in Fig.~\ref{fig:convStudySiF4}. The value of $\mathfrak{f}^0$ computed from equation~\eqref{fit} is 0.107000~Ha/Bohr, which is used to compute the relative errors in the force. Further, we note that the magnitude of atomic force obtained using plane-wave basis with ABINIT~\cite{ABINIT} package is 0.107002~Ha/Bohr. The rates of convergence of the finite-element approximation, given by the slopes of the linear fit to the data, is close to $\order(h^{2k-1})$.
\vspace{-0.7in}

In order to verify the variational nature of the computed forces, the force on the fluorine atom is also computed from finite-differencing the energies.  The $l^2$ norm of the difference between the finite-difference force and the computed force is $8.6 \times 10^{-6}$ Ha/Bohr. Further, the Kohn-Sham DFT ground-state energies computed for various Si-F bond lengths are fit to a sixth order polynomial (cf. Fig. 4 in the Supplemental Material for the energy curve) and the negative derivative of this polynomial fit to the ground-state energies is compared with the atomic force acting on the fluorine atom along the Si-F bond direction, which is shown in Fig. \ref{fig:forceBondlengthPSP}. These results confirm the variational nature of the computed forces.

Finally, we perturb the atomic positions such that the four Si-F bond lengths differ by up to 20 percent from the equilibrium bond length, and conduct geometry optimization to find the equilibrium position of atoms. The equilibrium bond length of Si-F achieved from atomic relaxation is 2.9130 a.u. which is found to be in very good agreement with the equilibrium bond length of 2.9134 a.u. obtained using plane-wave basis with ABINIT.
\vspace{-0.7in}


\subsection{Periodic systems}
In this subsection, we present all-electron and pseudopotential Kohn-Sham DFT calculations on periodic material systems involving aluminum (Al) face-centered (fcc) and lithium (Li) body-centered cubic (bcc) unit cells. The rates of convergence of the finite-element approximation in the configurational forces acting on a perturbed atom in Al face-centered cubic unit-cell and Li body-centered cubic unit-cell are examined. Further, the rates of convergence of the finite-element approximation in the hydrostatic stress obtained using the proposed configurational force approach is also studied. 

A sequence of quadratic (HEX27) and quartic (HEX125SPECT) spectral finite-element meshes with increasingly smaller mesh sizes are used to study the numerical rates of convergence of forces and stresses in the periodic systems considered here. The extrapolation procedure adopted in the previous section is used to estimate the force and stress in the limit $h \rightarrow 0$.  To this end, the force $\mathfrak{f}^h$ and the hydrostatic stress $\sigma^h$ computed from the sequence of meshes using HEX125SPECT finite-elements are fit to 
\begin{align}\label{perfit}
|\mathfrak{f}^h - \mathfrak{f}^0| &= C_f\left(\frac{1}{N_{el}}\right)^{p/3} \,,\\
|\sigma^h - \sigma^0| &= C_{\sigma}\left(\frac{1}{N_{el}}\right)^{p/3} \,,
\end{align}
in order to determine $C_f$, $C_{\sigma}$, $p$, $\mathfrak{f}^{0}$  and $\sigma^0$. The relative errors in force  $|\mathfrak{f}^h - \mathfrak{f}^0|/|\mathfrak{f}^0|$ and hydrostatic stress $|\sigma^h - \sigma^0|/|\sigma^0|$ are plotted as a function of $(1/N_{el})^{1/3}$. The reference values $\mathfrak{f}^{0}$ and $\sigma^0$ are benchmarked with the values obtained from ABINIT~\cite{ABINIT} for pseudopotential calculations. Further, in order to verify the variational nature of the computed forces and stress, we use a finite-difference test. In addition, as in the case of non-periodic calculation, we fit the ground-state energy data computed at various volumetric strains using a polynomial fit and compare the computed stresses with the derivative of the equation of state energy curve. In all the studies below, we choose a finite-element mesh with relative discretization error of around $10^{-5}$ in forces and stresses to verify the variational nature of the computed force and stresses.

We use the unshifted $2 \times 2 \times 2$ Monkhorst-Pack grid~\cite{mpgrid} for sampling the Brillouin zone in the all the studies reported below. Further, we employ the norm-conserving Troullier-Martins pseudopotential in the Kleinman-Bylander form~\cite{tm91,bylander82} in the case of pseudopotential calculations.
\vspace{-0.1in}
\subsubsection*{\normalsize{\textbf{All-electron calculations}}}
\begin{figure}[h]
\centering
\includegraphics[width=0.45\textwidth]{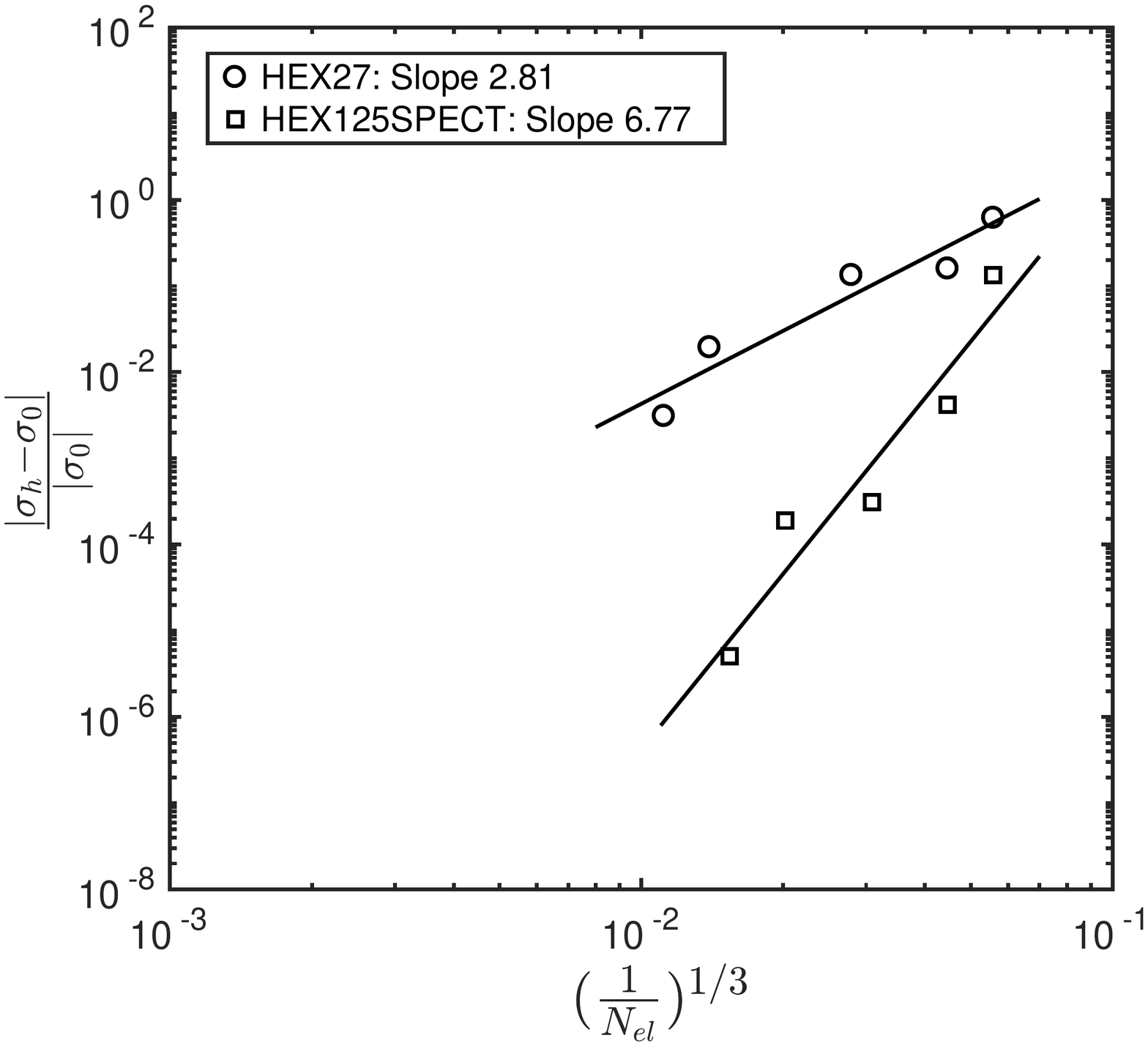}
\vspace{-0.8in}
\caption{\small{Convergence of the finite-element approximation in the hydrostatic stress of a fcc Al unit cell with lattice constant $a = 7.2$ Bohr (All-electron  study).}}\label{fig:convStudyStressAlAllElec}
\vspace{-0.3in}
\includegraphics[width=0.45\textwidth]{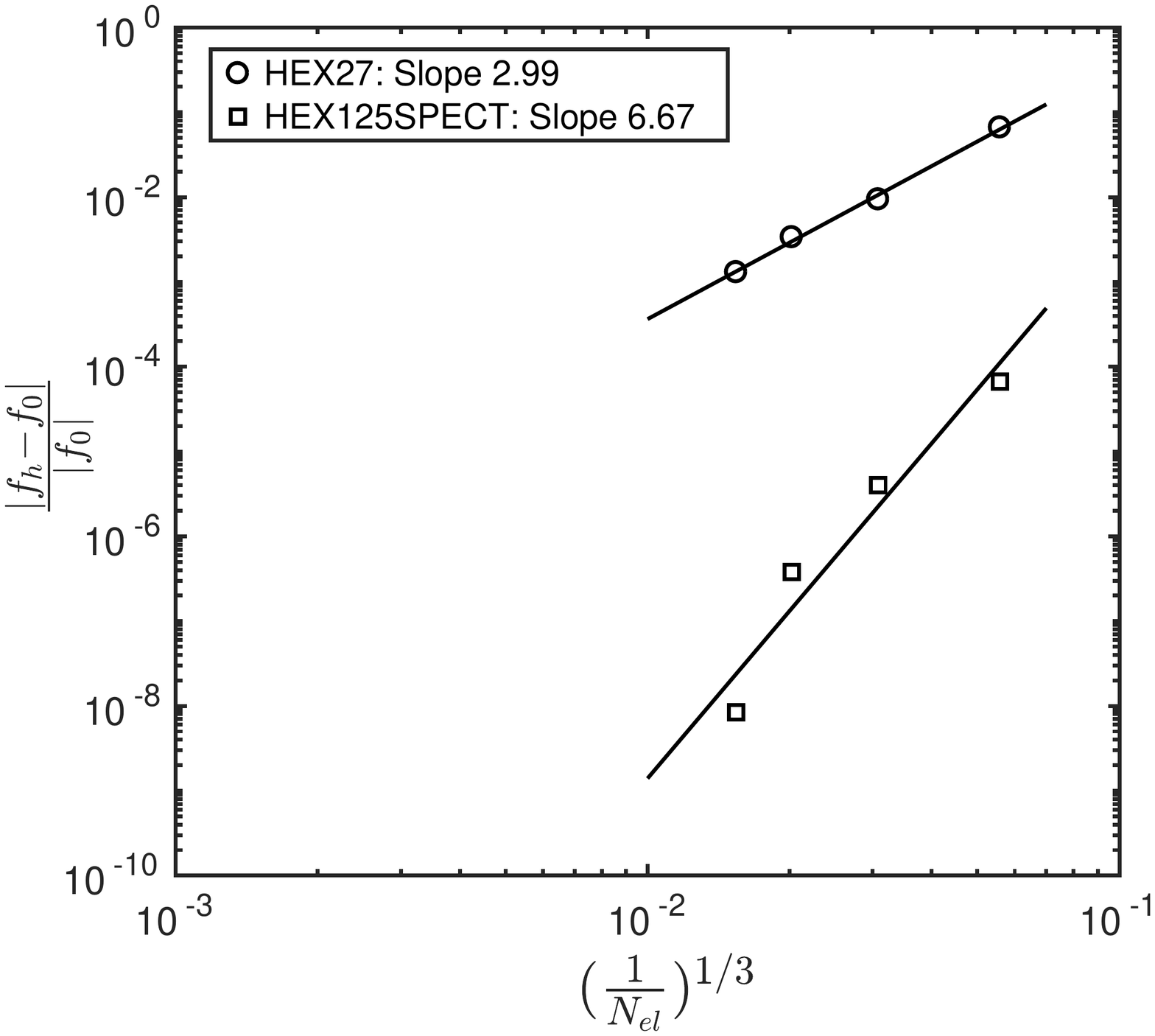}
\vspace{-0.8in}
\caption{\small{Convergence of the finite-element approximation in the magnitude of force computed on a perturbed atom in a fcc Al unit cell with lattice constant $a = 7.2$ Bohr (All-electron study).}}\label{fig:convStudyForceAlAllElec}
\end{figure}

\paragraph{Aluminum FCC unit-cell}\hspace{0.01in}
We consider all-electron Kohn-Sham DFT calculations on aluminum face-centered cubic unit cell with a lattice constant of 7.2 a.u., and study the convergence of the finite-element approximation in the hydrostatic stress using quadratic (HEX27) and quartic (HEX125SPECT) spectral finite-elements. Figure~\ref{fig:convStudyStressAlAllElec} shows the relative errors in the stress plotted against $(\frac{1}{N_{el}})^{1/3}$. The value of $\sigma^{0}$ computed from equation~\eqref{perfit} is $-4.51255993\times$ 10$^{-4}$~Ha/Bohr$^3$, and is used to compute the relative errors.  We next perturb an Al atom in the unit cell by 0.72 Bohr in [0 1 0] direction, and study the finite-element convergence of the atomic force acting on the perturbed atom. Figure~\ref{fig:convStudyForceAlAllElec} shows the relative errors in the forces. The value of $\mathfrak{f}^0$ computed from equation~\eqref{perfit} is 0.0624841~Ha/Bohr and is used to compute the relative errors. We note that the rates of convergence of the finite-element approximation in both stress and force are close to $\order(h^{2k-1})$.

\vspace{-0.7in}

\begin{figure}[h]
\centering
\includegraphics[width=0.45\textwidth]{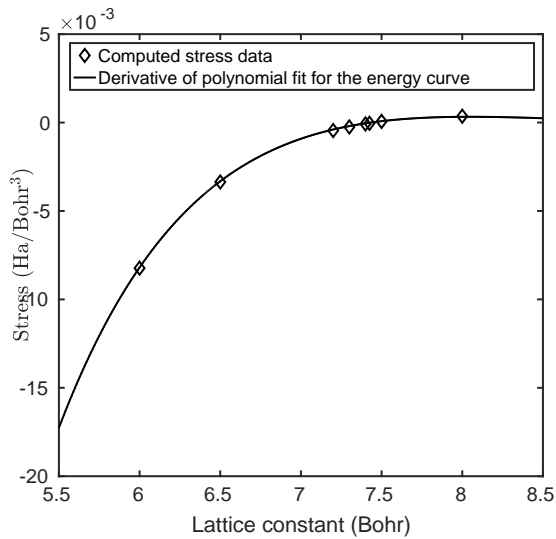}
\vspace{-0.7in}
\caption{\small{Variation in the stress as a function of lattice constant - Case study: Al fcc unit cell (All-electron study).}}\label{fig:stresslattParAlFCCAllElec}
\end{figure}

We also asses the variational nature of the computed stress and force using the finite difference test. The computed hydrostatic stress differs from the one obtained via finite difference of energies by $2.1\times10^{-7}$ Ha/Bohr$^3$, which is less than $0.01$ GPa. Similarly, the force on the perturbed Al atom is computed from finite-differencing the energies, and the $l^2$ norm of the difference between the finite-difference force and the computed force is $8.9 \times 10^{-6}$ Ha/Bohr. In addition, the hydrostatic stress is plotted as a function of lattice parameter in Fig. ~\ref{fig:stresslattParAlFCCAllElec}, and is compared with the derivative of the quartic polynomial fit of the ground-state energy computed for various lattice parameters (cf. Fig. 5 in the Supplemental Material for the energy curve). The comparison between the computed atomic force and the negative derivative of the quartic polynomial fit to the ground-state energy plot (cf. Fig. 6 in the Supplemental Material for the energy curve) is shown in Fig. ~\ref{fig:forcedispAllElec}. These results confirm the variational nature of the computed stresses and forces.
\vspace{-0.7in}
\vspace{0.1in}

\begin{figure}[h]
\centering
\includegraphics[width=0.45\textwidth]{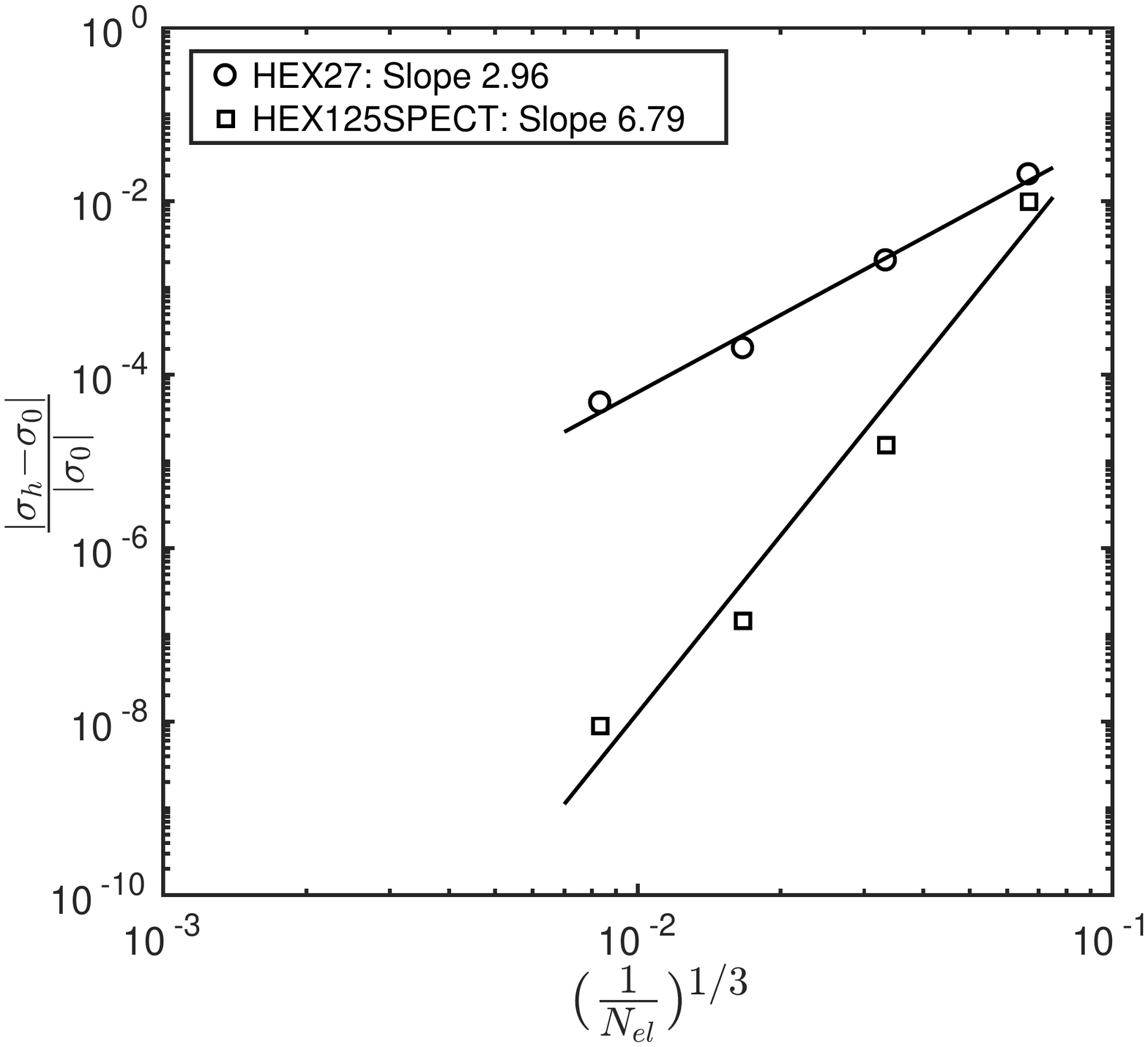}
\vspace{-0.8in}
\caption{\small{Convergence of the finite-element approximation in the hydrostatic stress of a bcc Li unit cell with lattice constant $a = 6.2$ Bohr (All-electron study).}}\label{fig:convStudyStressLiAllElec}
\vspace{-0.25in}
\includegraphics[width=0.45\textwidth]{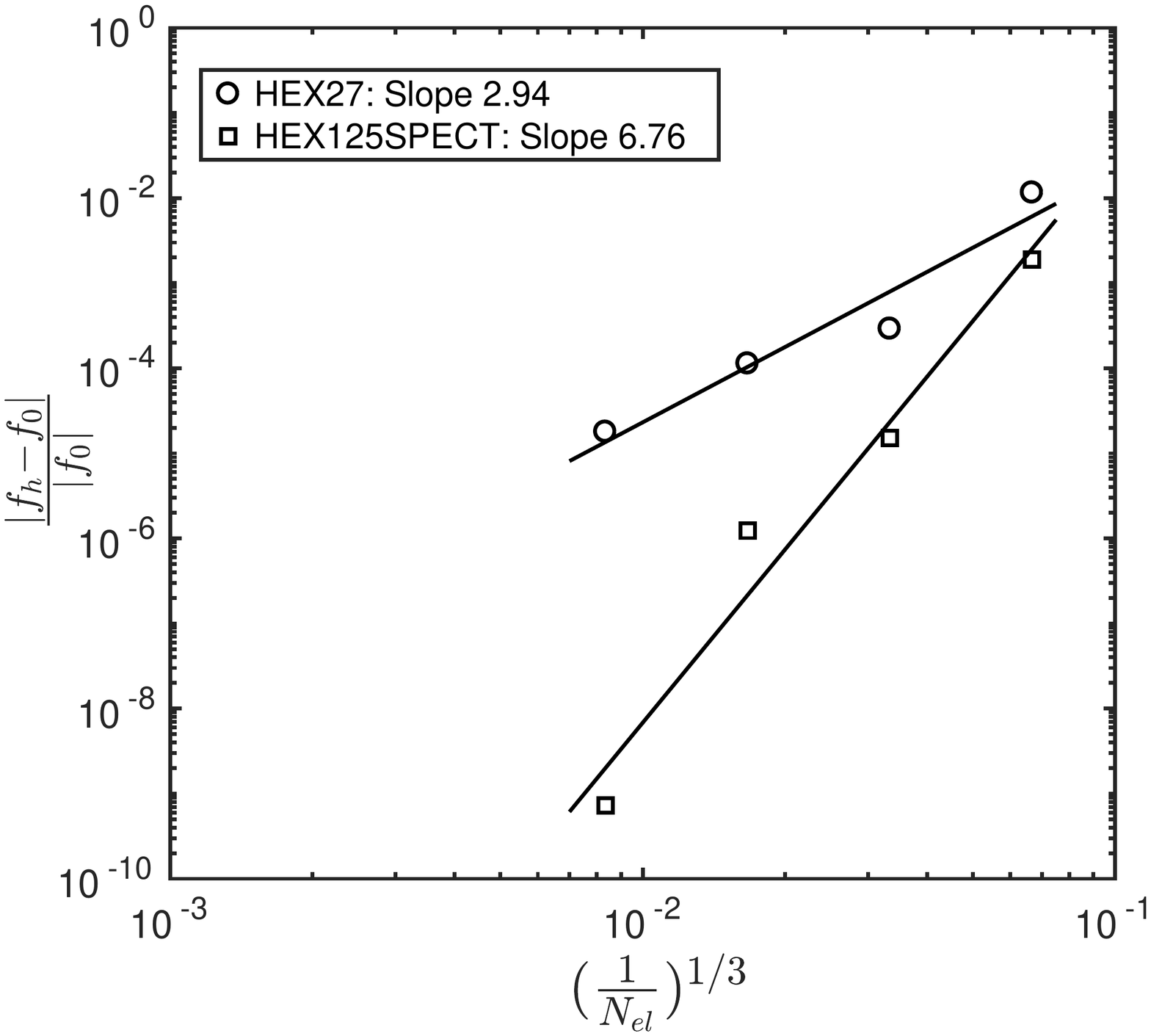}
\vspace{-0.79in}
\caption{\small{Convergence of the finite-element approximation in the magnitude of force computed on a perturbed atom in a bcc Li unit cell with lattice constant $a = 6.2$ Bohr (All-electron study).}}\label{fig:convStudyForceLiAllElec}
\end{figure}

\begin{figure}[h]
\vspace{-0.5in}
\includegraphics[width=0.45\textwidth]{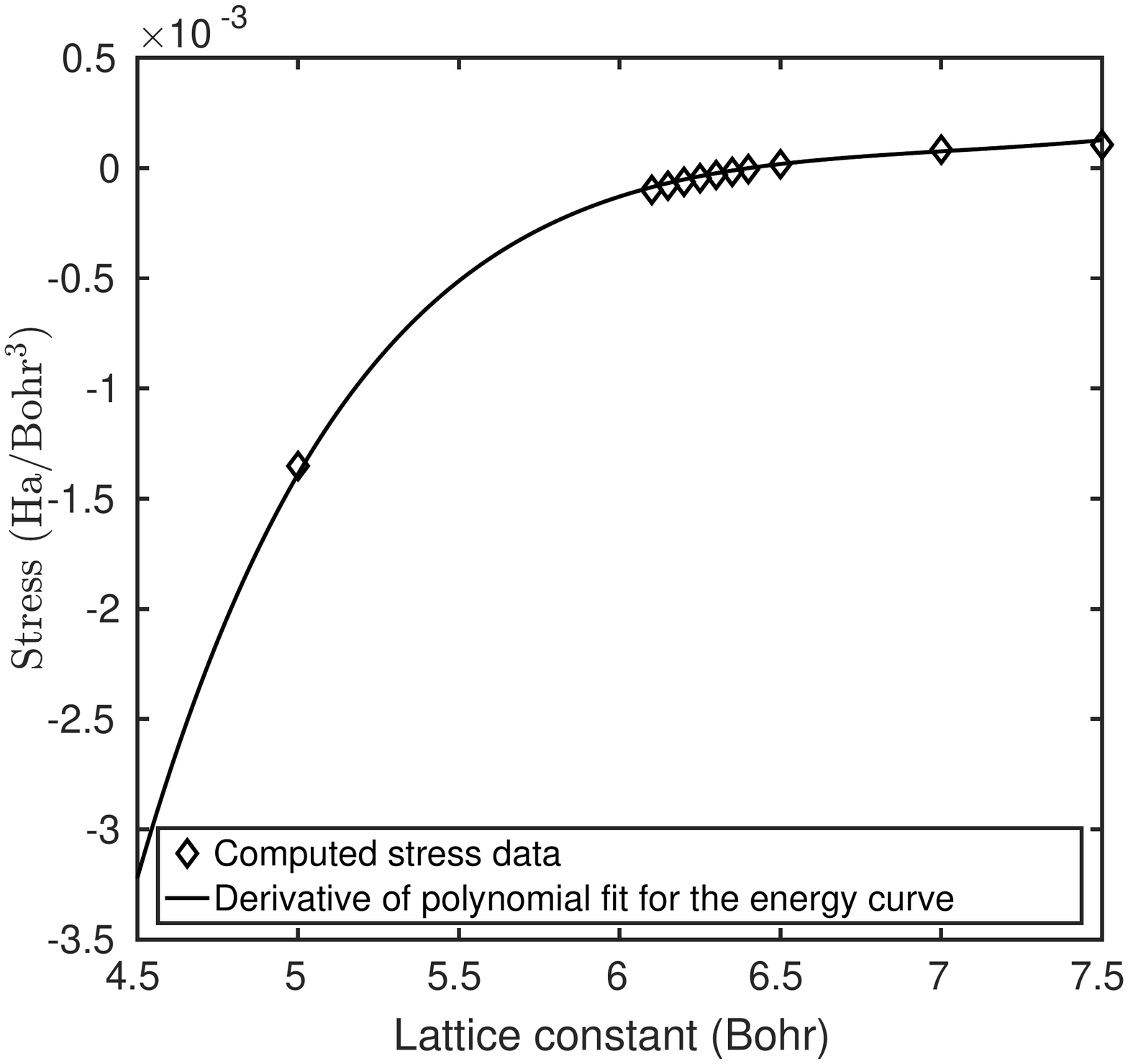}
\vspace{-0.79in}
\caption{\small{Variation in the stress as a function of lattice constant - Case study: Li bcc unit cell (All-electron study).}}\label{fig:stresslattParLiBCCAllElec}
\vspace{-0.2in}
\centering
\includegraphics[width=0.45\textwidth]{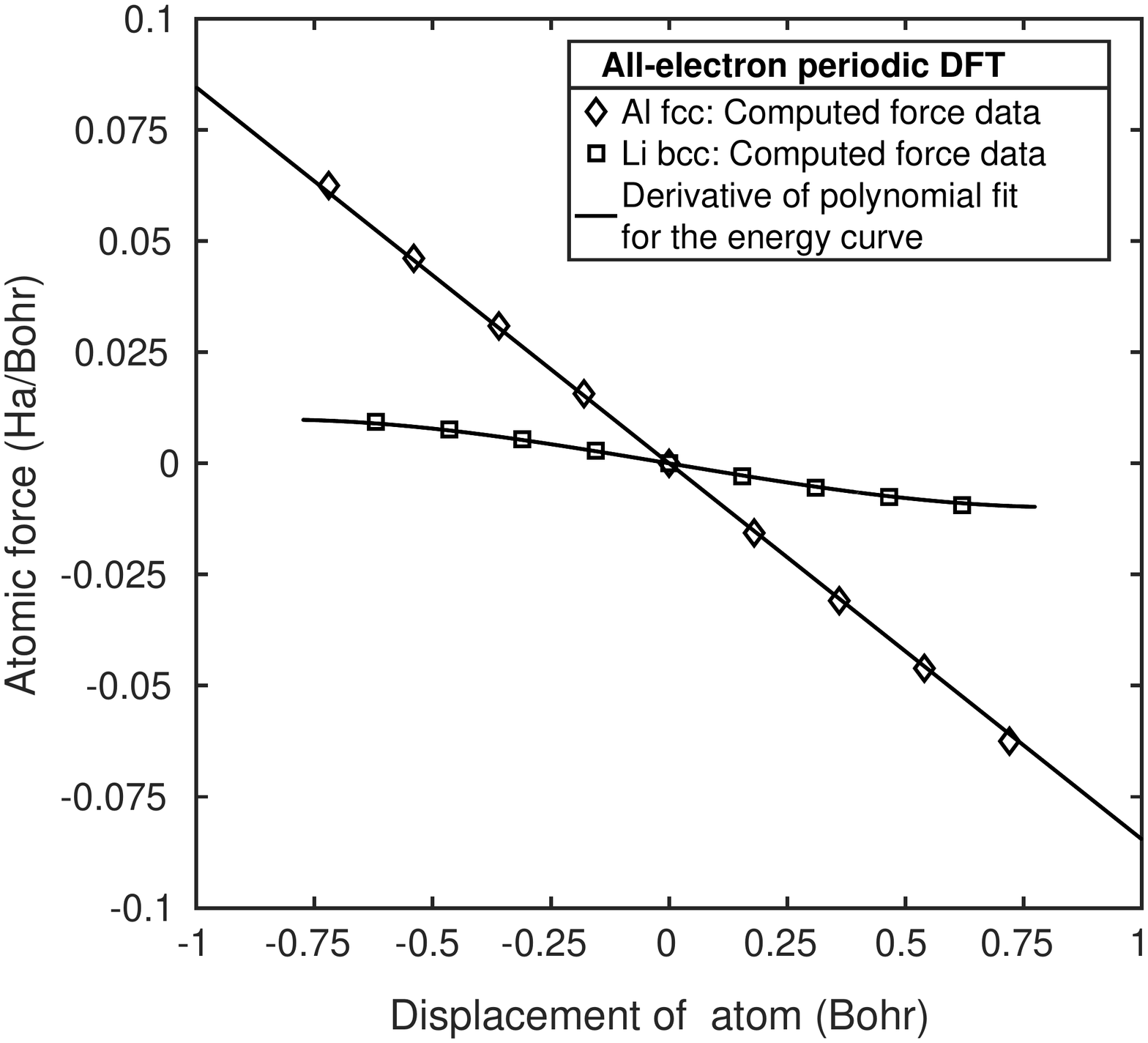}
\vspace{-0.79in}
\caption{\small{Comparison of the computed atomic force and the derivative of the ground-state energy plot as a function of atomic displacement. Case study: Al fcc unit cell and Li bcc unit cell (All-electron study).}}\label{fig:forcedispAllElec}
\end{figure}

\paragraph{Lithium BCC unit-cell}\hspace{0.01in}
We consider all-electron Kohn-Sham DFT calculations on lithium body-centered cubic unit cell with a lattice constant of 6.2 a.u., and conduct a similar study as in the case of aluminum fcc unit cell. Figure~\ref{fig:convStudyStressLiAllElec} shows the relative errors in the stress for quadratic and quartic spectral finite-elements, with $\sigma^{0} = -6.254757 \times 10^{-5}$~Ha/Bohr$^3$. Figure~\ref{fig:convStudyForceLiAllElec} shows the relative errors in the force on an atom which is perturbed by 0.62 Bohr in [1 0 0] direction ($\mathfrak{f}^0 = 0.009810153$~Ha/Bohr). These results suggest a close to $\order(h^{2k-1})$ convergence of the finite-element approximation for stresses and forces.

Further, the difference in computed hydrostatic stress and that obtained from finite differencing the energy differ by $3.6 \times 10^{-7}$ Ha/Bohr$^3$ ($\sim0.01$ GPa). The $l^2$ norm of the difference between the computed force and the finite-difference force on the perturbed Li atom is $6.4 \times 10^{-6}$ Ha/Bohr. In addition, Fig.  ~\ref{fig:stresslattParLiBCCAllElec} shows the comparison between the computed stress and the derivative of the quartic polynomial fit to the ground-state energy plot (cf. Fig. 7 in the Supplemental Material for the energy curve), and Fig. ~\ref{fig:forcedispAllElec} shows the comparison of the computed atomic force and the negative derivative of the quartic polynomial fit to the ground-state energy plot (cf. Fig. 8 in the Supplemental Material for the energy curve). These results validate variational nature of the computed stresses and forces.
\vspace{-0.5in}

\vspace{-0.2in}
\subsubsection*{\normalsize{\textbf{Pseudopotential calculations}}}
\begin{figure}[h]
\centering
\includegraphics[width=0.45\textwidth]{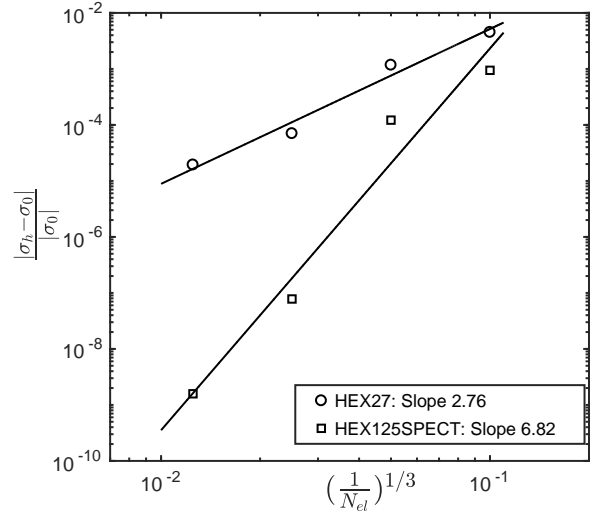}
\vspace{-0.8in}
\caption{\small{Convergence of the finite-element approximation in the hydrostatic stress of a fcc Al unit cell with lattice constant $a = 7.2$ Bohr (Pseudopotential study).}}\label{fig:convStudyStressAlPSP}
\end{figure}

\begin{figure}[h]
\includegraphics[width=0.45\textwidth]{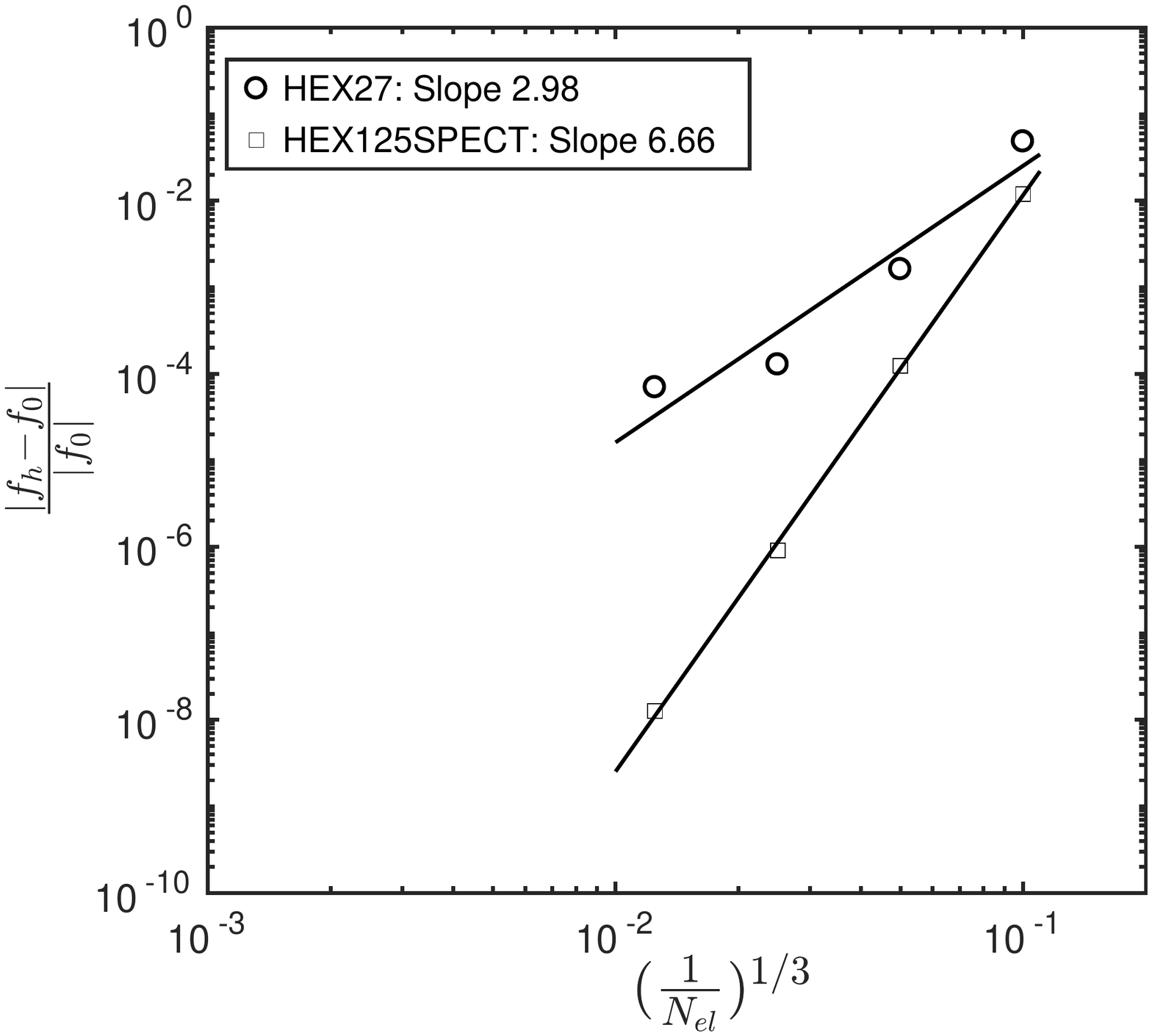}
\vspace{-0.8in}
\caption{\small{Convergence of the finite-element approximation in the magnitude of force computed on perturbed face atom of a fcc Al unit cell with lattice constant $a = 7.2$ Bohr (Pseudopotential study).}}\label{fig:convStudyForceAlPSP}
\vspace{-0.25in}
\centering
\includegraphics[width=0.45\textwidth]{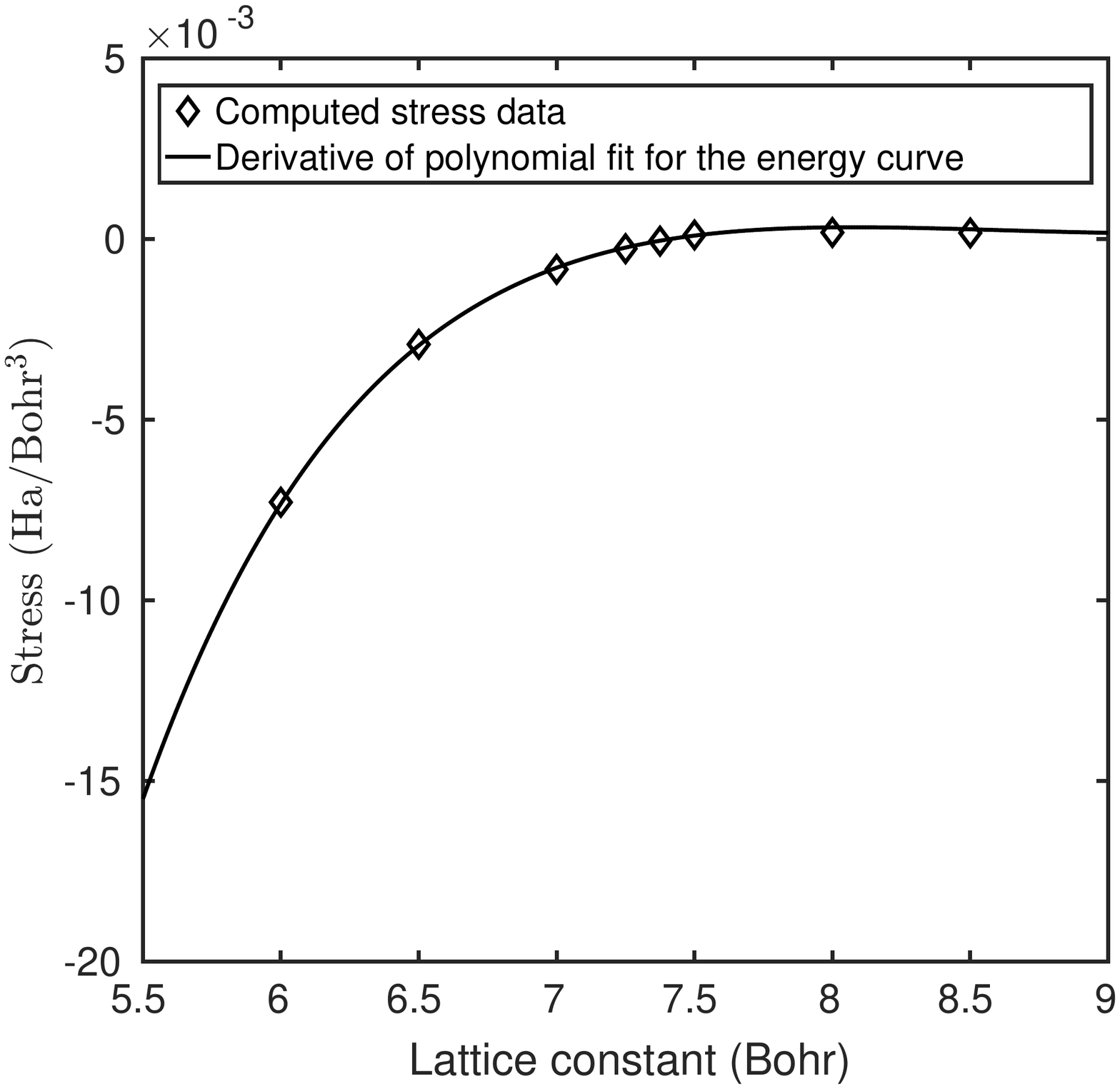}
\vspace{-0.8in}
\caption{\small{Variation in the stress as a function of lattice constant - Case study: Al fcc unit cell (Pseudopotential study).}}\label{fig:stresslattParAlFCCPSP}
\end{figure}

\paragraph{Aluminum FCC unit-cell}\hspace{0.01in}
We consider norm-conserving pseudopotential Kohn-Sham DFT calculations on aluminum face-centered cubic unit cell with a lattice constant of 7.2 a.u.. We consider the $3s$, $3p$, $3d$ angular momentum components to compute the projectors, while the $4f$ component is chosen to be the local part of the non-local pseudopotential expressed in Kleinman-Bylander form. Default cut-off radii  are chosen for $3s$, $3p$ and $3d$ components, which are 1.8 a.u., 2.0 a.u. and 2.15 a.u., respectively. Figure~\ref{fig:convStudyStressAlPSP} shows the relative errors in stress for finite-element approximations using quadratic (HEX27) and quartic (HEX125SPECT) spectral finite-elements, with $\sigma^{0}$ estimated to be $-3.6459132\times 10^{-4}$~Ha/Bohr$^3$. The hydrostatic stress obtained using plane-wave basis with ABINIT~\cite{ABINIT} package is $-3.6458596\times10^{-4}$~Ha/Bohr$^3$. Next, we perturb an atom in the unit cell by 0.72 Bohr in [0 1 0] direction, and Fig.~\ref{fig:convStudyForceAlPSP} shows the relative errors in the force on the perturbed atom ($\mathfrak{f}^0 = 0.0583693$~Ha/Bohr). The atomic force obtained using plane-wave basis with ABINIT~\cite{ABINIT} package is 0.0583694~Ha/Bohr. The rates of convergence obtained from the convergence study are close to $\order(h^{2k-1})$, as observed in the case of all-electron calculations. 

Further, the hydrostatic stress computed from finite-differencing the energies differs from the computed value by $1.3 \times 10^{-7}$ Ha/Bohr$^3$. The $l^2$ norm of the difference between the computed force on the perturbed Al atom and that obtained from finite differencing the energy is $7.5 \times 10^{-6}$ Ha/Bohr. This confirms the variational nature of the computed stress and force. Additionally, the hydrostatic stress is plotted against the lattice parameter in Fig. ~\ref{fig:stresslattParAlFCCPSP}, and is compared with the derivative of the quartic polynomial fit to the ground-state energy data (cf. Fig. 9 in the Supplemental material). The equilibrium lattice parameter obtained is 7.413 a.u., which is in good agreement with the equilibrium lattice parameter of 7.414 a.u. obtained using ABINIT. The atomic force on the atom perturbed along [0 1 0] direction is provided in Fig~\ref{fig:forcedispPSP}, and compared with the negative derivative of the quartic polynomial fit to the ground-state energy (cf. Fig. 10 in the Supplemental Material for the energy plot). These results further confirm the variational nature of the computed forces and stresses.
\vspace{-0.5in}

\begin{figure}[h]
\centering
\includegraphics[width=0.45\textwidth]{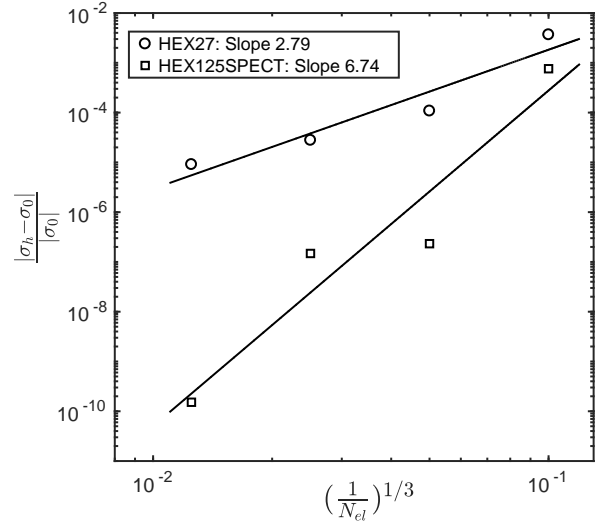}
\vspace{-0.8in}
\caption{\small{Convergence of the finite-element approximation in the hydrostatic stress of a bcc Li unit cell with lattice constant $a = 6.2$ Bohr (Pseudopotential study).}}\label{fig:convStudyStressLiPSP}
\end{figure}

\begin{figure}[h]
\includegraphics[width=0.45\textwidth]{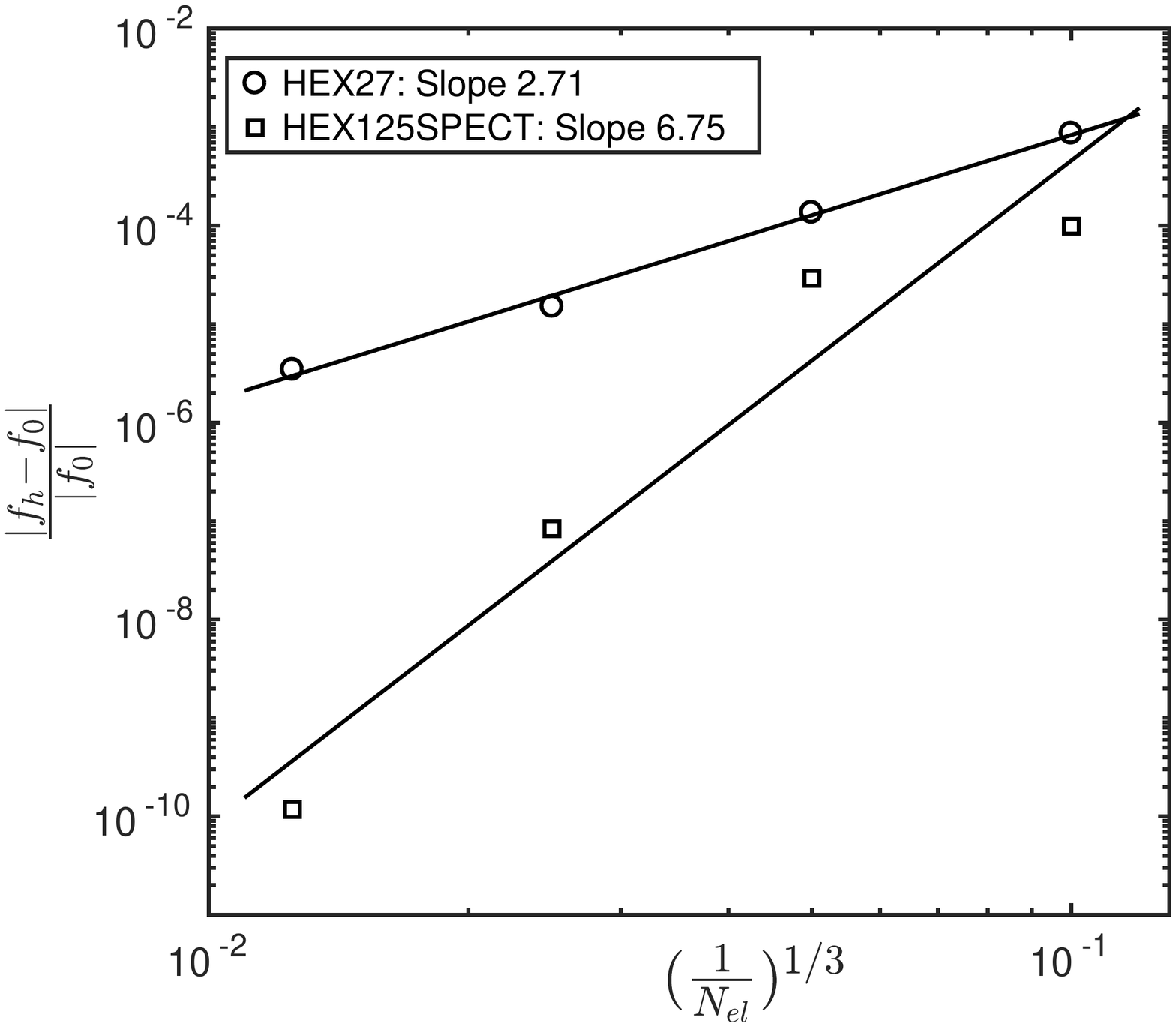}
\vspace{-0.79in}
\caption{\small{Convergence of the finite-element approximation in the magnitude of force computed on perturbed face atom of a bcc Li unit cell with lattice constant $a = 6.2$ Bohr (Pseudopotential study).}}\label{fig:convStudyForceLiPSP}
\vspace{-0.25in}
\includegraphics[width=0.45\textwidth]{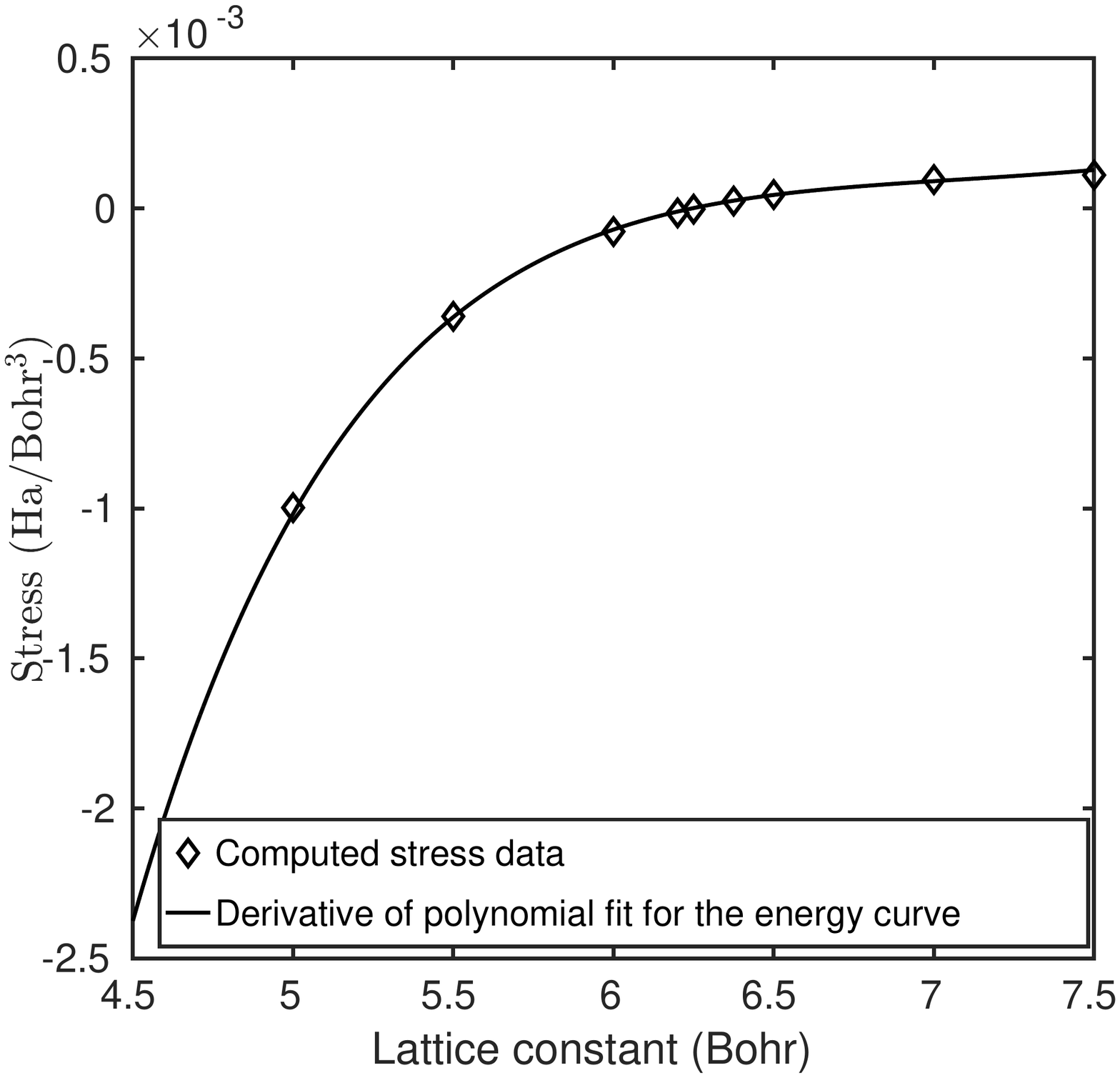}
\vspace{-0.79in}
\caption{\small{Variation in the stress as a function of lattice constant - Case study: Li bcc unit cell (Pseudopotential study).}}\label{fig:stresslattParLiBCCPSP}
\end{figure}

\begin{figure}[h]
\centering
\includegraphics[width=0.45\textwidth]{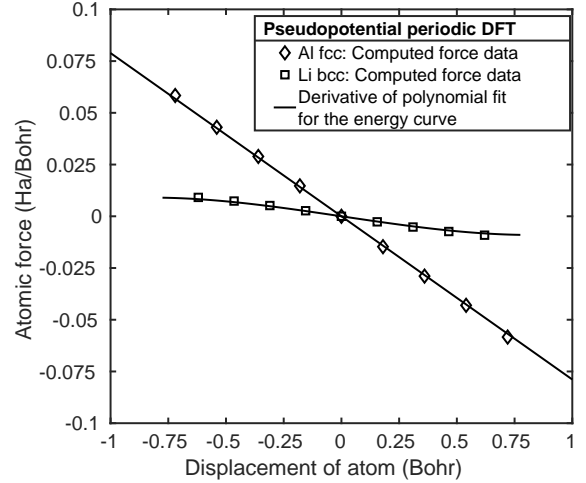}
\vspace{-0.79in}
\caption{\small{Comparison of the computed atomic force and the derivative of the ground-state energy plot as a function of atomic displacement. Case study: Al fcc unit cell and Li bcc unit cell (Pseudopotential study).}}\label{fig:forcedispPSP}
\end{figure}

\paragraph{Lithium BCC unit-cell}\hspace{0.01in}
Pseudopotential Kohn-Sham DFT calculations on lithium body-centered cubic unit cell with a lattice constant of 6.2 a.u are performed. We consider the $2s$, $2p$, $3d$ angular momentum components to compute the projectors, while the $4f$ component is chosen to be the local part of the non-local pseudopotential expressed in Kleinman-Bylander form. We use the default cut-off radii for $2s$, $2p$ and $3d$ components, which are 2.2 a.u., 2.2 a.u. and 2.5 a.u., respectively. Figure~\ref{fig:convStudyStressLiPSP} shows the relative errors in the stress for quadratic (HEX27) and quartic (HEX125SPECT) spectral finite-elements. The value of $\sigma^{0}$ computed from equation~\eqref{perfit} is $-1.521171 \times10^{-5}$ ~Ha/Bohr$^3$ and is used to compute the relative errors. We note that the hydrostatic stress obtained using plane-wave basis with ABINIT~\cite{ABINIT} package is $-1.521142 \times 10^{-5}$~Ha/Bohr$^3$. Further, we perturb a Li atom in the unit cell by 0.62 Bohr in all the three spatial directions, and study the convergence with respect to the finite-element discretization, which is presented in Fig.~\ref{fig:convStudyForceLiPSP}. The value of $\mathfrak{f}^0$ computed from equation~\eqref{perfit} is 0.02030927~Ha/Bohr, and is used to compute the relative errors. We note that the magnitude of atomic force obtained using plane-wave basis with ABINIT~\cite{ABINIT} package is 0.02030937~Ha/Bohr. Figures ~\ref{fig:convStudyStressLiPSP} and~\ref{fig:convStudyForceLiPSP} show close to $\order(h^{2k-1})$ rates of convergence for the finite element discretization, which is consistent will all previous studies.

Further, the computed hydrostatic stress differs from the value computed via finite differencing the energy by $7.3 \times 10^{-8}$ Ha/Bohr$^3$. Also, the $l^2$ norm of the difference between the computed force and that obtained from finite-differencing the energies is $5.8 \times 10^{-6}$ Ha/Bohr. Moreover, the computed hydrostatic stresses are plotted as a function of lattice parameter in Fig. ~\ref{fig:stresslattParLiBCCPSP}, and compared against the derivative of the quartic polynomial fit to the ground-state energy plot (cf. Fig. 11 in the Supplemental Material for the energy curve). The equilibrium lattice parameter obtained is 6.262 a.u., which is in good agreement with the equilibrium lattice parameter of 6.260 a.u. obtained using ABINIT. Figure ~\ref{fig:forcedispPSP} shows the comparison between the computed force for various values of the displacement of an atom along the [0 0 1] direction and the negative derivative of the quartic polynomial fit to the ground-state energy (cf. Fig. 12 in the Supplemental Material for the energy plot). These results validate the variational nature of the computed stresses and forces.
\vspace{-0.5in}

\vspace{-0.2in}
\subsection{Large material systems}
In this section, we investigate the accuracy of the proposed configurational force approach formulated within the framework of non-orthogonal Kohn-Sham wavefunctions using the subspace projection method (SubPJ-FE) proposed in ~\cite{motam2014}. In SubPJ-FE approach, the Kohn-Sham electronic ground-state is computed by solving the discrete Kohn-Sham eigenvalue problem ~\eqref{femEigen}, where each self-consistent field iteration (SCF) begins with a Chebyshev filter acting on a given subspace spanned by localized functions to compute an approximation to the occupied eigenspace. A localization procedure is subsequently employed to construct nonorthogonal localized wavefunctions spanning the Chebyshev filtered space, which are truncated using an adaptive truncation tolerance. The Kohn-Sham Hamiltonian is then projected into this localized basis, and a Fermi-operator expansion in terms of the projected Hamiltonian is employed to compute the electron-density. We note that in SubPJ-FE, the locality of the non-orthogonal wavefunctions in conjunction with the locality of the finite-element basis reduces the computational complexity of Kohn-Sham DFT, and has enabled large-scale electronic structure calculations using spectral finite-element discretization with reduced-order scaling as demonstrated in ~\cite{motam2014}. 

Upon computing the Kohn-Sham electronic ground-state for a given position of atoms using the SubPJ-FE approach, the atomic forces can be computed using the discrete form of ~\eqref{Eshelbyforce} which involves non-orthogonal localized truncated wavefunctions, as discussed in Section~\ref{sec:ConfigForces}. In order to assess the accuracy of the forces computed using the SubPJ-FE formulation, we compute the error between these forces and the atomic forces obtained using ChFSI-FE~\cite{motam2013} involving orthogonal wavefunctions. 

To this end, we consider representative benchmark systems involving an aluminum nanocluster, representative of a metallic system, containing $5 \times 5 \times 5$ unit-cells ($666$ atoms) with lattice spacing of 7.45 Bohr, and an alkane chains C$_{300}$H$_{602}$ ($902$ atoms), representative of an insulating system, with individual repeating units of CH$_2$ having C-C and C-H bond lengths to be 2.91018 and 2.0598 a.u, respectively. The H-C-H and C-C-C bond lengths are taken to be $109.47^{o}$. Finite-element meshes with fifth-order spectral finite-elements (HEX216SPECT) are chosen such that the discretization errors in ground-state energy are less than 5 meV per atom. SubPJ-FE approach is used to evaluate the Kohn-Sham ground-state as discussed in ~\cite{motam2014}. The compact support of non-orthogonal wave functions after the localization procedure is achieved by using truncation tolerances for localized wavefunctions which are chosen adaptively, with initial SCF iterations employing looser tolerances and progressively tightening these during the course of the SCF convergence. The adaptive truncation tolerances used in the current benchmark examples are identical to what has been proposed in ~\cite{motam2014}. 
Table~\ref{tab:comparison} tabulates $||\bff_i^c - \bff_i^s||_{\infty}$ where  $\bff_i^s$ denotes a vector of forces on all the atoms in the material system in the $i^{th}$ spatial direction where the Kohn-Sham electronic ground-state is obtained using SubPJ-FE approach, and $\bff_i^c$ denotes the vector of forces on all atoms where the Kohn-Sham ground state is obtained using ChFSI-FE. The results show that the maximum difference in the atomic forces between the two approaches is well within the chemical accuracy of 1 mHa/Bohr, which establishes the accuracy of SubPJ-FE in computing forces.
\begin{table}[htbp]
\caption{\small{Comparison of forces where the Kohn-Sham electronic ground state is computed using the SubPJ-FE technique~\cite{motam2014} and ChFSI-FE~\cite{motam2013}.}}
 \begin{center}
 \begin{tabular}{|c|c|c|c|}
   \hline
System & $||\bff_x^c - \bff_x^s||_{\infty}$  & $||\bff_y^c - \bff_y^s||_{\infty}$ & $||\bff_z^c - \bff_z^s||_{\infty}$  \\ \hline\hline
5x5x5 Al cluster & $8.9 \times 10^{-6}$  & $8.45 \times 10^{-6} $ & $ 8.11 \times 10^{-6}  $    \\ \hline
 C$_{300}$H$_{602}$  & $2.07 \times 10^{-5}$  & $1.48 \times 10^{-5}$  & $1.52 \times 10^{-5}$   \\ \hline
\end{tabular}
\end{center}\label{tab:comparison}
\end{table}

\vspace{-0.3in}
\section{Summary}
In the present work, we derived and implemented a configurational force approach to conduct geometry optimization using Kohn-Sham density functional theory (DFT) within the framework  of higher-order spectral finite-element discretization. The approach provides a unified expression for the generalized force corresponding to geometry of the simulation cell---atomic positions, as well as the simulation domain. The derived configurational force is variational, and inherently accounts for Pulay corrections arising from incomplete basis set errors. Further, the developed expressions treat both pseudopotential and all-electron DFT calculations for both periodic and non-periodic calculations in a single framework.

The development of the proposed approach involved the following main ideas. First, we employed a local variational real-space formulation of Kohn-Sham DFT functional  involving non-orthogonal wavefunctions that is amenable to reduced-order scaling. In particular, we have extended the local reformulation of the electrostatic interaction energy of prior works to treat both all-electron and pseudopotential calculations in a single framework. Using the local variational formulation, we derived the expressions for the generalized force corresponding to the perturbations of the underlying space. We note that the final form of the configurational force expression involves integrals of Eshelby tensors contracted with the gradient of the generator ($\bdir$) associated with the underlying deformation. We then employed a higher-order spectral finite-element basis, with atomic nuclei located at the nodes of the triangulation, to solve for the Kohn-Sham electronic ground state and used the discretized configurational force expression to evaluate the forces and stresses. The forces on atomic nuclei are evaluated by choosing an appropriate generator compactly supported around the atom of interest, while the stresses in periodic systems are evaluated by restricting the generator to affine deformations.

The accuracy of the proposed approach was investigated on representative benchmark examples. We first chose non-periodic systems involving all-electron calculations on CO, CH$_4$, and pseudopotential calculations on N$_2$, SiF$_4$ molecules. In these examples, we examined the numerical rates of convergence of higher-order finite-element approximation (quadratic and quartic orders) in computing forces, and the results  indicate a close to $\order(h^{2k-1})$ convergence in the forces for all systems. Further, a comparison between derivative of energy curves with computed forces ascertain the variational nature of the computed forces. We further extended these studies to periodic all-electron and pseudopotential calculations on Al fcc and Li bcc unit cells. As in the non-periodic case, we observed a close to $\order(h^{2k-1})$ convergence in stresses and atomic forces for both pseudopotential and all-electron calculations. We further confirmed the variational nature of the computed stresses and forces for these periodic systems.  Wherever applicable, we benchmarked the accuracies of forces and stresses obtained from the proposed configurational force approach with Kohn-Sham DFT calculations employing plane-waves (for pseuodopotential calculations) and Gaussian basis sets (for all-electron calculations), and found excellent agreement. Finally, we examined the accuracy of the forces computed when the Kohn-Sham electronic ground state is solved using a reduced-order scaling subspace projection technique~\cite{motam2014} on large materials systems. To this end, we chose metallic aluminum nanocluster containing 666 atoms and an alkane chain containing 902 atoms. The accuracy of computed atomic forces using the reduced-order scaling formulation is well within the required chemical accuracy.

The present work demonstrates a methodology for efficient computation of forces on nuclei and elastic stresses on periodic unit-cells by means of a configurational force approach within the framework of spectral finite-element discretization. To the best of our knowledge, expressions for computing both stresses and forces that are variationally consistent have not been developed for the finite-element discretization, and this work fills this important gap to enable large-scale electronic structure calculations using finite-element basis. Noting that the discrete Kohn-Sham ground-state energy is not only a function of position of atoms but also a function of position of nodes of the underlying finite-element triangulation, the configurational force approach also has the potential to be used as an \textit{a posteriori} mesh adaption strategy, which is a useful direction for future work. Further, developing and assessing the accuracy of configurational forces in the context of enriched finite-element basis~\cite{bikash2017} is currently being pursued, which can enable large-scale real-space all-electron DFT calculations on materials systems.

\vspace{-0.1in}
\appendix \label{sec:appendix}
\section{Derivation of Configurational force expressions}
\vspace{-0.1in}
We describe here the detailed derivation of the configurational force expressions given in Section~\ref{sec:ConfigForces} for both non-periodic and periodic DFT calculations. The notations followed here are consistent with that of Sections~\ref{sec:RS} \&~\ref{sec:ConfigForces}.
\vspace{-0.2in}
\subsection{Non-periodic DFT calculations}
\vspace{-0.1in}
In the case of non-periodic problems, $\Omega$ denotes a bounded domain large enough so that the values of wavefunctions are negligible outside of $\Omega$, i.e., we assume the wavefunctions have a compact support on $\Omega$. If $\bx$ denotes a point in $\Omega$, whose image in $\Omega' = \chieps(\Omega)$ is $\bx' = \chieps(\bx)$, the ground-state energy on $\Omega'$ is given by  $\mathcal{F}_0(\chieps)$ in equation ~\eqref{freenergynonper}. We now evaluate the configurational force by computing the Gateaux derivative of $\mathcal{F}_0(\chieps)$,  i.e., $\derveps \mathcal{F}_0(\chieps) \atzero = \derveps (\widetilde{\mathcal{L}}^{\eps} + \mathcal{L}_c^{\eps} + \mathcal{L}^{\eps}_{\text{el}}) \atzero$. We first focus on evaluating $\derveps \widetilde{\mathcal{L}}^{\eps} \atzero$ = $\derveps(T^{\eps}_{\text{s}} + E^{\eps}_{\text{xc}} + E^{\eps}_{\text{ent}}) \atzero$. In order to evaluate $\derveps T^{\eps}_{\text{s}} \atzero$, we first transform the integral in $T^{\eps}_{\text{s}}$,  which is defined on domain $\Omega'$ to domain $\Omega$ as:
\begin{widetext}
\begin{align}
T^{\eps}_{\text{s}}(\bGambar^{\phi^{\eps}},\bPhibar^{\eps}) &= \sum_{i,j,k=1}^{N} \int\displaylimits_{\Omega'} {\Gammabar^{\phi^{\scriptstyle \eps}}_{ij}}\, \invSeps {jk}\; \del_{\bx'} {\phiepsconj k}(\bx')\cdot\del_{\bx'} \phibar^{\eps}_i(\bx') \dx' \notag\\
&=\sum_{i,j,k=1}^{N} \int\displaylimits_{\Omega}\Bigl[{ \gammaphieps {ij}}\, \invSeps {jk}\left(\del_{\bx} {\phiepsconj k}(\chiepsx)\cdot\parder {\bx} {\bx'}\right)\cdot \left(\del_{\bx} \phibar^{\eps}_{i}(\chiepsx)\cdot\parder {\bx} {\bx'}\right) \text{det}(\parder {\bx'} {\bx})\Bigr] \dx
\end{align}
Now, the G\^{a}teaux derivative of $T^{\eps}_{\text{s}}$ is given by: 
\begin{gather}\label{derTs}
\derveps T^{\eps}_{\text{s}} \atzerob = A_1 + A_2 + A_3 + A_4; \\
A_1 =  \sum_{i,j,k=1}^{N}\int\displaylimits_{\Omega} \derveps ({ \gammaphieps {ij}})\atzerob \invSzero {jk}\, \del {\phizeroconj k}(\bx)\cdot\del {\phibar^{0}_i}(\bx) \dx\notag\;\;,\;\;
A_2 = \sum_{i,j,k=1}^{N}\int\displaylimits_{\Omega} {\gammaphizero {ij}} \derveps(\invSeps {jk})\atzerob\, \del {\phizeroconj k}(\bx)\cdot\del {\phibar^{0}_i}(\bx) \dx\notag\,,\\
A_3 = \sum_{i,j,k=1}^{N}\int\displaylimits_{\Omega} {\gammaphizero {ij}} \invSzero {jk}\derveps \left[\del {\phiepsconj k}(\chiepsx)\cdot\del {\phibar^{\eps}_i}(\chiepsx)\right]\atzerob d\bx\notag\,,\\
A_4 =  \sum_{i,j,k=1}^{N}\int\displaylimits_{\Omega} {\gammaphizero {ij}} \invSzero {jk}\Big\{\left[\del {\phizeroconj k}(\bx) \otimes \del \phibar^{0}_i(\bx) + \del {\phibar^{0}_i}(\bx) \otimes \del {\phizeroconj k}(\bx)\right]
: \derveps \parder {\bx} {\bx'} \atzerob+  \del {\phizeroconj k}(\bx).\del \phibar^{0}_i(\bx) \derveps \left(\text{det}(\parder \bx \bx')\right)\atzerob \Bigr\} \dx\notag\,.
\end{gather}
To evaluate $\derveps(\invSeps {jk})\atzero$ in $A_2$ in equation~\eqref{derTs}, we consider the variational derivative of the relation $\bS^{\eps}\invbSeps = \bI$ giving rise to $\derveps \invbSeps= - \invbSeps (\derveps \bS^{\eps}) \invbSeps$. To this end, we have the following expression for $\derveps(\invSeps {jk})\atzerob$:
\begin{equation}\label{derinvS}
\derveps \invSeps {jk}\atzerob = -\sum_{p,q=1}^{N}\left[\invSeps {jp}\derveps\left[\int\displaylimits_{\Omega}\left({\phiepsconj p }(\chiepsx)\phibar^{\eps}_q(\chiepsx)\detb{\parder {\bx'} {\bx}}\right)\dx\right]\invSeps {qk} \right]\atzerobb\,.
\end{equation}
Hence $A_2$ in equation ~\eqref{derTs} can be rewritten as $A_2 = A_{2a} + A_{2b}$ with $A_{2a}$ and $A_{2b}$ given by the following expressions:
\begin{align*}
A_{2a} &= \sum_{i,j,k=1}^{N}\int\displaylimits_{\Omega} {\gammaphizero {ij}} \derveps \left(S^{-1}_{jk}(\bPhibar^{\eps})\right)\atzerob\del {\phizeroconj k}(\bx)\cdot\del \phibar^{0}_i(\bx) \dx\,, \\A_{2b} &= -\sum_{i,j,k=1}^{N}\;\sum_{p,q=1}^{N}\int\displaylimits_{\Omega} {\gammaphizero {ij}} \invSzero {jp} \left(\int\displaylimits_{\Omega} {\phizeroconj p}(\bx) \phi^0_q(\bx)\derveps \left[\detb{\parder {\bx'} {\bx}}\right]\atzerob\dx\right)\invSzero {qk}\del {\phizeroconj k }(\bx)\cdot\del \phibar^{0}_i(\bx) \dx\,.
\end{align*}
We now evaluate $\derveps E^{\eps}_{\text{xc}} \atzero$ of $\derveps \widetilde{\mathcal{L}}^{\eps} \atzero$ in the following way:
\begin{align}\label{dervexc}
\derveps E^{\eps}_{\text{xc}} \atzerob &= \derveps \left[\int\displaylimits_{\Omega} F(\rhobar^{\eps}(\chieps)) \detb{\parder {\bx'} {\bx}} \dx\right]\atzerob = B_1 + B_2\,;
\end{align}
\text{with}\vspace{-0.1in}
\begin{align*}
B_1 = \int\displaylimits_{\Omega} \derveps \rhobar^{\eps}(\chiepsx) \atzerob \parder {F} {\rhobar^{\eps}}\atzerob \dx =\int\displaylimits_{\Omega} \derveps \rhobar^{\eps}(\chiepsx) \atzerob V_{\text{xc}}(\rhobar^{0}) \dx  \;\;,\;\;
B_2 = \int\displaylimits_{\Omega} F(\rhobar^0)\derveps \left(\detb{\parder {\bx'} {\bx}} \right)\atzerobb \dx\,.
\end{align*}
We now compute $\derveps E^{\eps}_{\text{ent}} \atzero$ and $\derveps \mathcal{L}_c^{\eps} \atzero$ as follows:
\begin{align}
\derveps E^{\eps}_{\text{ent}} \atzerob &= -2\sigma \left[\derveps {\bGambar^{\phi}}^{\eps} \atzerob \ln\left(\frac{{\bGambar^{\phi}}^{0}}{\bI-{\bGambar^{\phi}}^{0}}\right)\right] \label{entroDer}\,,\\
\derveps  \mathcal{L}_c^{\eps} \atzerob &= -2\mu \,\text{tr}\left(\derveps {\bGambar^{\phi}}^{\eps} \atzerob \right)\,. \label{constrDer}
\end{align}
Finally, we turn our attention to  $\derveps \mathcal{L}_{\text{el}}^{\eps} \atzero = \derveps (\mathcal{L}_{\text{el}}^{\text{ALL},\eps} + \mathcal{L}_{\text{el}}^{\text{PSP},\eps}) \atzero $. We first evaluate $\derveps \mathcal{L}_{\text{el}}^{\text{ALL},\eps} \atzero$. To this end, we first transform the integrals in $\mathcal{L}_{\text{el}}^{\text{ALL},\eps}(\rhobar^{\eps},\varphibar^{\eps},\bar{\mathcal{V}}^{\eps}_{{\dirac}},\bR^{\eps}) $  defined on domain $\Omega'$ to domain $\Omega$. We have
\begin{equation}\label{electroOmega'}
\begin{split}
\mathcal{L}^{\text{ALL},\eps}_{\text{el}}(\rhobar^{\eps},\varphibar^{\eps},\bar{\mathcal{V}}^{\eps}_{{\dirac}},\bR^{\eps}) = 
\int\displaylimits_{\Omega'}\Bigl[-\frac{1}{8\pi} |\del_{\bx'} \varphibar^{\eps}(\bx')|^2 &+ (\rhobar^{\eps}(\bx') + b(\bx',\chieps(\bR))\,)\varphibar^{\eps}(\bx')\Bigr]\dx'\\ 
&+ \sum_{I} \int\displaylimits_{\Rthree} \left[\frac{1}{8\pi} |\del_{\bx'} {\Vbar_{\dirac}^{I^{\eps}}}(\bx')|^2  +  Z_I \tilde{\delta}(|\bx' - \tau^{\eps}(\bR_I)|)\,\Vbar_{\dirac}^{I^{\eps}}(\bx') \right]\dx'\,.
\end{split}
\end{equation}
In the above, we note that $b(\bx',\chieps(\bR)) = \sum_{I}Z_I\dirac(|\bx' - \chieps(\bR_I)|) = \sum_{I}Z_I\dirac(|\bx - \bR_I|) = b(\bx,\bR)$, which follows from the restriction that $\chieps$ corresponds to rigid body deformations in the compact support of $b$. Transforming the integral in equation ~\eqref{electroOmega'} to domain $\Omega$, we obtain
\begin{equation}
\begin{split}
\mathcal{L}^{\text{ALL},\eps}_{\text{el}}(\rhobar^{\eps},\varphibar^{\eps},\bar{\mathcal{V}}^{\eps}_{{\dirac}},\bR^{\eps}) &= 
\int\displaylimits_{\Omega}\left[-\frac{1}{8\pi} \bigg|\del_{\bx} \varphibar^{\eps}(\chiepsx) \cdot \parder {\bx} {\bx'}\bigg|^2 + (\rhobar^{\eps}(\chiepsx) + b(\bx,\bR)\,)\varphibar^{\eps}(\chiepsx)\right] \detb{\parder {\bx'} {\bx}}\dx\\ 
&+ \sum_{I} \int\displaylimits_{\Rthree} \left[\frac{1}{8\pi} \bigg|\del_{\bx} \Vbar_{\dirac}^{I^{\eps}}(\chiepsx)\cdot \parder {\bx} {\bx'}\bigg|^2  +  Z_I \tilde{\delta}(|\bx - \bR_I|)\,\Vbar_{\dirac}^{I^{\eps}}(\chiepsx) \right]\detb{\parder {\bx'} {\bx}}\dx\,.
\end{split}
\end{equation}
Using a similar procedure as above, the G\^{a}teaux derivative of $\mathcal{L}^{\text{ALL},\eps}_{\text{el}}$ is evaluated as:  
\begin{equation}\label{dervelectro}
\derveps \mathcal{L}^{\text{ALL},\eps}_{\text{el}}(\rhobar^{\eps},\varphibar^{\eps},\bar{\mathcal{V}}^{\eps}_{{\dirac}},\bR^{\eps}) \atzerob = C_1 + C_2 + C_3\,;
\end{equation}
where
\begin{gather}
C_1 = \int\displaylimits_{\Omega} \derveps \rhobar^{\eps}(\chiepsx)\atzerob \varphibar^{0}(\bx)\,\dx \notag\,,\\
C_2 = \int\displaylimits_{\Omega}\Biggl\{\left[-\frac{1}{8\pi} |\del \varphibar^{0}(\bx)|^2 + (\rhobar^{0}(\bx) + b(\bx,\bR))\varphibar^{0}(\bx)\right]\derveps \left(\detb{\parder {\bx'} {\bx}} \right)\atzerob + \frac{1}{4\pi}\left[\del \varphibar^{0}(\bx) \otimes \del \varphibar^{0}(\bx) \boldsymbol{:} \left( \derveps \parder{\bx} {\bx'}\right)\atzerob \right]\Biggr\} \dx\notag\,,\\
C_3 = \sum_{I}\int\displaylimits_{\Rthree}\Biggl\{ \left[\frac{1}{8\pi} |\del \Vbar_{\dirac}^{I^{0}}(\bx)|^2 + Z_I \tilde{\delta}(|\bx - \bR_I|)\Vbar_{\dirac}^{I^0}(\bx) \right]\derveps \left(\detb{\parder {\bx'} {\bx}} \right)\atzerob - \frac{1}{4\pi}\left[\del \Vbar_{\dirac}^{I^{0}}(\bx) \otimes \del \Vbar_{\dirac}^{I^{0}} (\bx) \boldsymbol{:} \left( \derveps \parder{\bx} {\bx'}\right)\atzerob\right]\Biggr\} \dx \notag\,.
\end{gather}
In the above equation~\eqref{dervelectro}, the terms arising from the inner variations of  $\mathcal{L}^{\text{ALL},\eps}_{\text{el}}$ with respect to $\varphibar^{\eps}$ and $\Vbar_{\dirac}^{I^{\eps}}$ do not appear in the above equation as $\varphibar^{0}$ and $\Vbar_{\dirac}^{I^0}$ are the solutions of the variational problem \eqref{electroLocal} satisfying the Euler-Lagrange equations associated with the functional $\mathcal{L}_{\text{el}}^{\text{ALL},0}(\tau^{0})$. We now turn our attention to evaluate $\derveps \mathcal{L}_{\text{el}}^{\text{PSP},\eps} \atzero$. To this end, the integrals in $\mathcal{L}_{\text{el}}^{\text{PSP},\eps}({\bGambar^{\phi}}^{\eps},\bPhibar^{\eps},\bR^{\eps})$ defined on domain $\Omega'$ are transformed to domain $\Omega$, and we get
\begin{equation}
\begin{split}
&\mathcal{L}_{\text{el}}^{\text{PSP,}\eps}({\bGambar^{\phi}}^{\eps},\bPhibar^{\eps},\bR^{\eps}) = \sum_{J}\int\displaylimits_{\Omega} \Bigl(V^{J}_{\text{loc}}\left(|\chiepsx-\chieps(\bR_J)|\right) - \vself {J^\eps}\left(|\chiepsx-\chieps(\bR_J)|\right)\Bigr)\rhobar^{\eps}(\chiepsx) \detb{\parder {\bx'} {\bx}} \dx \\+\,& 2 \sum_{i,j,k=1}^{N} \int\displaylimits_{\Omega}  \gammaphieps {ij} \, \invSeps {jk} \,{\phiepsconj k} (\chiepsx) \Biggl( \int\displaylimits_{\Omega}V_{\text{nl}}\left(\chiepsx,\chieps(\by),\chieps(\bR)\right)\, \phibar^{\eps}_i(\chieps(\by)) \detb{\parder {\by'} {\by}}\dy\Biggr)  \detb{\parder {\bx'} {\bx}}\dx\,,
\end{split}
\end{equation}
with
\begin{equation*}
\begin{split}
&\int\displaylimits_{\Omega}V_{\text{nl}}\left(\chiepsx,\chieps(\by),\chieps(\bR))\, \phibar^{\eps}_i(\chieps(\by)\right) \detb{\parder {\by'} {\by}}\dy = \sum_J \sum_{lm} V_{lm}^{J} \zeta^{J}_{lm}(\chiepsx,\chieps(\bR_J)) \Delta V_{l}^{J}(|\chiepsx-\chieps(\bR_J)|)\\
&\times\!\int\displaylimits_{\Omega}\zeta^{J}_{lm}(\chiepsy,\chieps(\bR_J)) \Delta V_{l}^{J}(|\chiepsy-\chieps(\bR_J)|)\phibar^{\eps}_i(\chiepsy) \detb{\parder {\by'} {\by}}d\by.
\end{split}
\end{equation*}
The G\^{a}teaux derivative of $\mathcal{L}_{\text{el}}^{\text{PSP},\eps}$ can now be evaluated as
\begin{gather}\label{nonlocpsp}
\derveps \mathcal{L}^{\text{PSP},\eps}_{\text{el}}({\bGambar^{\phi}}^{\eps},\bPhibar^{\eps},\bR^{\eps}) \atzerob = D_1 + D_2 + D_3 + D_4 + D_5 + D_6 +D_7;\\
D_1 = \sum_{J} \int\displaylimits_{\Omega} \derveps\rhobar^{\eps}(\chiepsx)\atzerob \left(V^{J}_{\text{loc}}(|\bx-\bR_J|) - \vself {J^0}(|\bx-\bR_J|)\right)\dx \,,\notag\\
D_2 = \sum_{J} \int\displaylimits_{\Omega} \rhobar^{0}(\bx)\left(\del V^{J}_{\text{loc}}(|\bx-\bR_J|) - \del \vself {J^0}(|\bx-\bR_J|)\right)\cdot \left(\frac{d\chiepsx}{d\epsilon}\atzerob - \frac{d\chieps {(\bR_J)}}{d\epsilon}\atzerob\right)\dx \,, \notag \\
D_3 = \sum_{J} \int \displaylimits_{\Omega}\rhobar^{0}(\bx) \left(V^{J}_{\text{loc}}(|\bx-\bR_J|) - \vself {J^0}(|\bx-\bR_J|)\right)\derveps \left(\detb{\parder {\bx'} {\bx}} \right)\atzerob \dx \,,\notag\\
D_4 = 2\sum_{i,j,k=1}^{N}\int\displaylimits_{\Omega}\derveps ({ \gammaphieps {ij}})\atzerob {S^{0^{{\scriptstyle -1}}}_{jk}}{\phizeroconj k}(\bx) \Biggl(\int\displaylimits_{\Omega} V_{\text{nl}}(\bx,\by,\bR) \phibar^{0}_i(\by)\dy\Biggr)\dx \,,\notag\\
D_5 = 2\sum_{i,j,k=1}^{N}\int\displaylimits_{\Omega}{\gammaphizero {ij}} \derveps(\invSeps {jk})\atzerob {\phizeroconj k}(\bx) \Biggl(\int\displaylimits_{\Omega} V_{\text{nl}}(\bx,\by,\bR) \phibar^{0}_i(\by)\dy\Biggr)\dx \,,\nonumber\\
D_6 =   2\sum_{i,j,k=1}^{N} {\gammaphizero {ij}}{S^{0^{{\scriptstyle -1}}}_{jk}} (D_{ki} + D_{ik}^{*}) \,,\notag
\end{gather}
\vspace{-0.1in}
with
\begin{equation}
\begin{split}
D_{ki} &=   \sum_{J} \sum_{lm}  V_{lm}^{J}\Bigg[\int\displaylimits_{\Omega} {\phizeroconj k}(\bx)\biggl\{ \del \left(\zeta^{J}_{lm}(\bx,\bR_J)\Delta V_{l}^{J}(|\bx-\bR_J|)\right)\cdot\left(\frac{d\chiepsx}{d\epsilon}\atzerob - \frac{d\chieps {(\bR_J)}}{d\epsilon}\atzerob \right)\notag\\
& + \zeta^{J}_{lm}(\bx,\bR_J)\Delta V_{l}^{J}(|\bx-\bR_J|) \derveps \left(\detb{\parder {\bx'} {\bx}} \right)\atzerob\biggr\}\,\dx\Biggr] \Biggl[\int\displaylimits_{\Omega}\zeta^{J}_{lm}(\by,\bR_J)\Delta V_{l}^{J}(|\by-\bR_J|)\phibar^{0}_{i}(\by) \dy\Biggr] \,,\notag
\end{split}
\end{equation}
and, finally,
\begin{gather}
D_7 = 2\sum_{i,j,k=1}^{N}\int\displaylimits_{\Omega}{\gammaphizero {ij}}{S^{0^{{\scriptstyle -1}}}_{jk}} \left[\derveps {\phiepsconj k}(\chiepsx) \atzerob \Biggl(\int\displaylimits_{\Omega} V_{\text{nl}}(\bx,\by,\bR) \phibar^{0}_i(\by)\dy\Biggr) + {\phizeroconj k}(\bx)\Biggl(\int\displaylimits_{\Omega}V_{\text{nl}}(\bx,\by,\bR)\derveps\phibar^{\eps}_i(\chiepsy)\atzerob\dy\Biggr)\right]\dx \,,\notag
\end{gather}
where $()^{*}$ denotes the complex conjugate of $()$ in all the above equations. In all the above equations, we assume that the pseudo wavefunctions $\zeta^{J}_{lm}$ are real. Using the arguments presented in equation ~\eqref{derinvS} for evaluating the G\^{a}teaux derivative of $\invSeps {jk}$, $D_5$ in the above equation can be written as $D_5 = D_{5a} + D_{5b}$ with $D_{5a}$ and $D_{5b}$ given by the following expressions:
\begin{gather}
D_{5a} = 2\sum_{i,j,k=1}^{N}\int\displaylimits_{\Omega} {\gammaphizero {ij}} \derveps \left(S^{-1}_{jk}(\bPhibar^{\eps})\right)\atzerob {\phizeroconj k}(\bx)  \Biggl(\,\int\displaylimits_{\Omega} V_{\text{nl}}(\bx,\by,\bR) \phibar^{0}_i(\by)\dy\Biggr)\dx \,,\notag\\
D_{5b} = -2\sum_{i,j,k=1}^{N}\;\sum_{p,q=1}^{N}\int\displaylimits_{\Omega} {\gammaphizero {ij}} \invSzero {jp} \left(\int\displaylimits_{\Omega} {\phizeroconj p}(\bx) \phibar^0_q(\bx)\derveps \left[\detb{\parder {\bx'} {\bx}}\right]\atzerob\dx\right)\invSzero {qk} {\phizeroconj k}(\bx) \Biggl(\int\displaylimits_{\Omega} V_{\text{nl}}(\bx,\by,\bR) \phibar^{0}_i(\by)\dy\Biggr)\dx \,.\notag
\end{gather}
Using the expression for $\rho(\bx)$ in equation ~\eqref{rhoexp}, we now rewrite the sum of terms $B_1$, $C_1$ and $D_1$ in equations ~\eqref{dervexc}, ~\eqref{dervelectro} and ~\eqref{nonlocpsp} in terms of $V^{\text{loc}}_{\text{eff}}(\rhobar^{0}) = V_{\text{xc}}(\rhobar^{0}) + \varphibar^{0} +  \sum_{J} \left(V^{J}_{\text{loc}}(|\bx-\bR_J|) - \vself {J^{0}}(|\bx-\bR_J|)\right)$ and the wavefunctions $\bPhibar^{\eps}$ as:
\begin{align}\label{derrhoTimesloc}
&\int\displaylimits_{\Omega} \derveps \rhobar^{\eps}(\chiepsx)\atzerob  V^{\text{loc}}_{\text{eff}}(\rhobar^{0}) \dx  
= 2 \sum_{i,j,k=1}^{N} \int\displaylimits_{\Omega} \derveps \left({ \gammaphieps {ij}} \,\invSeps {jk} \,{\phiepsconj k}(\chiepsx)\, \phibar^{\eps}_i(\chiepsx) \right)\atzerob V_{\text{eff}}^{\text{loc}}(\rhobar^0) \dx 
 = F_1+F_2+F_{3a} + F_{3b} 
\end{align}
where
\begin{gather}
F_1 = 2\sum_{i,j,k=1}^{N}\int\displaylimits_{\Omega} \derveps ({ \gammaphieps {ij}})\atzerob {S^{0^{{\scriptstyle -1}}}_{jk}}{\phizeroconj k}(\bx)\, \phibar^{0}_i(\bx)\,V^{\text{loc}}_{\text{eff}}(\rhobar^0)\dx \,,\notag\\
F_2 = 2\sum_{i,j,k=1}^{N}\int\displaylimits_{\Omega} {\gammaphizero {ij}} \invSzero {jk}\derveps \left[{\phiepsconj k}(\chiepsx){\phibar^{\eps}_i}(\chiepsx)\right]\atzerob \,V^{\text{loc}}_{\text{eff}}(\rhobar^0) \dx\notag\,,\\
F_{3a} = 2\sum_{i,j,k=1}^{N}\int\displaylimits_{\Omega} {\gammaphizero {ij}} \derveps \left(S^{-1}_{jk}(\bPhibar^{\eps})\right)\atzerob{\phizeroconj k}(\bx)\phibar^{0}_i(\bx) V^{\text{loc}}_{\text{eff}}(\rhobar^0)\dx\notag\,,\\
F_{3b} = -2\sum_{i,j,k=1}^{N}\;\sum_{p,q=1}^{N}\int\displaylimits_{\Omega} {\gammaphizero {ij}} \invSzero {jp} \left(\int\displaylimits_{\Omega} {\phizeroconj p} (\bx) \phibar^0_q(\bx)\derveps \left[\detb{\parder {\bx'} {\bx}}\right]\atzerob\dx\right)\invSzero {qk} {\phizeroconj k}(\bx) \phibar^{0}_i(\bx) V^{\text{loc}}_{\text{eff}}(\rhobar^0)\dx\,.\notag
\end{gather}
We further note that the sum of terms $A_{2a}+A_3 + D_{5a} + D_{7}+F_2+F_{3a} = 0$ as $\bPhibar^{0}$ is the solution of the saddle point variational problem corresponding to $\mathcal{F}_{0}(\tau^{0})$ and hence satisfy the Euler-Lagrange equations corresponding to equation ~\eqref{saddlepoint}. Furthermore, the sum of terms $A_1 + D_4 + F_1+\derveps E^{\eps}_{\text{ent}} \atzerob + \derveps  \mathcal{L}_c^{\eps} \atzerob = 0$  since ${\bGambar^{\phi}}^{0}$ satisfy the Fermi-Dirac distribution (Euler-Lagrange equation of $\mathcal{F}_0$ with respect to $\bGam^{\phi}$).
We now note the following identities:
\begin{gather}
\derveps \left\{\frac{\partial x_i}{\partial x_j'}\right\}\atzerob  = -\frac{\partial x_i}{\partial x_k'}\left(\derveps\frac{\partial x_k'}{\partial x_l} \right)\frac{\partial x_l}{\partial x_j'}\atzero = -\frac{\partial \dir_i(\bx)}{\partial x_j},\notag\\
\derveps \left\{\det\left(\frac{\partial x_i'}{\partial x_j}\right)\right\}\atzerob = \derveps \left\{\det\left(\delta_{ij} + \varepsilon \frac{\partial \dir_i(\bx)}{\partial x_j}\right)\right\}\atzerob 
= \derveps \left\{1 + \frac{\partial \dir_j(\bx)}{\partial x_j}\varepsilon + \order(\varepsilon^2)\right\}\atzerob = \frac{\partial \dir_j(\bx)}{\partial x_j}.\label{identities}
\end{gather}
Using these identities and rearranging the non-zero terms $A_{2b}$, $A_4$, $B_2$, $C_2$, $C_3$, $D_2$, $D_3$, $D_{5b}$, $D_6$, $F_{3b}$  we arrive at
\begin{equation}\label{EshelbyforceApp}
\frac{d\mathcal{F}_0(\chieps)}{d\varepsilon}\atzerob = \int\displaylimits_{\Omega} \bE:\del \bdir (\bx) \dx + \sum_{I} \int\displaylimits_{\Rthree} \bE'^I:\del \bdir (\bx) \dx \,+\, \text{F}^{\text{PSP}}
\end{equation}
where $\bE$ and $\bE'$ denote Eshelby tensors whose expressions in terms of the solutions of the saddle point problem ~\eqref{saddlepoint} on the original space  ($\bGambar^{\phi^0}$, $\bPhibar^{0}$, $\varphibar^0$, $\bar{\mathcal{V}}_{\dirac}^{0}$) are given in Section IV of the paper.
\end{widetext}

\subsection{Periodic DFT calculations}
 We follow the similar procedure as described  in the case of non-periodic calculations. As defined before,  $\bx$ denotes a point in $\Omegaper$, whose image in $\Omegaper' = \chieps(\Omegaper)$  is $\bx' = \chieps(\bx)$ with $\bdir = \derveps \chieps\atzero$ being the generator of the underlying deformation. Further let  $\bk' = \boldsymbol{\kappa}^{\eps}(\bk)$ correspond to the bijective mapping representing the infinitesimal perturbation of the  reciprocal space due to the underlying deformation of the real-space. 

The expression for configurational force in the present case will differ from  equation~\eqref{EshelbyforceOrth} derived in the case of non-periodic calculations in terms of the contributions arising from the kinetic energy term $T_s(\bU,\bff)$ in equation ~\eqref{kinEngyBloch2} and the nonlocal pseudopotential energy term $\mathcal{L}^{\text{nl}}(\bU,\bff)$ in equation~\eqref{Vnl}. To this end, we evaluate $\derveps (T_s^{\eps} + \mathcal{L}_{\text{nl}}^{\eps})\atzero$ where $T_s^{\eps}$ and $\mathcal{L}_{\text{nl}}^{\eps}$ are the functionals defined on $\Omegaper'$.  To evaluate $\derveps (T_s^{\eps} +  \mathcal{L}_{\text{nl}}^{\eps}) \atzero$, we transform the integrals in $T_s^{\eps}$ and $\mathcal{L}_{\text{nl}}^{\eps}$ defined on domain $\Omegaper'$ to  domain $\Omegaper$ as follows.
\begin{widetext}
\vspace{-0.1in}
We first consider the functional $T_s^{\eps}$ in equation~\eqref{kinEngyBloch2}. To this end, we have,
\begin{equation*}
T_s^{\eps}(\bffbar^{\eps},\bUbar^{\eps}) = \sum_{n=1}^{N}\fintd_{BZ}\Biggl(\intd_{\Omegaper}g(\fbar_{n\bk}^{\eps}, \ubarn^{\eps}_{n\bk}(\chiepsx),\uepsconj {n\bk}(\chiepsx), \del_{\bx} \ubarn^{\eps}_{n\bk}(\chiepsx) \cdotp \parder {\bx}{\bx'}, \del_{\bx} \uepsconj {n\bk}(\chiepsx) \cdotp \parder {\bx}{\bx'},\,\bkappa^{\eps}(\bk))\text{det}\left(\parder {\bx'} {\bx}\right)\dx\Biggr)\dk \,,
\end{equation*}
where
\begin{equation*}
\begin{split}
&g(\fbar_{n\bk}^{\eps}, \ubarn^{\eps}_{n\bk}(\chiepsx),\uepsconj {n\bk}(\chiepsx), \del_{\bx} \ubarn^{\eps}_{n\bk}(\chiepsx) \cdotp \parder {\bx}{\bx'}, \del_{\bx} \uepsconj {n\bk}(\chiepsx) \cdotp \parder {\bx}{\bx'},\,\bkappa^{\eps}(\bk))\\
&=\fbar_{n\bk}^{\eps}\,\,\Bigl( \left|\del_{\bx} \ubarn^{\eps}_{n\bk}(\chiepsx) \cdotp \parder {\bx}{\bx'} \right|^2 - 2\, i \,\uepsconj {n\bk}(\chiepsx) \,\bkappa^{\eps}(\bk)\cdotp \left(\del_{\bx} \ubarn_{n\bk}^{\eps}(\chiepsx)\cdotp \parder {\bx}{\bx'}\right) \,
+ |\bkappa^{\eps}(\bk)|^2\, |\ubarn^{\eps}_{n\bk}(\chiepsx)|^2 \Bigr)\,.
\end{split}
\end{equation*}
Subsequently, the G\^{a}teaux derivative of  $T_s^{\eps}(\bU^{\eps},\bff^{\eps})$ is given as
\begin{gather}
\derveps T_s^{\eps}(\bffbar^{\eps},\bUbar^{\eps})  \atzerob = G_1 + G_2 + G_3 + G_4 \,;\\
\begin{split}
G_1 &= \sum_{n=1}^{N}\fintd_{BZ}\intd_{\Omegaper}\fbar_{n\bk}^{0} \Biggl\{\Bigl[\del \uzeroconj {n\bk}(\bx) \otimes \del \ubarn^{0}_{n\bk}(\bx) + \del \ubarn^{0}_{n\bk}(\bx) \otimes \del \uzeroconj {n\bk}(\bx)- 2\,i \,\uzeroconj {n\bk}(\del \ubarn^{0}_{n\bk} \otimes \bk)\Bigr]: \derveps \parder {\bx} {\bx'} \atzerob \notag\\
&+ \Bigl(|\del \ubarn_{n\bk}^{0}(\bx)|^2 - 2\,i\, \uzeroconj {n\bk}(\bx) \,\bk\cdotp \del \ubarn_{n\bk}^{0}(\bx) + |\bk|^2 \,|\ubarn^{0}_{n\bk}(\bx)|^2\Bigr)\derveps \left(\text{det}\left(\parder \bx \bx'\right)\right)\atzerob \Biggr\}\dx\dk \,,\notag
\end{split}\\
G_2 = \sum_{n=1}^{N}\fintd_{BZ} \int\displaylimits_{\Omegaper}  \fbar^{0}_{n\bk}\Biggl(-2\,i\,\uzeroconj {n\bk}(\bx)\derveps\bkappa^{\eps}(\bk)\atzerob\cdotp \del \ubarn^{0}_{n\bk}(\bx) + \derveps|\bkappa^{\eps}(\bk)|^2\atzerob |\ubarn_{n\bk}^{0}(\bx)|^2 \Biggr)\dx \dk \,,\notag
\end{gather}
\begin{gather}
G_3 = \sum_{n=1}^{N}\fintd_{BZ}\intd_{\Omegaper} \parder {g}{\fbar_{n\bk}^{\eps}} \derveps \fbar_{n\bk}^{\eps}\atzerob \dx \dk \,,\notag\\
\begin{split}
G_4 =   \sum_{n=1}^{N}\fintd_{BZ}\intd_{\Omegaper} &\Biggl(\parder {g}{\ubarn^{\eps}_{n\bk}}  \derveps (\ubarn_{n\bk}^{\eps}(\chiepsx))+\parder {g}{\uepsconj {n\bk}}  \derveps (\uepsconj {n\bk}(\chiepsx)) \\
&+ \parder {g}{\del \ubarn^{\eps}_{n\bk}} \derveps (\del \ubarn_{n\bk}^{\eps}(\chiepsx)) + \parder {g}{\del \uepsconj {n\bk}} \derveps (\del \uepsconj {nk} (\chiepsx))\Biggr)\atzerobb \dx \dk \,.\notag
\end{split}
\end{gather}
Further, the functional corresponding to the non-local pseudopotential energy~\eqref{Vnl} is treated as described below. We have
\begin{equation*}
\begin{split}
&\mathcal{L}^{\text{nl},\eps}(\bffbar^{\eps},\bUbar^{\eps}) = \\
&\sum_{n=1}^N \fintd_{BZ} \Biggl[\intd_{\Omegaper}\;\Biggl(\;\intd_{\Rthree} h(\fbar_{n\bk}^{\eps},\ubarn_{n\bk}^{\eps}(\chiepsx),\uepsconj {n\bk}(\chiepsx),\bkappa^{\eps}(\bk),V_{\text{nl}}(\chiepsx,\chieps(y),\chieps(\bR)))\,\text{det}\left(\parder {\by'} {\by}\right)\dy\Biggr)\text{det}\left(\parder {\bx'} {\bx}\right)\dx\Biggr]\dk\,,
\end{split}
\end{equation*}
where
\begin{equation*}
\begin{split}
&\intd_{\Rthree} h(\fbar_{n\bk}^{\eps},\ubarn_{n\bk}^{\eps}(\chiepsx),\uepsconj {n\bk}(\chiepsx),\bkappa^{\eps}(\bk),V_{\text{nl}}(\chiepsx,\chieps(y),\chieps(\bR)))\text{det}\left(\parder {\by'} {\by}\right)\dy \\
&= 2\,\fbar_{n\bk}^{\eps}\,\uepsconj {n\bk}(\chiepsx)\,e^{-i\bkappa^{\eps}(\bk)\cdotp\chiepsx} \intd_{\Rthree} V_{\text{nl}}(\chiepsx,\chieps(\by),\chieps(\bR))\,e^{i\bkappa^{\eps}(\bk)\cdotp\chieps(\by)}\,\ubarn_{n\bk}^{\eps}(\chieps(\by))\,\text{det}\left(\parder {\by'} {\by}\right)\dy\,.
\end{split}
\end{equation*}
Now, the G\^{a}teaux derivative of  $\mathcal{L}^{\text{nl},\eps}(,\bffbar^{\eps},\bUbar^{\eps})$ within the Kleinman-Bylander setting can be evaluated as:
\begin{equation}
\derveps \mathcal{L}^{\text{nl},\eps}(\bffbar^{\eps},\bUbar^{\eps}) \atzerob = H_{\text{nl}} + H_{\text{nl}}^{*} + H^{K}_{\text{nl}} + H_1 + H_2\,;\label{nonlocperiodic}
\end{equation}
\begin{align*}
H_{\text{nl}} 
&= 2 \sum_{n=1}^{N} \sum_{a,l,m}\fintd_{BZ} \intd_{\Omegaper}   \fbar^{0}_{n\bk} V_{lm}^{a} \sum_{r}\uzeroconj {n\bk}(\bx)e^{-i\bk\cdotp(\bx - \bL_r)}\Biggl\{\del (\zeta_{lm}^{a}(\bx, \bR_a + \bL_{r}) \Delta V_{l}^{a}(|\bx - (\bR_a + \bL_{r})|))\boldsymbol{\cdotp}\frac{d}{d\epsilon}\left(\chiepsx -\chieps (\bR_a + \bL_r)\right)\atzerob\\
&+ \zeta_{lm}^{a}(\bx,\bR_a + \bL_{r}) \Delta V_{l}^{a}(|\bx - (\bR_a + \bL_{r})|)\left(\derveps \left(\detb{\parder {\bx'} {\bx}}\right)\atzerob - i\bk\cdotp \frac{d}{d\epsilon}\Bigl(\chiepsx - (\chieps (\bR_a + \bL_r) - \chieps(\bR_a))\Bigr)\atzerob\right)\Biggr\}\dx\notag\\
&\times\Biggl[\intd_{\Omegaper}\sum_{s}e^{i\bk\cdotp(\by - \bL_s)}\zeta^{a}_{lm}(\by,\bR_a + \bL_{s})\Delta V_{l}^{J}(|\by - (\bR_a + \bL_{s})|)\ubarn^{0}_{n\bk}(\by) \dy\Biggr] \dk\,,
\end{align*}
\vspace{-0.1in}
\begin{gather}
\begin{split}
H^{K}_{\text{nl}} &= 2\,\sum_{n=1}^N \;\sum_{a,l,m}\fintd_{BZ} \fbar^{0}_{n\bk}V_{lm}^{a} \derveps \Biggl\{\Biggl[\int\displaylimits_{\Omegaper} \sum_{r} \uzeroconj {n\bk}(\bx)\,e^{-i\bkappa^{\eps}(\bk).(\bx - \bL_{r})}\, \zeta^{a}_{lm}(\bx, \bR_a+\bL_r) \,\Delta V_l^{a}(|\bx - (\bR_a + \bL_r)|) \dx\Biggr]\\
&\times \Biggl[\int \displaylimits_{\Omegaper}\sum_{s}e^{i\bkappa^{\eps}(\bk).(\by - \bL_{s})} \,\zeta^{a}_{lm}(\by,\bR_a+\bL_{s}) \,\Delta V_l^{a}(|\by - (\bR_a + \bL_{s})|)\,\ubarn^{0}_{n\bk}(\by) \dy\Biggr]\Biggr\}\atzerobb \dk\,,\notag
\end{split}\\
H_1 = \sum_{n=1}^{N}\fintd_{BZ}\intd_{\Omegaper} \intd_{\Rthree} \parder {h}{\fbar_{n\bk}^{\eps}} \derveps \fbar_{n\bk}^{\eps}\atzerob \dy \dx \dk \,,\notag\\
H_2 =  \sum_{n=1}^{N}\fintd_{BZ}\intd_{\Omegaper} \intd_{\Rthree} \Biggl(\parder {h}{\ubarn^{\eps}_{n\bk}}  \derveps (\ubarn_{n\bk}^{\eps}(\chiepsx))+\parder {h}{\uepsconj {n\bk}}  \derveps (\uepsconj {n\bk}(\chiepsx)) \Biggr)\atzerobb \dy \dx \dk \,.\notag
\end{gather}
We note that the evaluation of configurational force for terms other than $T_s$ and $\mathcal{L}^{\text{nl}}$ in the Kohn-Sham functional ~\eqref{minmaxper} is similar to the procedure described in the case of non-periodic calculations. To this end, we note that the sum of terms $G_4$, $H_2$ along with the the term $F_2$ arising in equation ~\eqref{derrhoTimesloc} (restricted to the case of orthonormal wavefunctions) vanish as $\bUbar^0$ is the solution of the saddle point variational problem corresponding to $\mathbb{E}_0(\btau^{0})$, and, hence, satisfy the Euler-Lagrange equations corresponding to equation ~\eqref{minmaxper}. Furthermore, the sum of the terms $G_3$, $H_1$ along with the terms $F_1$ arising in equation ~\eqref{derrhoTimesloc} , $\derveps E^{\eps}_{\text{ent}} \atzerob$ in equation ~\eqref{entroDer} and  $\derveps  \mathcal{L}_c^{\eps} \atzerob = 0$ in equation ~\eqref{constrDer} (restricted to the case of orthonormal wavefunctions) vanish, since $\fbar_{n\bk}^{0}$ satisfy the Fermi-Dirac distribution (Euler-Lagrange equation of $\mathbb{E}_0$  with respect to $f_{nk}$). Using the identities given in ~\eqref{identities} and rearranging the non-zero tems $G_1$, $G_2$, $H_{\text{nl}}$, $H_{\text{nl}}^{*}$, $H_{\text{nl}}^{K}$, $B_2$,$C_2$, $C_3$, $D_2$, $D_3$ and orthonormal counterparts of $A_{2b}$, $D_{5b}$ and $F_{3b}$, the G\^{a}teaux derivative of $\mathbb{E}_0(\chieps)$ can be written as:
\begin{equation}\label{EshelbyforceOrthPer1}
\begin{split}
 \frac{d\,\mathbb{E}_0(\chieps)}{d\varepsilon}\atzerob = \int\displaylimits_{\Omega_p} \bE:\del \bdir (\bx) \dx + \sum_{I} \int\displaylimits_{\Rthree} \bE'^I:\del \bdir (\bx) \dx \,
+\, \text{F}^{\text{PSP}} +  \text{F}^{K}\,,
\end{split}
\end{equation}
where $\bE$ and $\bE'$ denote Eshelby tensors whose expressions in terms of the solutions of the saddle point problem ~\eqref{minmaxper} ( $\bffbar^0$, $\bUbar^0$, $\varphibar^0$, $\mathcal{V}_{\dirac}^{0}$) are given in Section IV of the paper.
\end{widetext}

\begin{acknowledgments}
We gratefully acknowledge the support from the Department of Energy, Office of Basic Energy Sciences, under Award numbers DE-SC0008637 and DE-SC0017380 for supporting the pseudopotential and all-electron development, respectively. We are also grateful for the support from Toyota Research Institute, under the auspices of which the reduced order scaling work was performed. This work, in its initial stages, was supported by the National Science Foundation under Grant number 1053145. This work used the Extreme Science and Engineering Discovery Environment (XSEDE), which is supported by National Science Foundation Grant number ACI-1053575. This research used resources of the National Energy Research Scientific Computing Center, a DOE Office of Science User Facility supported by the Office of Science of the U.S. Department of Energy under Contract No. DE-AC02-05CH11231. We also acknowledge the Advanced Research Computing at University of Michigan for providing additional computing resources through the Flux computing platform.
\end{acknowledgments}


\end{document}